\documentclass[prd,superscriptaddress,nofootinbib]{revtex4-1}
\newcommand{\flip}{{\mathrm{flip}}}
\newcommand{\IR}{{\mathrm{\scriptscriptstyle IR}}}

\let\oldIm\Im
\renewcommand{\Im}{\mathop{\oldIm m}}
\usepackage{paralist}
\usepackage{float}
\usepackage{color}
\usepackage{amsmath,bm,bbm}
\usepackage{amssymb}
\usepackage{enumitem}
\usepackage{textcomp}
\usepackage{feynmp}
\usepackage{graphicx}
\usepackage{dsfont}
\usepackage{color}
\graphicspath{{./figures/}}
\usepackage[makeroom]{cancel}
\usepackage{hyperref}
\hypersetup{
	colorlinks=true,
	linkcolor=blue,
	filecolor=magenta,      
	urlcolor=blue,
	citecolor=blue,
}

\definecolor{purple}{rgb}{0.8,0,0.6}

\newcommand{\llangle}{\langle\!\langle}
\newcommand{\rrangle}{\rangle\!\rangle}
\newcommand{\Llangle}{\big<\!\big<}
\newcommand{\Rrangle}{\big>\!\big>}
\newcommand{\LLangle}{\Big<\!\Big<}
\newcommand{\RRangle}{\Big>\!\Big>}

\makeatletter
\renewcommand*{\thesection}{\arabic{section}}
\renewcommand*{\thesubsection}{\thesection.\arabic{subsection}}
\renewcommand*{\p@subsection}{}
\renewcommand*{\thesubsubsection}{\thesubsection.\arabic{subsubsection}}
\renewcommand*{\p@subsubsection}{}
\renewcommand*{\theparagraph}{\thesubsubsection.\arabic{paragraph}}
\renewcommand*{\p@paragraph}{}
\renewcommand*{\thesubparagraph}{\theparagraph.\arabic{subparagraph}}
\renewcommand*{\p@subparagraph}{}
\makeatother

\renewcommand\appendix{\par
  \setcounter{section}{0}%
  \setcounter{subsection}{0}%
  \makeatletter
  \renewcommand*{\thesection}{\Alph{section}}
  \renewcommand*{\thesubsection}{\thesection.\arabic{subsection}}
  \renewcommand*{\thesubsubsection}{\thesubsection.\arabic{subsubsection}}
  \renewcommand*{\theparagraph}{\arabic{paragraph}}
  \renewcommand*{\thesubparagraph}{\theparagraph.\arabic{subparagraph}}
  \makeatother
}

\begin{document}
\title{Equilibration of the chiral asymmetry due to finite electron mass \\ in electron-positron plasma}

\author{A.~Boyarsky}
\affiliation{Instituut-Lorentz for Theoretical Physics, Universiteit Leiden, Niels Bohrweg 2, 2333 CA Leiden, The Netherlands}

\author{V.~Cheianov}
\affiliation{Instituut-Lorentz for Theoretical Physics, Universiteit Leiden, Niels Bohrweg 2, 2333 CA Leiden, Netherlands}

\author{O.~Ruchayskiy}
\affiliation{Niels Bohr Institute, University of Copenhagen, Blegdamsvej 17, DK-2100 Copenhagen, Denmark}

\author{O.~Sobol}
\affiliation{Institute of Physics, Laboratory for Particle Physics and Cosmology, \'{E}cole Polytechnique F\'{e}d\'{e}rale de Lausanne, CH-1015 Lausanne, Switzerland}
\affiliation{Physics Faculty, Taras Shevchenko National University of Kyiv, 64/13, Volodymyrska Str., 01601 Kyiv, Ukraine}
\email{oleksandr.sobol@epfl.ch}

\begin{abstract}
  We calculate the rate of collisional decay of the axial charge in an ultrarelativistic electron-positron plasma, also known as the chirality flipping rate.
  We find that contrary to the existing estimates, the chirality flipping rate appears already in the first order in the fine-structure constant $\alpha$ and is therefore orders of magnitude greater than previously believed. The main channels for the rapid relaxation of the axial charge are the collinear emission of a weakly damped photon and the Compton scattering. The latter contributes to the $\mathcal{O}(\alpha)$ result because of the infrared divergence in its cross section, which is regularized on the soft scale $\sim eT$ due to the thermal corrections.
  Our results are important for the description of the early Universe processes (such as leptogenesis or magnetogenesis) that affect differently left- and right-chiral fermions of the Standard Model, as discussed in more details in the companion Letter.
\end{abstract}

\maketitle

\section{Introduction}
\label{sec:introduction}

An axially-charged electron-positron plasma coupled to a large-scale
magnetic field provides a remarkable example of a system whose macroscopic collective motion evinces a purely quantum phenomenon, the axial gauge anomaly. 
Due to the presence of the anomaly, the hydrodynamics of such a plasma contains 
an unusual collective degree of freedom, which is not locally connected to the 
thermodynamic variables characterizing the local equilibrium \cite{Joyce:1997uy,Boyarsky:2011uy,Rogachevskii:2017uyc,Giovannini:2013oga,DelZanna:2018dyb}. Such a deformation of the equations of hydrodynamics 
gives rise to new types of macroscopic behavior such as the chiral magnetic effect 
or the inverse magnetic cascade \cite{Joyce:1997uy,Boyarsky:2011uy,Tashiro:2012mf,Hirono:2015rla,Dvornikov:2016jth,Gorbar:2016klv,Brandenburg:2017rcb,Schober:2018wlo}.
These phenomena can play an important role in the context of leptogenesis and cosmic magnetogenesis \cite{Joyce:1997uy,Frohlich:2000en,Frohlich:2002fg,Semikoz:2004rr,Semikoz:2009ye,Boyarsky:2011uy,Boyarsky:2012ex,Semikoz:2012ka,Tashiro:2012mf,Dvornikov:2011ey,Dvornikov:2012rk,Dvornikov:2013bca,Dvornikov:2016jth,Manuel:2015zpa,Gorbar:2016klv,Pavlovic:2016mxq,Pavlovic:2016gac}, for magnetic field evolution in primordial plasma \cite{Joyce:1997uy,Boyarsky:2011uy,Tashiro:2012mf,Dvornikov:2016jth,Gorbar:2016klv,Brandenburg:2017rcb,Schober:2018wlo}, as well as in the heavy-ion collisions and quark-gluon plasma \cite{Akamatsu:2013pjd,Taghavi:2013ena,Hirono:2015rla} and in the neutron stars \cite{Ohnishi:2014uea,Dvornikov:2014uza,Dvornikov:2015lea,Sigl:2015xva,Yamamoto:2016xtu,Dvornikov:2016cmz}. Analogous effects are also known in condensed matter theory; see Ref.~\cite{Miransky:2015ava} for a review. 

Unlike the electric charge, the axial charge is not fundamentally conserved. Typically, 
there exist scattering events which lead to chirality flipping and equilibration 
of the opposite chirality components. For this reason the applicability conditions of 
chiral hydrodynamics depend on the kinetic rate describing chirality flipping.
In quantum electrodynamics (QED) and other weakly coupled quantum theories transport coefficients and kinetic rates are, in principle, amenable to calculation by means 
of perturbation theory. Computations of transport coefficients in high-temperature QED (more generally, in Standard Model) plasmas have been the subject of active research for many years (see, e.g., Refs.~\cite{Hosoya:1983xm,Chakrabarty:1986xx,vonOertzen:1990ad,Blaizot:1992gn,Heiselberg:1994vy,Baym:1995fk,Ahonen:1996nq,Baym:1997gq,Ahonen:1998iz,Arnold:2000dr,Arnold:2003zc,Hou:2017szz};
see Refs.~\cite{Arnold:2000dr,Arnold:2003zc,Thoma:2008my} for review and discussion). Notwithstanding the broad scope of this theoretical effort, rigorous methods have never 
been applied to the calculation of the kinetic coefficient associated with the collisional decay of the axial charge---the chirality flipping rate, $\Gamma_{\flip}.$ 

In our accompanying Letter~\cite{PaperI}, we use Boltzmann's kinetic theory to investigate the collisional decay of the axial charge in an electron-positron plasma at temperatures below the electroweak crossover and much greater than the electron mass. We demonstrate the existence of scattering channels which contribute to the chirality flipping rate in the first order of the fine-structure constant $\alpha,$ despite their nominal perturbative order being $\alpha^2.$ 
This happens due to the power law infrared divergence in the process's matrix element, which gets regularized on the soft scale $q_{\IR}\sim eT$. We also provide qualitative arguments as to why apart from the $2\leftrightarrow 2$ processes, there also might exist nearly collinear $1\leftrightarrow 2$ processes also contributing to the same order. In the present paper, we perform a comprehensive investigation of all scattering mechanisms
contributing to the chirality flipping rate to the first order in $\alpha.$ The main result of our analysis is 
the leading-order (in $\alpha$) 
asymptotic expression for the chirality flipping rate summarized in Eq.~\eqref{gamma-flip-final} and the rate equation~\eqref{rate-eq-final}, where we used 
the numerical value of the fine structure constant in 
all logarithmic terms.

The structure of the article is as follows.
In Sec.~\ref{sec-statement} we state the problem and qualitatively discuss the scattering processes contributing to chirality flipping rate at leading order.
In Sec.~\ref{sec-linear-response} we introduce the diagrammatic formalism which we then use for the perturbative calculation of the chirality flipping rate. 
We investigate the structure of infrared divergences 
in the leading-order terms of the perturbative 
expansion and then perform partial resummation of the leading singularity in all orders of perturbation theory.
In Sec.~\ref{sec-concl} we summarize all our results.
Appendices~\ref{app-4-fermion}--\ref{app-details} provide the details of the computations.

\section{Statement of the problem and preliminary discussion}
\label{sec-statement}

We consider a hot 
QED plasma in which the typical kinetic energy of 
an electron is ultrarelativistic. In this regime,
each electron can be characterized by an almost 
conserved chirality quantum number, which is defined as 
an eigenvalue of the operator $\gamma_5.$\footnote{We define $\gamma_5 = i\gamma^0\gamma^1 \gamma^2 \gamma^3$ and $\gamma^{0,\dots, 3}$ are Dirac gamma-matrices, see any QED textbook for details.}
In purely massless QED the conservation of the chirality of each individual particle is reflected in the existence of the conserved gauge invariant charge operator \cite{Alekseev:1998ds,Frohlich:2000en,Frohlich:2002fg}
\begin{equation}
\label{relation-Q5-N5}
\hat{Q}_{5}=\int\! d^{3}\mathbf{x}\,
\hat{j}^{0}_{5} +\frac{e^{2}}{4\pi^{2}}\int\! d^{3}\mathbf{x}\, \hat{\mathbf{A}}\cdot\hat{\mathbf{B}},
\end{equation}
where the first term is constructed from the axial current
\begin{equation}
\hat{j}^{\mu}_{5}=\ :\bar{\psi}\gamma^{\mu}\gamma^{5}\psi :,
 \label{eq:j5}
\end{equation}
while the second term arises from the axial gauge anomaly \cite{bertlmann2000anomalies}. 

For a nonvanishing fermion mass the charge $\hat{Q}_{5}$ is no longer conserved and therefore its expectation value tends to relax from any initial nonequilibrium state to its equilibrium value $Q_{5}=0$. 
In a weakly nonequilibrium situation it obeys the general near-equilibrium decay law
\begin{equation}
\label{eq-flip}
\dot Q_5 = - \Gamma_{\rm flip} Q_5,
\end{equation}
where $Q_{5}=\llangle\hat{Q}_{5} \rrangle$ and $\llangle\ldots \rrangle$ stands for the quantum statistical ensemble average of the corresponding operator. The kinetic rate $\Gamma_{\rm flip}$ which appears in Eq.~(\ref{eq-flip}) is the main subject of the present work.

The chirality flipping rate has never been calculated 
from first principles, however the following simple-minded estimate has been used in a number of works (see, e.g., Refs.~\cite{Boyarsky:2011uy,Grabowska:2014efa,Manuel:2015zpa,Pavlovic:2016mxq}). Consider an ultrarelativistic particle moving with a certain momentum in a given helicity state. Note, that the eigenstate of helicity coincides with the eigenstate of chirality up to $(m_e/E)$ corrections. 
The helicity can only change in an act of collision, which changes the direction of the particle's momentum, therefore
\begin{equation}
  \label{eq:flip_naive}
  \Gamma_\flip^{\rm naive} \propto \frac{m_e^2}{\langle p^2\rangle} \Gamma_{\rm scat} \propto \alpha^2 \frac{m_e^2}{9T}
\end{equation}
where $\Gamma_{\rm scat} \propto \alpha^2 T$ is the (chirality-preserving) electron scattering rate in plasma with $T \gg m_e$ and $\alpha = e^2/(4\pi)$ is the fine structure constant. As we shall show momentarily, this crude estimate neglects some crucial aspects of the analytic structure of the scattering amplitudes and for that reason falls orders of magnitude short of the correct 
prediction.

\begin{figure}[t!]
	\centering
	$(a)$\includegraphics[height=3.5cm]{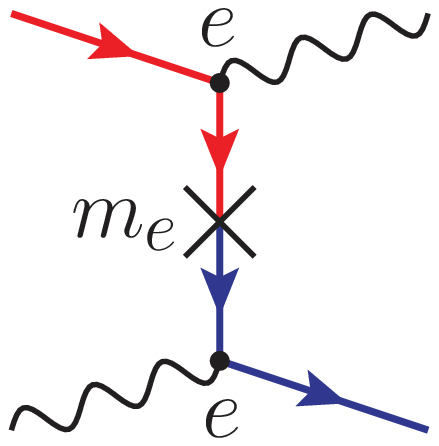} \hspace*{1cm}
	$(b)$\includegraphics[height=3.5cm]{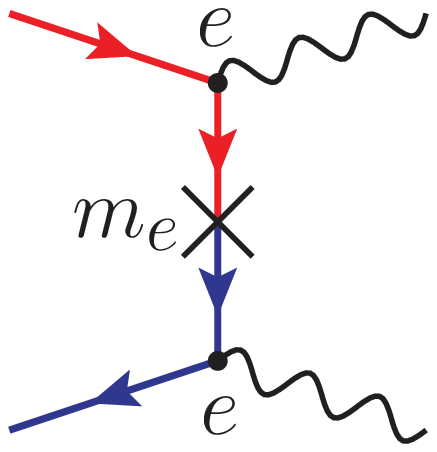}
	\caption{The $t$-channel Compton scattering (a) and electron-positron annihilation (b) with the chirality flip in the intermediate state contributing to the chirality equilibration rate. Although naively they are of the second order in $\alpha$, their amplitudes contain infrared singularities. Regularization of these singularities leads to the result which is of the first order in $\alpha$. \label{fig-Compton}}
\end{figure}

\textit{Infrared singularity in $2\leftrightarrow 2$ chirality flipping processes.}---The problem of the decay of the axial charge from the perspective of elementary scattering processes was addressed in our companion Letter~\cite{PaperI}. Here we briefly re-iterate 
its argument. Consider $2\leftrightarrow 2$ chirality changing processes, starting from the massless QED and treating mass as perturbation
(see Fig.~\ref{fig-Compton}).
Such processes have a nonintegrable infrared singularity at small momentum transfer~\cite{Lee:1964is,Dolgov:1971ri}.
This signals out that a partial resummation of the perturbative 
series is required, which should result in the answer
\begin{equation}
  \Gamma_\flip \propto \alpha^2 T \frac{m_e^2}{q_\IR^2},
  \label{eq:1}
\end{equation}
where $q_\IR$ is the infrared regulator scale.
In vacuum, the only possible resummation is the $m_{e}$ series, cutting the infrared divergence at $q_{\IR}\sim m_{e}$.
This would cancel the $m_{e}^{2}$ factor in the numerator and lead to a strange result that the chirality flipping rate is mass-independent.
However in a hot medium we also have to take into account the thermal corrections to the dispersion relations of (quasi)particles.
Performing hard thermal loops (HTL) resummation (see Appendix~\ref{app-self-energy}) we find that these corrections appear at the order $eT$.
If the temperature is high enough, $T\gg m_{e}/e$, their effect is much more significant than that of the zero-temperature mass.
With $q_\IR \sim eT$ we arrive to
\begin{equation}
  \label{eq:flip_PaperI}
  \Gamma_\flip \propto \alpha^2 T\frac{m_e^2}{(eT)^2} = C \times \alpha \frac{m_e^2}T  
\end{equation}
where the numerical coefficient $C \approx 0.24$.
Below, we will see that the $2\leftrightarrow 2$ processes are not the only ones that can lead to $\mathcal{O}(\alpha)$  chirality flipping rate.

\begin{figure}[h!]
	\centering
	$(a)$\includegraphics[width=5.cm]{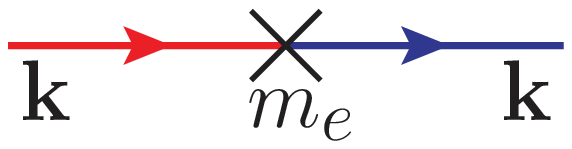} \hspace*{1cm}
	$(b)$\includegraphics[width=5.5cm]{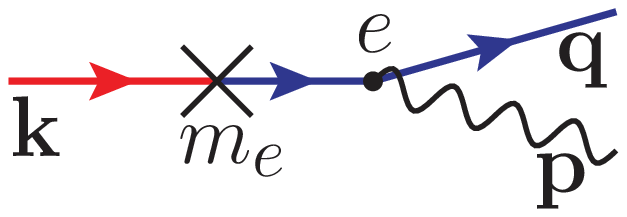}
	\caption{Processes with the flip of chirality in the lowest orders of the double asymptotic expansion in $m_{e}/T$ and $e$ (states with different chiralities are shown in different colors): (a) $1\leftrightarrow 1$ process with chirality flip of a free electron which is forbidden by the angular momentum conservation unless the particle's momentum vanishes; (b) one of $1\leftrightarrow 2$ collinear processes with chirality flip of the incoming electron. Although for massless particles the process (b) has a finite phase space, it is very sensitive to any modification of the particles' dispersion relations. This leads to an uncertainty in the probability of such a process which will be resolved in Sec.~\ref{sec-linear-response}. \label{fig-lowest}}
\end{figure}

\textit{$1\leftrightarrow 1$ processes with chirality flip.}---The only process of the order $\mathcal{O}(\alpha^0)$ is the $1\leftrightarrow 1$ process shown 
in Fig.~\ref{fig-lowest}(a). In order to better understand its contribution, we first recall the properties of propagating states of a free massless chiral fermion.
For given chirality, a propagating state is described by a 2-component spinor satisfying the Weyl equation. The corresponding propagator is a $2\times 2$ matrix which can be decomposed into a sum of two helicity projectors. For instance, the propagator of the right fermion reads as
\begin{equation}
\label{propagator-free}
    \mathcal{S}_{0\,R}(k^{0},\mathbf{k})=\frac{1+\boldsymbol{\sigma}\cdot\hat{\mathbf{k}}}{2}\frac{1}{k^{0}-|\mathbf{k}|}+\frac{1-\boldsymbol{\sigma}\cdot\hat{\mathbf{k}}}{2}\frac{1}{k^{0}+|\mathbf{k}|},
\end{equation}
while for the left fermion one must interchange the helical projectors. Here $\hat{\mathbf{k}}=\mathbf{k}/|\mathbf{k}|$. As usual, the zeros of the denominator give the dispersion relations (on-shell conditions) for the real propagating particles. It is straightforward to see that for the propagating positive-energy states, $k^{0}=|\mathbf{k}|,$ the chirality of an electron is correlated with its helicity, i.e., the right-handed electron has positive helicity and the left-handed one has the negative helicity. Furthermore, for the negative-energy states the helicity is anticorrelated
with chirality. In contrast, as is obvious from Eq.~\eqref{propagator-free}, e.g., a right-handed electron having negative helicity cannot propagate at positive energy. It can only exist as a virtual state. 
We shall henceforth refer to such virtual states obeying the $k^2=0$ condition as off-shell states. 

We now consider the mass vertex in Fig.~\ref{fig-lowest}. The mass term imposes 
the following selection rules: (a) The momentum of 
the in-state coincides with the momentum of the out-state. (b) The chirality of the out-state 
is opposite to the chirality of the in-state. (c)
The helicity of the out-state is the same as the helicity of the in-state. 
In view of the explanations given above, the outgoing particle appears to be off shell, because its energy is equal to $k^{0}=|\mathbf{k}|$ while the on-shell condition imposed by its spin state is $k^{0}=-|\mathbf{k}|$. The only special case is 
the chirality flip of the $\mathbf k = 0$ state. 
In this case, negative energy states are 
degenerate with the positive energy states, 
therefore a chirality flip is permitted as a real 
process. We conclude
that while the chirality flip process in the $1\to 1$ channel is possible, its contribution to the chirality relaxation rate vanishes because the 
momentum space for such a process has vanishing measure.

\textit{$1\leftrightarrow 2$ processes contributing to the flipping rate.}---The situation gets more subtle for $1\leftrightarrow 2$ processes, one of which is shown in Fig.~\ref{fig-lowest}(b). 
Naively, such processes are forbidden in massive QED as they do not allow for simultaneous energy and momentum conservation. However as we discuss the relaxation of chiral asymmetry,
we use the massless QED as zero approximation and take mass into account perturbatively.
 The QED interaction vertex respects the chiral symmetry therefore 
the minimal diagram has to include one mass vertex insertion in the incoming or outgoing electron leg. 
At the tree level, the conservation of the four-momentum imposes the constraint that the momenta of the incoming and the outgoing particles be perfectly collinear. At the same time, selection rules impose two consecutive flips: chirality flip at the mass vertex and helicity flip at the 
electron-photon vertex. This results in a valid on-shell out state of an electron with consistent signs of chirality and helicity, even though in the intermediate virtual state between the vertices 
such a consistency is absent.
Of course, in vacuum, we would have to take into account  the corrections that are higher order in mass. This would restore massive theory and make $1\leftrightarrow 2$ processes impossible again. The effects of medium change the situation as the thermal contributions are 
larger than mass. 

Despite the very stringent constraint of 
perfect collinearity of the momenta of all involved particles, the phase space for such a process has a finite volume
\begin{equation}
    \Omega \propto \int d^3\mathbf{q} \,d^3 \mathbf{p}\, \delta^{(3)}(\mathbf{k} - \mathbf{q} - \mathbf{p}) 
\delta(\epsilon_k- \epsilon_q-\omega_p)
= 
\frac{\pi k^2}{3}, 
\end{equation}
where $\epsilon_k=|\mathbf{k}|$, and $\omega_p = |\mathbf{p}|$. This result, however, is physically unstable. 
Even the slightest deformation of the dispersion relations 
of the particles $\epsilon\to \epsilon +\delta \epsilon$,
$\omega \to \omega + \delta \omega$, 
for example due to the mean-field effects of the environment, 
can completely wipe out the phase space available for the process. It is worth noting that due to the infrared divergence of the propagator describing 
the intermediate virtual state, the probability of the process shown 
at Fig.~\ref{fig-lowest}(b) diverges at small momentum of the incoming particle. 
This infrared singularity is not integrable 
therefore its resolution requires a unitarity-restoring resummation of the 
perturbation theory series. Such a resummation cannot be done consistently without modifying the dispersion relations of the particles and the resulting suppression of the phase space.
Thus, we arrive at an uncertainty of the $0\times \infty$ type which we shall resolve in the following sections.

\begin{figure}[h!]
	\centering
	$(a)$\includegraphics[width=5.cm]{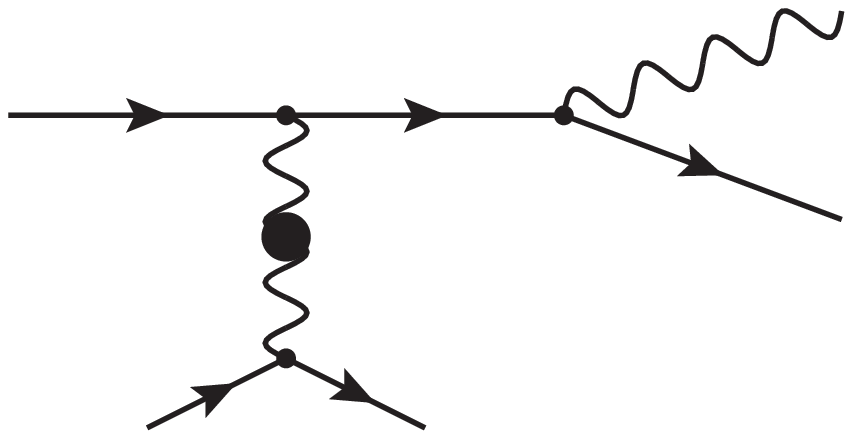} \hspace*{1cm}
	$(b)$\includegraphics[width=5.5cm]{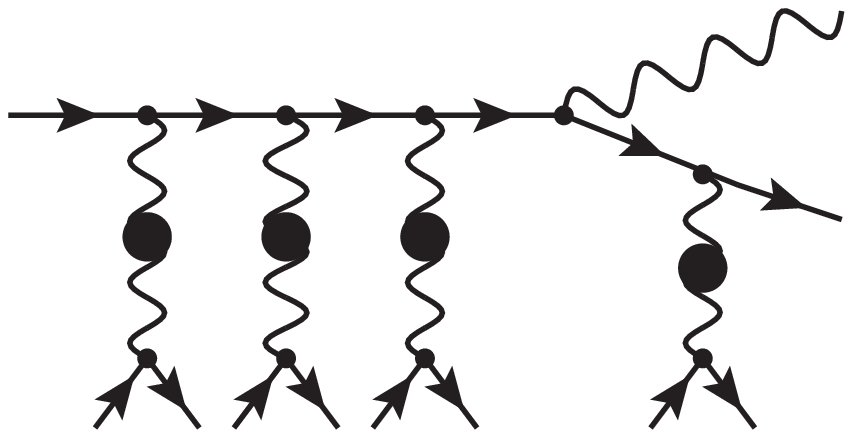}
	\caption{(a) Process of collinear bremsstrahlung in massless QED plasma at leading order in $\alpha$. The blob denotes the virtual photon with soft momentum $q\sim eT$. This soft scattering is essential for angular momentum conservation. (b) Multiple soft scatterings accompanying the collinear bremsstrahlung. The formation time of a hard nearly collinear photon is of the same order as the characteristic soft scattering time $\tau\propto (e^{2}T)^{-1}$. That is why the calculation of the full photon emission amplitude requires the coherent resummation of the infinite series of such multiple-scattering graphs (the Landau-Pomeranchuk-Migdal effect). \label{fig-LPM}}
\end{figure}

It is instructive to  compare
the $1\to 2$ chirality flipping process 
discussed here with the nearly collinear 
photon emission in a QED plasma of massless fermions. The latter process is an example of bremsstrahlung caused by the accelerated motion of an electron due to its scattering 
off the thermally fluctuating electromagnetic background. At the tree level the rate of such processes is nominally on the order $\alpha^3$ [see Fig.~\ref{fig-LPM}(a)], however due to a nonintegrable divergence of the matrix element at small scattering angles, the actual order of this 
process is $\alpha^2$. 
The calculation of the correct numerical value of the photon emission rate is a complicated task due to the accumulation of divergent contributions representing multiple low-angle scattering events shown in Fig.~\ref{fig-LPM}(b) and due to the interference between such contributions 
known as the Landau-Pomeranchuk-Migdal effect Refs.~\cite{Baier:2000mf,Kovner:2003zj,Aurenche:2000gf,Arnold:2001ba,Arnold:2002ja,Arnold:2002zm}. 
In contrast to bremsstrahlung, the $1\to 2$ chirality flipping process discussed here is an example of resonant Cherenkov-like emission from a particle moving at the speed of light. 
Unlike the massless QED plasma, where the perfectly collinear emission is prohibited by the 
combination of the angular momentum conservation law and chirality conservation, in a theory which treats mass as a small perturbation
collinear emission has a nonvanishing tree-level rate that has the nominal order $m_{e}^{2} \alpha.$ Perturbative corrections to the tree-level result contain two types of effects: (a) Deformation of the 
dispersion relations of the colliding particles due to the mean-field 
thermal background; (b) finite lifetime of each involved particle due to real scattering events. While the effects of type (a) tend to suppress the phase volume of collinear emission, the 
effects of type (b) compensate for that suppression allowing for a weak violation of energy conservation in Fermi's golden rule.
It turns out that contrary to the bremsstrahlung case both types of effects can be completely absorbed into the self-energy corrections to the single-particle propagators, at least to the leading order in $\alpha.$ The effects of the LPM interference only emerge in the first subleading $\alpha$ order. On a technical level, this 
difference between the nearly collinear bremsstrahlung and the photon-assisted chirality 
flip is due to the presence of a virtual intermediate state of an electron in the latter case, while an electron always remains near 
its classical trajectory in the former. 
A detailed comparison of the perturbative computation of the photon emission rate vs the chirality flipping rate with particular attention to the multiple soft scatterings in plasma is given in Appendix~\ref{app-LPM}.

To conclude, the tree-level kinetic approach to the problem of chirality flipping is fraught with difficulties arising from an assortment of light-cone and phase space singularities. These singularities should presumably be resolved by a partial resummation 
of the perturbation theory series. We now proceed to the next 
section where we introduce a formalism which gives $\Gamma_{\rm flip}$ in the form of a diagrammatic perturbation theory series.
We then use the formalism to perform the required partial resummation. 

\bigskip

\section{Chirality flipping rate from the linear response formalism}
\label{sec-linear-response}

In this section, we use first principle quantum mechanical description to develop a perturbative expansion of 
the chirality flipping rate and identify in it 
the contributions from $1\leftrightarrow1$, $1\leftrightarrow2$ and $2\leftrightarrow2$ processes described in Sec.~\ref{sec-statement}.
We demonstrate again that these perturbative contributions 
are IR-divergent. Then, in subsection~\ref{ssec-thermal} we perform a partial resummation of the perturbative expansion, 
thus taking into account that fermion and photon propagators are dressed with thermal corrections and show that this results in resolution of IR-divergences.

We use massless electrodynamics as the zeroth-order approximation. 
In the massless case, the chiral charge $\hat{Q}_{5}$, Eq.~\eqref{relation-Q5-N5}, is conserved. This means that the expectation value of its time derivative (treated as the quantum operator) vanishes, $\llangle d\hat{Q}_{5}/dt \rrangle_{0}=0$.
In the presence of a finite mass the time evolution of the axial charge operator is described by the Heisenberg equation:
\begin{equation}
i\frac{d\hat{Q}_{5}}{dt}=\left[\hat{Q}_{5},\,H\right]=\left[\hat{Q}_{5},\,H_{m}\right],
\end{equation}
where $H_{m}=m_{e}\int d^{3}\mathbf{x}\ \bar{\psi}\psi$ is the chiral symmetry breaking term. Using the canonical anticommutation relations for spinor field $\{\psi_{\alpha}(t,\mathbf{x}),\psi^{\dagger}_{\beta}(t,\mathbf{y})\}=\delta_{\alpha\beta}\delta^{(3)}(\mathbf{x}-\mathbf{y})$ one can see that
\begin{equation}
\label{Heisenberg-eq}
\frac{d\hat{Q}_{5}}{dt}=2im_{e}\int\!d^{3}\mathbf{x}\, \bar{\psi}\gamma^{5}\psi.
\end{equation}
Thus, in the presence of mass, operator $d\hat{Q}_{5}/dt$ acquires the nonzero expectation value which can be calculated in the lowest order in perturbation by means of the linear response formalism (see, e.g., textbook in Ref.~\cite{Mahan-book}, Chap.~3).
It should 
be borne in mind, however, that massless electrodynamics
does not contain an intrinsic length scale, 
therefore the actual dimensionless parameter containing 
the electron mass is the ratio $m_e/T$, where $T$ is the temperature 
of the system. In the high-temperature regime considered here, 
this parameter is small therefore its use as an asymptotic 
parameter of perturbative expansion is appropriate.

Another small parameter of electrodynamics is the electromagnetic coupling constant $\alpha=e^{2}/(4\pi).$ As discussed above, the $\alpha\to 0$ limit of massive electrodynamics does not admit for chirality flipping processes, even though the axial charge is not formally conserved.
For this reason, we also need to consider the effects of QED interaction beyond the zeroth-order perturbation theory.
To summarize, in the analysis of the relaxation of axial charge it is natural to make use of the double asymptotic expansion in two small parameters, the fine structure constant and in the dimensionless parameter $m_{e}/T$.
Following through the standard steps described in detail in Appendix~\ref{app-4-fermion}, we arrive at the following expression for the rate of change of the axial charge per unit volume:
\begin{equation}
\label{chirality-flipping-rate}
\dot{q}_{5}\equiv \frac{1}{V}\dot{Q}_{5}=-4m_{e}^{2}\Im\left[G_{\rm ret}(\omega=-2\mu_{5},\mathbf{q}=0)\right],
\end{equation}
where $G_{\rm ret}(\omega,\mathbf{q})$ is the momentum space retarded Green's function which is defined as follows:
\begin{equation}
\label{retarded-Green-function}
G_{\rm ret}(\omega,\mathbf{q})=-i\int d^{4}x\, e^{i\omega t-i\mathbf{q}\cdot\mathbf{x}}\theta(t)\Llangle\!\left[\bar{\Psi}_{L}(t,\mathbf{x})\Psi_{R}(t,\mathbf{x}),\,\bar{\Psi}_{R}(0,0)\Psi_{L}(0,0)\right]\!\Rrangle.
\end{equation}
Here $\Psi_{R,L}=P_{R,L}\Psi$ are the field operators of right- or left-handed fermions in the Heisenberg representation of the Hamiltonian (\ref{Hamiltonian-great-canonical}) of massless QED in the great canonical ensemble, $P_{R,L}=(1\pm \gamma^{5})/2$ are the right and left chiral projectors. Note that the expectation value in Eq.~(\ref{retarded-Green-function}) is taken over the state of massless QED.
Using the near-equilibrium relation $q_{5}=\mu_{5}T^{2}/3$, we arrive to the rate equation for the axial charge density, similar to Eq.~(\ref{eq-flip}) with  
the microscopic expression for the kinetic coefficient $\Gamma_{\rm flip}$
given by
\begin{equation}
\label{eq-chirality-flip}
\boxed{\Gamma_{\rm flip}=\frac{12m_{e}^{2}}{T^{2}}\left[\left.\frac{\partial}{\partial \mu_{5}}\Im G_{\rm ret}(-2\mu_{5},0)\right]\right|_{\mu_{5}=0}.}
\end{equation}
We would like to note that here we consider the weakly nonequilibrium situation, $\mu_5 \ll T$, that is why it is appropriate to set $\mu_{5}=0$ after taking the derivative in Eq.~(\ref{eq-chirality-flip}).

The retarded Green's function at a finite temperature can be calculated by analytic continuation of the corresponding Matsubara function $\mathcal{G}(i\Omega_{n},\mathbf{q}=0)$ to the real axis $i\Omega_{n}\rightarrow \omega+i\delta$, with $\delta\to 0^{+}$. Therefore, in the next section we will consider the perturbative expansion for the following Matsubara correlation function:
\begin{equation}
\label{loop-correlator}
\mathcal{G}(i\Omega_{n},\mathbf{q}=0)=-\int_{0}^{\beta}\!d\tau\ e^{i\Omega_{n}\tau}\int d^{3}\mathbf{x}\Llangle T_{\tau}\bar{\Psi}_{L}(\tau,\mathbf{x})\Psi_{R}(\tau,\mathbf{x})\bar{\Psi}_{R}(0,0)\Psi_{L}(0,0)\Rrangle.
\end{equation}
Here $\Omega_{n}=2\pi n T$ with $n\in \mathds{Z}$ is the bosonic Matsubara frequency and $T_{\tau}$ denotes the chronological ordering in the imaginary time.

\subsection{Perturbative expansion for the chirality flipping rate \label{sec-perturbative}}

In this subsection we investigate the perturbative expansion in $\alpha$ for the chirality flipping rate~(\ref{eq-chirality-flip}). In particular, in Sec.~\ref{subsec-zero-order} we show that in the absence of QED interactions fast oscillations of the chirality of each given electron are statistically averaged in the ensemble leading to a constant chiral charge, i.e., the relaxation of the chiral imbalance is absent. Then, in Sec.~\ref{subsec-first-order} we show that in the first order in $\alpha$ there is a nontrivial contribution to the chirality flipping rate while the result contains infrared and collinear divergences. Further, in Sec.~\ref{subsec-second-order} we consider all topologically nontrivial graphs for the chirality flipping rate at the second order in $\alpha$ and determine the classes of diagrams which contain accumulating infrared divergences and, thus, require resummation. In Sec.~\ref{subsec-higher-orders} we use the power counting to find out how rapidly do these divergences accumulate in higher perturbative orders. We also show that all of them can be absorbed into the self-energy renormalization of the fermion propagator which is finally performed in Sec.~\ref{subsec-self-energy-resummation}. We would like to emphasize that the diagrams corresponding to vertex corrections do not accumulate divergences and do not have to be resummed (see also Appendix~\ref{app-LPM} for the analysis of this class of diagrams). Nevertheless, there is a finite contribution from the vertex correction diagram appearing at the first perturbative order which must be included in the final result for the chirality flipping rate which is done in Sec.~\ref{ssec-thermal}.

\subsubsection{Zeroth order}
\label{subsec-zero-order}
As was discussed in Sec.~\ref{sec-statement}, chirality relaxation does not occur in the $\alpha=0$ limit even 
though the mass term does not formally conserve the chirality quantum number. This can be understood on the basis 
of kinetic considerations, see our discussion of 
Fig.~\ref{fig-lowest}(a). Alternatively, one can 
notice that if a thermal electron is prepared in a given chirality state, the mass term will only 
lead to rapid low-amplitude oscillations of the expectation value of chirality around a mean value,
which is close to the initial value of chirality. Due to the frequency of the oscillations being a function of an electron's energy, thermal averaging over the momentum states of all particles wipes the oscillations out resulting in a time-independent mean value of the axial charge. This implies that the right-hand side of Eq.~\eqref{chirality-flipping-rate} should vanish in the zeroth order in $\alpha$ and so should 
$\Gamma_{\rm flip},$ Eq.~\eqref{eq-chirality-flip}.

It is instructive to see how this result follows 
directly from a diagrammatic calculation.
In the $\alpha =0 $ limit the correlation function~\eqref{loop-correlator} decomposes into a product of two electron Green's functions, one for a left-handed and one for a right-handed particle 
\begin{multline}
\Llangle T_{\tau}\bar{\Psi}(\tau,\mathbf{x})P_{R}\Psi(\tau,\mathbf{x})\bar{\Psi}(0,0)P_{L}\Psi(0,0)\Rrangle_{0}\\
=-\Llangle T_{\tau}\Psi_{\sigma}(0,0)\bar{\Psi}_{\alpha}(\tau,\mathbf{x})\Rrangle_{0}\Llangle T_{\tau}\Psi_{\beta}(\tau,\mathbf{x})\bar{\Psi}_{\rho}(0,0)\Rrangle_{0}\left(P_{R}\right)_{\alpha\beta}\left(P_{L}\right)_{\rho\sigma}\\
=-{\rm tr}\left[P_{L}\mathcal{S}_{0}(-\tau,-\mathbf{x})P_{R}\mathcal{S}_{0}(\tau,\mathbf{x})\right].
\end{multline}

\begin{figure}[h!]
	\centering
	\includegraphics[width=5cm]{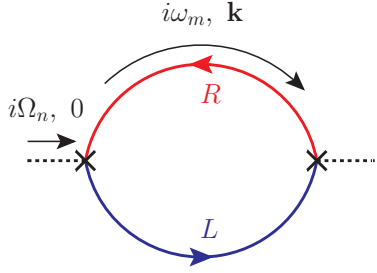}
	\caption{Feynman diagram for the chirality flipping rate in zeroth order in $\alpha$. Its imaginary part is expressed as an overlap of the free electron spectral densities with the same momentum and different chiralities. For nonzero momenta they never overlap so that the diagram vanishes. Cutting this diagram we get the amplitude of $1\leftrightarrow 1$ process with the chirality flip discussed in Sec.~\ref{sec-statement}. \label{fig-0-order}}
\end{figure}

In Fourier space, this corresponds to the Feynman diagram shown in Fig.~\ref{fig-0-order}, and the final result reads as 
\begin{equation}
\label{G-zero order}
\mathcal{G}(i\Omega_{n},\mathbf{q}=0)=-\frac{1}{2}\int\!\!\frac{d^{3}\mathbf{k}}{(2\pi)^{3}}\sum_{\lambda=\pm}\frac{{\rm tanh\,}\frac{\lambda k-\mu_{L}}{2T}-{\rm tanh\,}\frac{-\lambda k-\mu_{R}}{2T}}{i\Omega_{n}+2\lambda k+2\mu_{5}},
\end{equation}
where we used the expression for the free fermion propagator (\ref{propagator}) and performed the summation over the Matsubara frequencies by means of the formulas listed in Appendix~\ref{app-Matsubara}. Here $\mu_{R,L}$ are the chemical potentials for the right and left chiral fermions, respectively. In order to calculate the rate of change of the chiral charge (\ref{chirality-flipping-rate}), we perform the analytic continuation $i\Omega_{n}\to -2\mu_{5}+i0$ in Eq.~(\ref{G-zero order}) and take the imaginary part using the Sokhotski formula. We end up with
\begin{equation}
\label{dot-n5-0order}
\dot{q}_{5,(0)}=-\frac{m_{e}^{2}}{\pi}\int_{0}^{+\infty}\!\!k^{2}\,dk\sum_{\lambda=\pm}\left[{\rm tanh\,}\frac{\lambda k-\mu_{L}}{2T}+{\rm tanh\,}\frac{\lambda k+\mu_{R}}{2T}\right]\delta (2\lambda k)=0,
\end{equation}
because of the identity $k^{2}\delta(k)\equiv 0$. This result is a mathematical expression of the fact discussed in Sec.~\ref{sec-statement} that angular 
momentum conservation dictates that the mass term as such can only induce chirality flips at zero momentum, therefore the phase volume for such processes vanishes. 

\subsubsection{First order}
\label{subsec-first-order}

\begin{figure}[h!]
	\centering
	$(a)$\includegraphics[height=2.8cm]{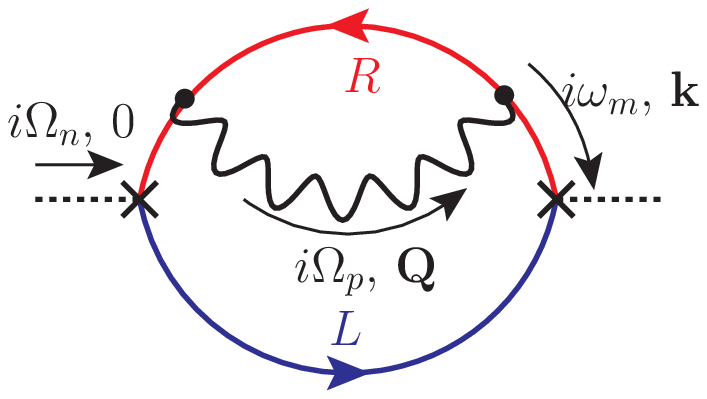}\hfill 
	$(b)$\includegraphics[height=2.8cm]{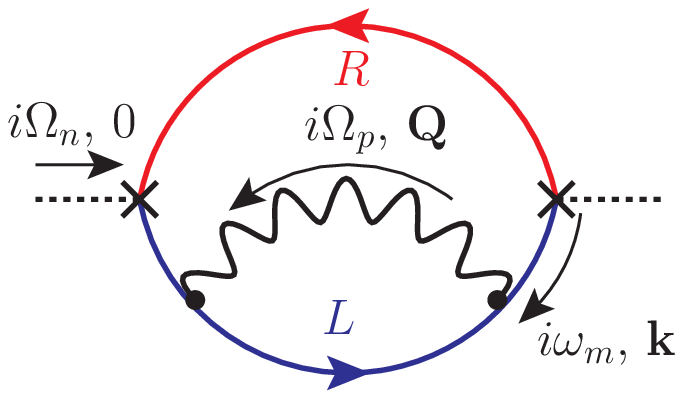}\hfill 
	$(c)$\includegraphics[height=2.8cm]{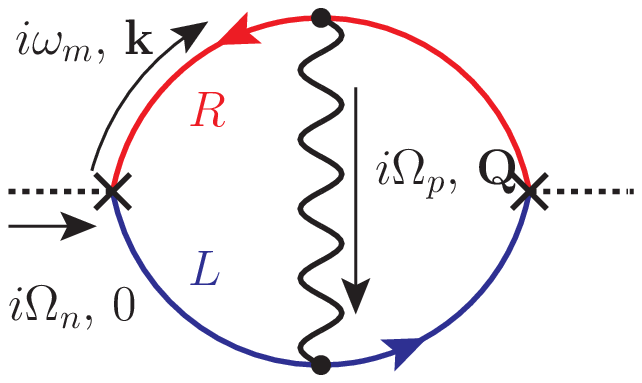}
	\caption{Feynman diagrams for chirality flipping rate in the first order in $\alpha$. Diagrams (a) and (b) contain logarithmic infrared divergence for the loop momentum $k\to 0$. It comes from the singular behavior of the self-energy insertions in these two diagrams as well as from multiple nearly-on-shell propagators. The divergence appears when all three particles in the loop are on shell and collinear. Cutting these diagrams, we get the matrix elements of the $1\leftrightarrow 2$ scattering processes discussed in Sec.~\protect\ref{sec-statement}, Fig.~\protect\ref{fig-lowest}. In addition, all three diagrams contain the unphysical divergence associated with vanishing photon momentum; however, it is canceled in the final result due to the gauge invariance.  \label{fig-1-order}}
\end{figure}

In the first order in $\alpha$ one gets three diagrams depicted in Fig.~\ref{fig-1-order}. We will see that the first two of them give the logarithmically divergent contribution to the chirality flipping rate. In Appendix~\ref{app-first-order} it is shown that this divergence occurs when all three lines in diagrams (a) and (b) are on shell. Cutting these lines, we immediately get the matrix elements of collinear $1\leftrightarrow 2$ scattering processes qualitatively discussed in Sec.~\ref{sec-statement}.
The subtlety with the first-order diagrams is that their divergence resulting from the collinear processes is extremely sensitive to the the dispersion relations of the involved particles. 
After performing resummation of leading divergences in higher order perturbation theory we will find that the dressing of the particles' dispersion relations leads to a collapse of the phase volume of the collinear processes, however this effect is counteracted by a finite lifetime of the quasiparticles resulting in a finite contribution to $\Gamma.$ 

Coming back to the diagrams in Fig.~\ref{fig-1-order}, we now calculate their contributions to the chirality flipping rate. The corresponding expressions are rather cumbersome and for the sake of convenience we list them in Appendix~\ref{app-first-order} together with the details of further manipulations.

When we try to perform the analytic continuation of the final expression for the Matsubara Green's function to the desired point on the real axis, $i\Omega_{n}\to -2\mu_{5}+i0$, the result appears to be logarithmically divergent. In order to extract the divergent part, we perform the continuation into the point shifted along the real axis, $i\Omega_{n}+2\mu_{5}\to i\delta+\epsilon$, and take the symmetric limit $\epsilon\to 0$. 
Using Eq.~(\ref{eq-chirality-flip}), we obtain the result
\begin{equation}
\label{gamma-flip-log-divergent}
\Gamma_{\rm flip}^{\rm 1\, order}=\frac{3}{4}\frac{m_{e}^{2}}{T}\alpha\left[\ln\frac{2T}{\epsilon}+(\text{finite})\right].
\end{equation}
It is worth noting that there are in fact two different logarithmic divergences arriving from two different regions of the phase space. The first one, which is associated with the zero photon momentum in the loop, appears in all three diagrams of Fig.~\ref{fig-1-order}, however it exactly cancels in their sum due to the gauge invariance of the theory. One can eliminate such a divergence by an 
appropriate gauge choice for the photon propagator, however, it is more convenient for us to work in the Feynman gauge. The second logarithmic divergence is caused by the fact that the particles in the loop are massless and there is a possibility for real collinear $1\leftrightarrow 2$ processes to take place.  Remarkably, this divergence is present only in the first two diagrams in Fig.~\ref{fig-1-order}.  More precisely, it comes from the phase space region where the momentum of lower electron line in diagram (a) [upper electron line in diagram (b)] as well as the momenta of both particles in the self-energy blobs are on shell and collinear. The remaining electron propagators are off shell because they have different chirality; however, they approach the shell when the loop momentum $k\to 0$ and this is where this divergence appears.

\begin{figure}[h!]
	\centering
	$(a)$\includegraphics[height=3cm]{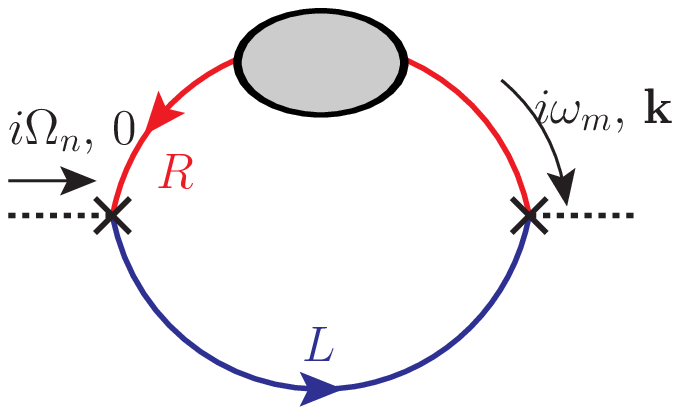}\hspace*{1cm} 
	$(b)$\includegraphics[height=3cm]{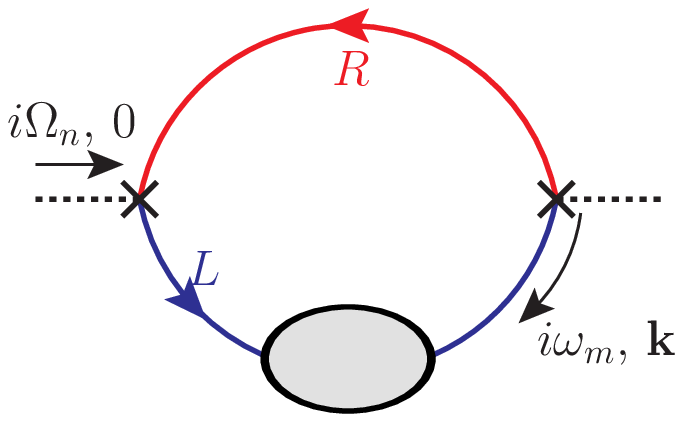}\hfill 
	\caption{First-order diagrams in Fig.~\ref{fig-1-order}(a) and \ref{fig-1-order}(b) represented in an equivalent form by introducing the self-energy insertions (grey blobs). For convenience, the self-energy operator is discussed in Appendix~\ref{app-self-energy}. Its singular behavior for small $k$ as well as the presence of two identical propagators in each diagram lead to the logarithmic divergence in the result. \label{fig-SE-ins}}
\end{figure}

Let us look more precisely at the reason why this divergence comes about. The first two diagrams correspond to the self-energy insertion into one of the fermionic lines in the diagram in Fig.~\ref{fig-0-order}. In fact, they can be represented as shown in Fig.~\ref{fig-SE-ins}.

In terms of the general self-energy correction $\Sigma$, the imaginary part of the retarded Green's function for the first two diagrams has the form
\begin{equation}
\label{diag-1-SE}
\Im G^{\rm ret}_{1,2}=\frac{1}{2}\int\frac{d^{3}\mathbf{k}}{(2\pi)^{3}}\sum_{\lambda=\pm}\left[{\rm tanh} \frac{\lambda k-\mu_{R}}{2T}-{\rm tanh} \frac{\lambda k-\mu_{L}}{2T}\right]\frac{\Im \Sigma_{R,L}^{\pm\lambda}(\lambda k+i0,\mathbf{k})}{(2\lambda k)^{2}},
\end{equation}
where
\begin{equation}
\Sigma_{L,R}^{\lambda}(k_{0},\mathbf{k})={\rm tr}\left[\Sigma_{L,R}(k_{0},\mathbf{k})\frac{1-\lambda \boldsymbol{\sigma}\cdot\mathbf{k}/k}{2}\right]
\end{equation}
and $\Sigma_{L,R}(k_{0}+i0,\mathbf{k})$ is the retarded self-energy of the chiral fermion which is given by one-loop diagram shown in Fig.~\ref{fig-ap-self-energy}. 

It is important to note that $k^{2}$ in the denominator comes from the fact that we have two identical fermion propagators surrounding the self-energy insertion in Fig.~\ref{fig-SE-ins}. The momentum $k$ is on shell for the fermion running in the undressed line. Because of the chirality flip in mass vertices, this momentum is off shell for the dressed line, however, for $k\to 0$ it approaches the shell and the divergence occurs.
Moreover, the imaginary part of the fermion self-energy in HTL approximation (\ref{self-energy}), is also singular for $k\to 0$:
\begin{equation}
\label{self-energy-asympt}
\Im \Sigma_{R,L}^{\pm\lambda}(\lambda k+i0,\mathbf{k})=-\frac{\pi e^{2}T^{2}}{16k}\propto \frac{1}{k}.
\end{equation}

Finally, substituting Eq.~(\ref{diag-1-SE}) into Eq.~(\ref{eq-chirality-flip}), we obtain
\begin{equation}
\Gamma_{\rm flip}^{4a,4b}=\frac{3}{4}\alpha\frac{m_{e}^{2}}{T}\int\frac{dk}{k}\frac{1}{{\rm cosh}^{2}\frac{k}{2T}},
\end{equation}
which is indeed logarithmically divergent and after the regularization leads to Eq.~(\ref{gamma-flip-log-divergent}). Thus, multiple propagators with the same momentum and singular behavior of the self-energy for $k\to 0$ are two features which lead to the logarithmically divergent result.

Let us now discuss the diagram in Fig.~\ref{fig-1-order}(c) which corresponds to  dressing of a vertex (not a QED interaction vertex but simply the mass insertion). This diagram contains two propagators with one momentum and two other propagators with the different momentum (any two of them cannot be simultaneously on shell because of the different chiralities). Combining this with the fact that the singular self-energy does not appear in this diagram, we conclude that this diagram does not contain the divergence at $k\to 0$. Nevertheless, as we mentioned earlier, it contains the fictitious infrared divergence at small photon momenta which is also contained in the diagrams (a) and (b) in Fig.~\ref{fig-1-order}. That is why we have to keep all three diagrams together and, moreover, regularize them in the similar way so that not to violate the cancellation of this unphysical divergence.


Before dealing with these divergences, let us check whether they appear also in higher perturbative orders.

\subsubsection{Second order}
\label{subsec-second-order}

\begin{figure}[h!]
	\centering
	$(a)$\includegraphics[width=4.5cm]{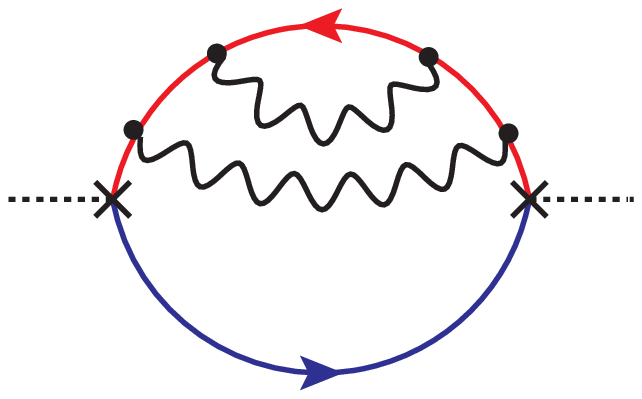}\hfill 
	$(b)$\includegraphics[width=4.5cm]{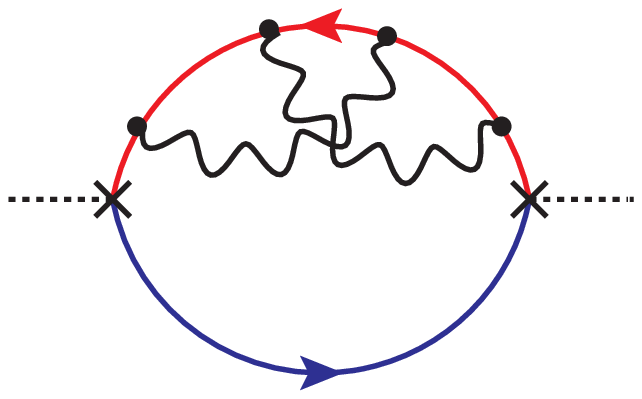}\hfill 
	$(c)$\includegraphics[width=4.5cm]{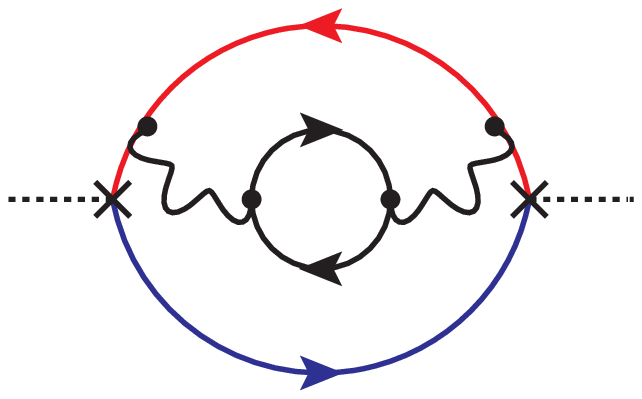}\\
	$(d)$\includegraphics[width=4.5cm]{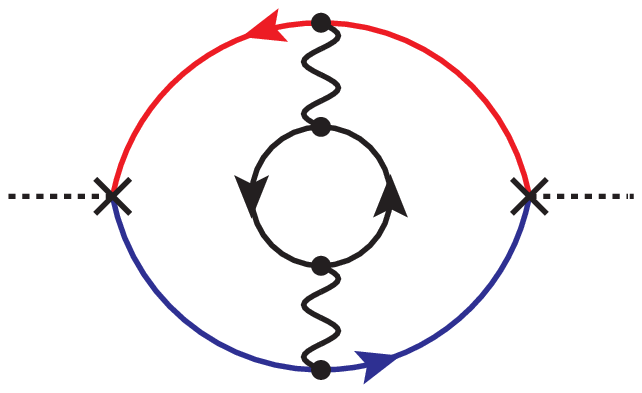}\hfill 
	$(e)$\includegraphics[width=4.5cm]{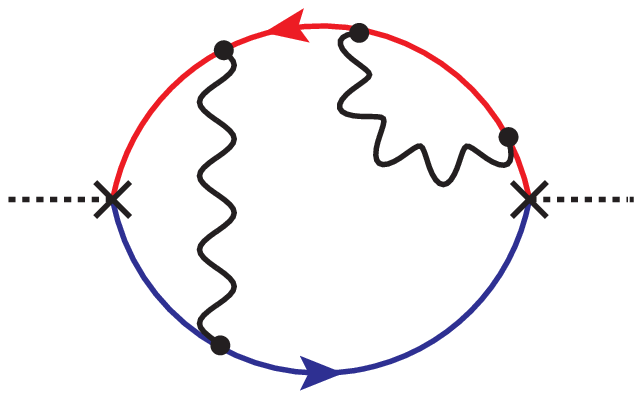}\hfill 
	$(f)$\includegraphics[width=4.5cm]{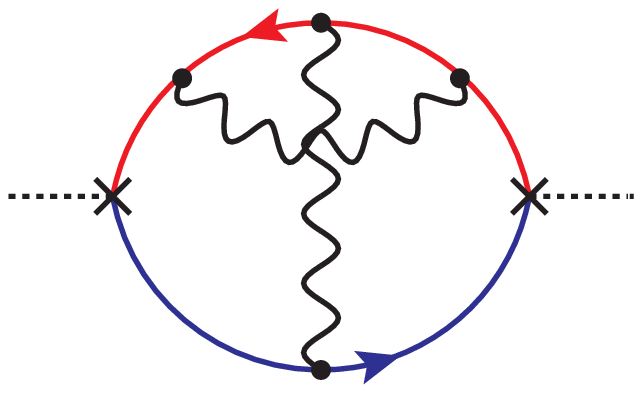}\\
	$(g)$\includegraphics[width=4.5cm]{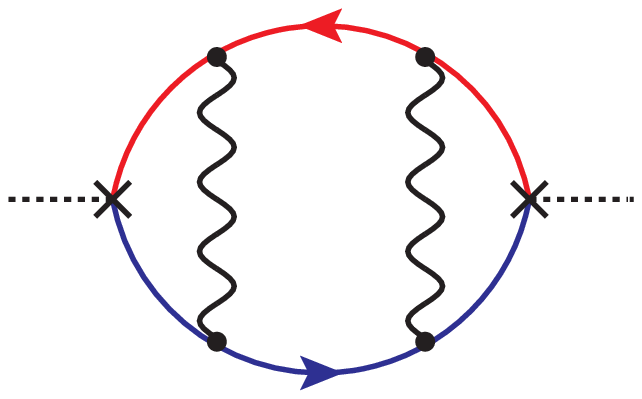}\hspace{1cm} 
	$(h)$\includegraphics[width=4.5cm]{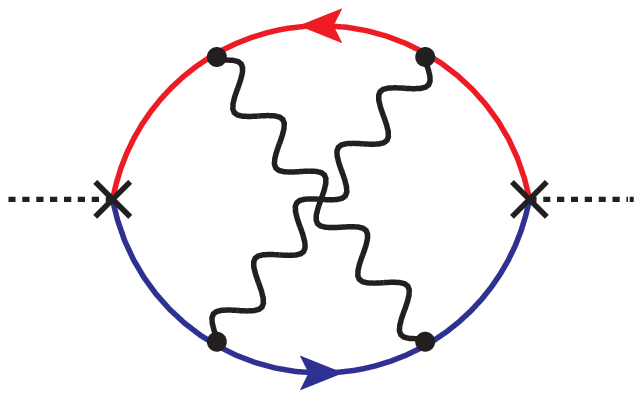}\\ 
	$(i)$\includegraphics[width=4.5cm]{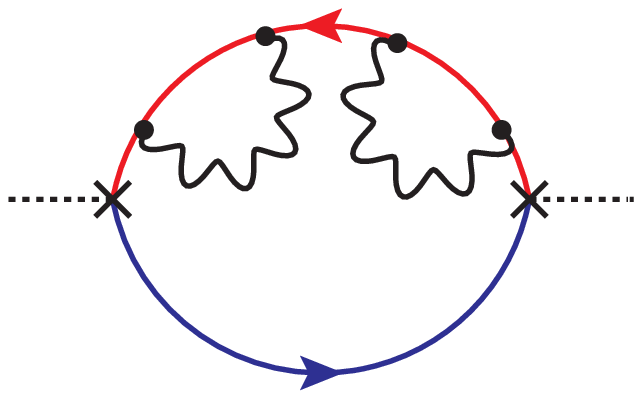}\hspace{1cm}
	$(j)$\includegraphics[width=4.5cm]{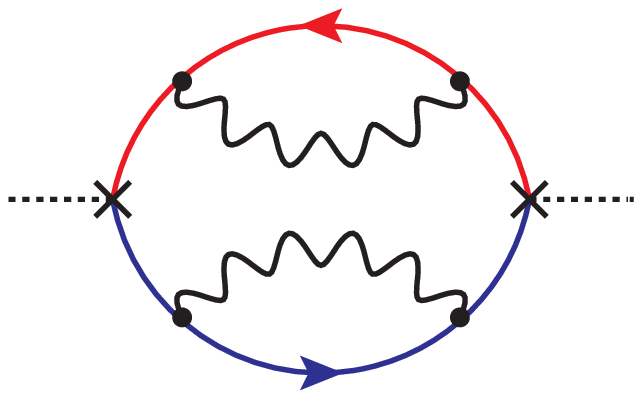}
	\caption{Ten topologically different classes of Feynman diagrams for chirality flipping rate in the second order in $\alpha$. Diagrams (a), (b), and (c) 
	represent two-loop self-energy corrections to one of the propagators and thus are of the same type as shown in Fig.~\ref{fig-SE-ins}. Diagrams (d), (e), and (f) contain one singular self-energy or QED vertex correction and at most two identical propagators; thus, their singularity is of the same type as for diagrams in Fig.~\ref{fig-1-order}(a) and (b). Because of the higher order in $\alpha$, they can be neglected. Diagrams (g) and (h) do not contain nontrivial singularities just as that in Fig.~\ref{fig-1-order}(c). Diagram (i) is more singular than the first-order graphs because it accumulates the singularity of two self-energy corrections and three identical propagators. This divergence is associated with on-shell loop momentum $k$ and quickly grows in higher orders. Diagram (j) additionally contains the singularity coming from off-shell (spacelike) loop momenta. Such a divergence appears for the first time in diagram (j) and it is also accumulated in higher orders. The behavior of the diagrams (i) and (j) show the necessity for the resummation.
	\label{fig-2-order}}
\end{figure}

In this subsection, we consider the diagrams of the second order in $\alpha$. We have many more diagrams than in the previous case which can be classified into 10 topologically different classes shown in Fig.~\ref{fig-2-order}. Analyzing their singularities, we will show that there are two types of divergences: the first one is of the same origin as discussed in the previous subsection, however, it is more severe at this order; the second one is new and it is associated with the off-shell loop momentum.

Let us discuss the singularity of each class. Diagrams of the first line, (a), (b), and (c) represent the higher order corrections to electron's self-energy operator inserted into one of the lines. In other words, these diagrams can be combined together with the diagram in Fig.~\ref{fig-1-order}(a) resulting in a diagram in Fig.~\ref{fig-SE-ins}(a). It is worth noting that the two-loop self-energy correction which is present in the diagram (c) is important because it provides the finite electron damping rate which is crucial for the calculation of the chirality flipping rate.


The diagrams (d), (e), and (f) contain at most two similar propagators and one singular self-energy blob (or vertex correction). So that they cannot be more singular as the diagram in Fig.~\ref{fig-1-order}(a), but they are of higher order in $\alpha$ and thus can be neglected.

The diagrams (g) and (h) are just finite for the same reason as the diagram in Fig.~\ref{fig-1-order}(c) is. There are no multiple nearly-on-shell propagators and no singular self-energy blobs.

Finally, we are left with two types of diagrams, (i) and (j), which need special attention. Obviously, they contain two singular self-energy insertions and more identical propagators in the loop. This leads to the accumulation of divergences which will be discussed for a general order in the next subsection.

\subsubsection{Higher orders in perturbation theory}
\label{subsec-higher-orders}

Here we concentrate our attention on the two classes of diagrams which accumulate divergences with the increase of the perturbative order. These classes were detected in previous subsections.

\begin{figure}[h!]
	\centering
	$(a)$\includegraphics[height=2.5cm]{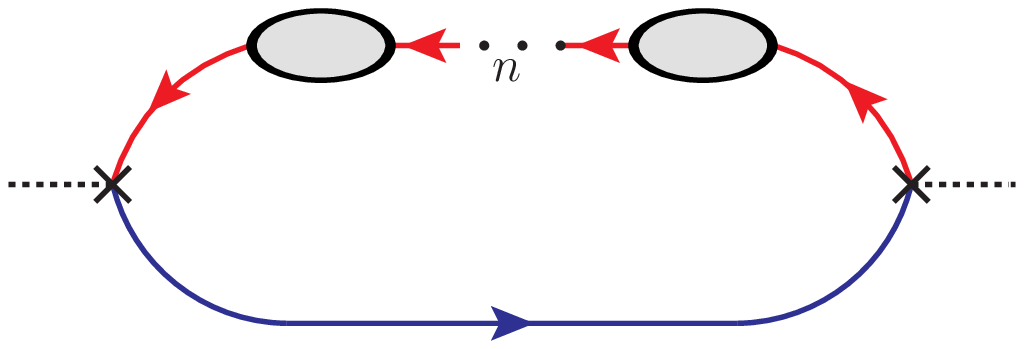}\hfill
	$(b)$\includegraphics[height=2.5cm]{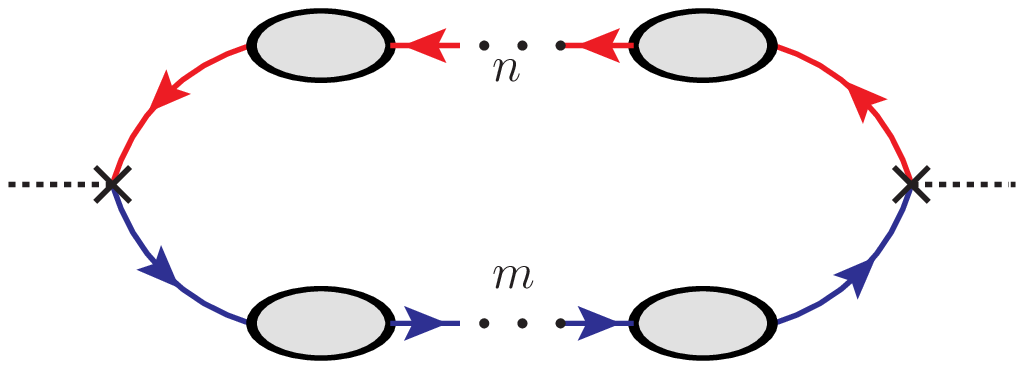}\\
	$(c)$\includegraphics[width=5cm]{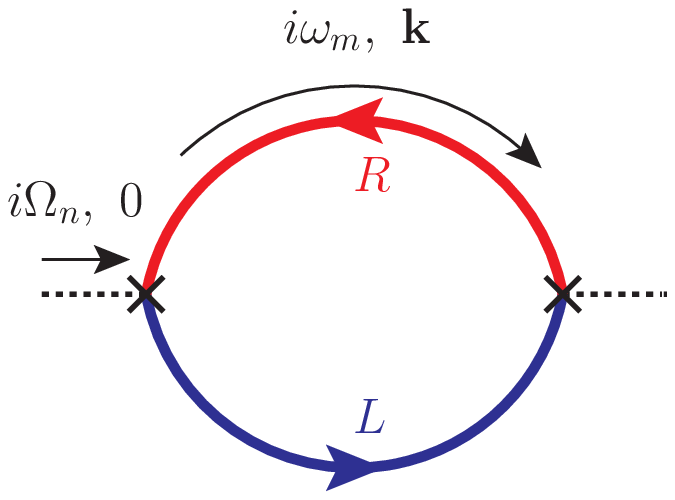}
	\caption{(a) General case of multiple self-energy insertions into one fermion line which illustrates the accumulation of the divergences associated with on-shell loop momentum. The lowest order diagram of this type is shown in Fig.~\ref{fig-1-order}(a) and is logarithmically divergent. The  resummation of this class of diagrams would lead to dressing of only one fermion propagator in the loop. (b) General representative of the diagrams with self-energy insertions in both lines. The lowest order diagram is shown in Fig.~\ref{fig-2-order}(j) and is quadratically infrared divergent; this divergence quickly increases in higher orders. This singularity comes from the incoherent part of the self-energy associated with on-shell particles inside the self-energy loops. The resummation of these diagrams would lead to dressing of both lines in the loop, which is shown by the diagram (c). Obviously, such a resummation absorbs also diagrams of the type (a). Since the vertex corrections do not lead to the accumulation of the divergences, the vertices in diagram (c) remain undressed.
	\label{fig-SE-ins-2}}
\end{figure}

The first one is represented by a diagram in Fig.~\ref{fig-SE-ins-2}(a). In such a case, one of the fermion lines remains undressed and the contribution comes from the on-shell loop momentum $k$. Performing the power counting of the previous subsections, we would have $n+1$ identical propagators giving $1/k$ factor each and $n$ divergent self-energy blobs. As a result the divergence of the integral would be powerlike in the infrared, $\int dk/k^{2n-1}$.

Another possibility arises when we dress both fermion lines as shown in Fig.~\ref{fig-2-order}(j), or in the general case in Fig.~\ref{fig-SE-ins-2}(b). 
As we will see explicitly below, the sought $\Im G_{\rm ret}$ is expressed as the overlap of the imaginary parts of the upper and lower lines in the loop. 
If we have both dressed lines, the imaginary part arises not only for on-shell loop momentum [as for diagram in Fig.~\ref{fig-SE-ins-2}(a)], but also for off-shell (spacelike) momentum $k$, where the incoherent part of the self-energy operator is nontrivial. The power counting is very similar to the previous one, because for any spacelike momentum $k_{0}^{2}-\mathbf{k}^{2}<0$ the behavior $\Im  \Sigma(k_0,\mathbf{k})\sim e^2T^2/k$ is valid. The propagators linking the self-energy blobs all behave $\sim 1/k$ for $k\to 0$. As a result, for the diagram shown in Fig.~\ref{fig-SE-ins-2}(b) we would have the powerlike divergence of the form $\sim \int dk/k^{2n+2m-1}$. As we see, this divergence is accumulated if we go to higher perturbative orders.

The two divergences described above have different physical meaning.  The ``on-shell'' contribution whose lowest order diagrams are given by the first two graphs in Fig.~\ref{fig-1-order}, can be identified with the effective $1\leftrightarrow 2$ processes in plasma. Indeed, such processes appear if we cut the on-shell lines of these diagrams. The lowest order diagram containing the ``off-shell'' divergence is shown in Fig.~\ref{fig-2-order}(j). Here, the imaginary parts of the self-energy insertions for the off-shell momentum $k$ appear when the particles inside the blobs are on shell. Cutting these lines we arrive at the
$2\leftrightarrow2$ Compton or annihilation diagrams shown in Fig.~\ref{fig-Compton}. Thus, we see the one-to-one correspondence of the different regions of the phase space with the elementary processes discussed in Sec.~\ref{sec-statement}.


As we saw above, when we add more self-energy insertions, the singularities are accumulated. Thus, in order to obtain the physically meaningful result, we need to perform the resummation of the self-energy corrections to the fermion propagators in the loop. This will be done in the next subsection.
We would like to note that at leading order there is no need for resummation of the vertex corrections [diagrams like in Fig.~\ref{fig-1-order}(c) and of higher orders]. That is because the divergences in those are much weaker and do not accumulate as quickly as in self-energy. 



\subsubsection{Resummation of leading divergences in the perturbative expansion}
\label{subsec-self-energy-resummation}

The divergences arising in our naive perturbative expansion signal out that the free fermions and photons are not the correct physical degrees of freedom in such a situation and we have to think of some other choice. The self-energy resummation gives us a correct choice replacing the free particles with the quasiparticles with modified dispersion relations due to thermal effects in plasma. In fact, in the kinetic approach considered in Ref.~\cite{PaperI} we used the resummed fermion propagator and this helped to regularize the divergence. We will use the same strategy in this subsection.

At a technical level, the resummation of the self-energy corrections is performed by considering the full fermion propagators instead of bold ones.  Since there is no need for the vertex resummation, we obtain the diagram shown in Fig.~\ref{fig-SE-ins-2}(c). The full propagator of the chiral fermion is the solution of the Schwinger-Dyson equation (\ref{SD-eq}) and in general can be decomposed into two  helicity components as follows:
\begin{equation}
\label{propagator-decomposition}
\mathcal{S}_{R,L}(k^{0},\mathbf{k})=\sum_{\lambda=\pm}\frac{1}{\Delta_{\lambda}(k^{0},\mathbf{k})}\frac{1\pm \lambda\boldsymbol{\sigma}\cdot \hat{\mathbf{k}}}{2}.
\end{equation}
Here $\Delta_{\pm}(k^{0},\mathbf{k})$ is the denominator whose zeros give the dispersion relations for the quasiparticles with positive(negative) ratio of chirality to helicity.

Then, the expression for the diagram in Fig.~\ref{fig-SE-ins-2}(c) takes the form
\begin{eqnarray}
\label{diag-dressed}
\mathcal{G}(i\Omega_{n})&=&T\sum_{m}\int\frac{d^{3}\mathbf{k}}{(2\pi)^{3}}{\rm tr}[\mathcal{S}_{L}(i\omega_{m}-i\Omega_{n},\mathbf{k})\mathcal{S}_{R}(i\omega_{m},\mathbf{k})]\nonumber\\
&=&T\sum_{m}\int\frac{d^{3}\mathbf{k}}{(2\pi)^{3}}\sum_{\lambda=\pm}\frac{1}{\Delta_{\lambda}(i\omega_{m}-i\Omega_{n}+\mu_{L},\mathbf{k})\Delta_{-\lambda}(i\omega_{m}+\mu_{R},\mathbf{k})},
\end{eqnarray}
It is natural that the functions with opposite signs $\lambda$ appear in our result.

The sum over the Matsubara frequencies can be rewritten as the integral in a complex plane $z$ along the contour spanning on the branch cuts at $z=k^{0}$ and $z=k^{0}+i\Omega_{n}$, $k^{0} \in \mathds{R}$. Further, we perform the analytic continuation $i\Omega_{n}\to -2\mu_{5}+i0$ and take the imaginary part to get the following expression for the rate of change of the chiral charge per unit volume:
\begin{eqnarray}
\dot{q}_{5}&=&-4m_{e}^{2}\int\frac{d^{4}k}{(2\pi)^{4}} {\rm tanh\,}\frac{k^{0}}{2T}\sum_{\lambda=\pm} \nonumber\\
&\times&\Bigg[\Im  \frac{1}{\Delta_{\lambda}(k^{0}+\mu_{R}-i0,\mathbf{k})}\Im  \frac{1}{\Delta_{-\lambda}(k^{0}+\mu_{R}+i0,\mathbf{k})}\nonumber\\
&+&\Im  \frac{1}{\Delta_{\lambda}(k^{0}+\mu_{L}+i0,\mathbf{k})}\Im  \frac{1}{\Delta_{-\lambda}(k^{0}+\mu_{L}+i0,\mathbf{k})}\Bigg].
\end{eqnarray}
Then, shifting the variable $k^{0}+\mu_{R}\to k^{0}$ in the first term and $k^{0}+\mu_{L}\to k^{0}$ in the second one and noting that two values of $\lambda$ give equal contributions, we have the final expression
\begin{equation}
\label{dot-n5-general}
\dot{q}_{5}=2m_{e}^{2}\int\frac{d^{4}k}{(2\pi)^{4}} \left({\rm tanh\,}\frac{k^{0}-\mu_{R}}{2T}-{\rm tanh\,}\frac{k^{0}-\mu_{L}}{2T}\right)\rho_{+}(k^{0},\mathbf{k})\rho_{-}(k^{0},\mathbf{k}),
\end{equation}
where 
\begin{equation}
\label{spectral-density-general}
\rho_{\pm}(k^{0},\mathbf{k})=-2\Im  \frac{1}{\Delta_{\pm}(k^{0}+i0,\mathbf{k})}
\end{equation}
is the spectral density of the fermions with positive-negative ratio between chirality and helicity.

Expanding the hyperbolic tangent in Eq.~(\ref{dot-n5-general}) and using Eq.~(\ref{eq-chirality-flip}), we obtain the part of the chirality flipping rate coming from the self-energy corrected diagram in Fig.~\ref{fig-SE-ins-2}(c) in the following form:
\begin{equation}
\label{chirality-flip}
\boxed{\Gamma_{\rm flip}^{7c}=\frac{6m_{e}^{2}}{T^{3}}\int\frac{d^{4}k}{(2\pi)^{4}} \frac{1}{{\rm cosh}^{2}\frac{k^{0}}{2T}}\rho_{+}(k^{0},\mathbf{k})\rho_{-}(k^{0},\mathbf{k}).}
\end{equation}
This equation is well known in the literature. It arises, for example, in condensed matter theory in the context of particle exchange between two reservoirs across 
a tunneling barrier  [see, e.g., Eq.~(9.3.11) and the corresponding diagram 9.9 in Ref.~\cite{Mahan-book}].

In the following subsection we will use Eq.~(\ref{chirality-flip}) in order to calculate the chirality flipping rate in the leading order in $\alpha$. 


\subsection{Calculation of the chirality flipping rate at leading order \label{ssec-thermal}}

Here, we calculate the chirality flip at leading order in $\alpha$. As we discovered in the previous subsection, there are two diagrams which must be taken into account. First of all, this is the dressed loop diagram in Fig.~\ref{fig-SE-ins-2}(c) which is a result of resummation of an infinite number of divergent graphs from all orders in perturbation theory containing only the self-energy corrections. 
The second one is the vertex correction diagram in Fig.~\ref{fig-1-order}(c) which is of the first order in $\alpha$. 

\subsubsection{Diagram in Fig.~\ref{fig-SE-ins-2}(c)}

The part of the chirality flipping rate coming from the diagram in Fig.~\ref{fig-SE-ins-2}(c) is given by Eq.~(\ref{chirality-flip}). In order to calculate this integral, we need to know the spectral densities of electrons with different chiralities (\ref{spectral-density-general}).
For free fermions, the spectral density is simply $\rho^{(0)}_{\pm}=2\pi \delta(k^{0}\mp k)$, and expression (\ref{dot-n5-general}) coincides with Eq.~(\ref{dot-n5-0order}) derived in the zeroth order in $\alpha$. Since the spectral densities with opposite chiralities do not overlap in this case, the chirality flipping rate vanishes. This is the manifestation of the fact that the free fermion cannot spontaneously flip its chirality because of the angular momentum conservation.

In the medium, however, the spectral function acquires a more complicated form
\begin{equation}
\label{spectral-function}
\rho_{\pm}(k^{0},\mathbf{k})=2\pi\left[Z_{\pm}(k)\delta_{\gamma_{e}}(k^{0}-\epsilon_{\pm}(k))+Z_{\mp}(k)\delta_{\gamma_{e}}(k^{0}+\epsilon_{\mp}(k))\right]+\rho_{\pm}^{({\rm LD})}(k^{0},\mathbf{k}).
\end{equation}
Here, the first two terms in brackets correspond to the quasiparticle poles with the energies $\epsilon_{\pm}(k)$, residues $Z_{\pm}(k)$, and decay width $\gamma_{e}$. The function $\delta_{\gamma_{e}}(k^{0}\pm \epsilon)$ is the Lorentz contour
\begin{equation}
\delta_{\gamma_{e}}(k^{0}\pm \epsilon)\equiv \frac{\gamma_{e}}{\pi}\frac{1}{(k^{0}\pm \epsilon)^{2}+\gamma_{e}^{2}}.
\end{equation}
The last term in Eq.~(\ref{spectral-function}) is the continuous (or incoherent) contribution to the spectral density. 

At leading order in $\alpha$, the thermal corrections can be described in the HTL approximation (see, e.g., the textbook \cite{LeBellac}). It accounts for the 1-loop self-energy correction in which the loop momentum is hard $p\sim T$ and much greater than the external momentum. In this approximation, the dispersion relations $\epsilon_{\pm}(k)$ and the residues $Z_{\pm}(k)$ are well known and for convenience they are listed in Appendix~\ref{app-self-energy}. The incoherent contribution is nontrivial in the region of spacelike momenta and corresponds to the Landau damping of the corresponding modes. Its explicit expression is given by Eq.~(\ref{incoherent-part}).

It is important to note that the HTL approximation does not predict the finite lifetime for the quasiparticles in plasma. Indeed, in the region of timelike momenta, the imaginary part of the HTL self-energy (\ref{sigma-0-complex}) is absent and, therefore, the dispersion relations are real. The finite lifetime comes from higher-order scattering processes, which correspond to 
two and higher loop orders in the self-energy diagram. A certain class of such higher order processes is known to contain accumulating infrared divergences arising from the emission of soft photons. Such accumulating divergences are known to reduce the perturbative order of a decay rate as compared to the nominal order of the corresponding tree-level process and generally require a partial resummation of the perturbation theory series \cite{Arnold:2001ba,Arnold:2002ja,Arnold:2002zm}. The analysis of an electron's lifetime 
was performed in
Refs.~\cite{Thoma:1995ju,Blaizot:1996hd,Blaizot:1996az}; see also Eq.~(5.15) of 
Ref.~\cite{Arnold:2001ba}. It was, in particular, 
found that for hard momenta $k\sim T$
which give rise to the dominant contribution to the integral (\ref{chirality-flip}) 
the width of an electron's on-shell peak is given by
\begin{equation}
\label{decay-width}
\gamma_{e}
\simeq 
\frac{1}{2}T\alpha\ln\alpha^{-1} + O(\alpha).
\end{equation}

Now, let us apply Eq.~(\ref{chirality-flip}) in order to calculate the part of the chirality flipping rate coming from the self-energy corrected diagram in Fig.~\ref{fig-SE-ins-2}(c). There are several contributions from different terms in the spectral function. We consider them separately in Appendix~\ref{app-details} in detail and here we list only the main results.

\paragraph{Contribution from the plasmino branch.}

First of all, we show that the contribution of the plasmino branch in the quasiparticle spectrum is of higher order in $\alpha$ and can be neglected. In fact, this branch only exists in the soft momentum region, $k\lesssim eT$, and for hard momenta the residue of the corresponding pole is exponentially suppressed, see Appendix~\ref{app-self-energy}. In this region, however, the energy of the plasmino is displaced from the incoherent part as well as from the normal branch by energy $\Delta\epsilon\sim eT$, which is much larger than the characteristic width (\ref{decay-width}) of the pole. That is why, the integral (\ref{chirality-flip}) is suppressed by higher power of $\alpha$ if at least one of the spectral densities $\rho_{\pm}$ corresponds to the plasmino branch, see the estimates in Appendix~\ref{subapp-plasmino}.

\paragraph{Overlap of the incoherent parts.}

If we take only the incoherent parts in both spectral functions in Eq.~(\ref{chirality-flip}), we will get the following result:
\begin{eqnarray}
\label{gamma-LD-LD}
\Gamma_{\rm flip}^{7c,({\rm LD})}&=&\frac{3m_{e}^{2}m_{\rm th}^{4}}{4\pi T^{3}}\int_{0}^{\infty}dk\int_{0}^{k} \frac{dk^{0}}{{\rm cosh}^{2}\frac{k^{0}}{2T}}\nonumber\\
&\times& \frac{1-\frac{k_{0}}{k}}{\left[(k^{0}-k)\left(1+ \frac{m_{\rm th}^{2}}{4k^{2}}\ln\left|\frac{k+k^{0}}{k-k^{0}}\right|\right)-\frac{m_{\rm th}^{2}}{2k}\right]^{2}+\left[\frac{\pi m_{\rm th}^{2}}{4k}\left(1-\frac{k^{0}}{k}\right)\right]^{2}}\nonumber\\
&\times&\frac{1+\frac{k_{0}}{k}}{\left[(k^{0}+k)\left(1- \frac{m_{\rm th}^{2}}{4k^{2}}\ln\left|\frac{k+k^{0}}{k-k^{0}}\right|\right)+\frac{m_{\rm th}^{2}}{2k}\right]^{2}+\left[\frac{\pi m_{\rm th}^{2}}{4k}\left(1+\frac{k^{0}}{k}\right)\right]^{2}},
\end{eqnarray} 
where the expression (\ref{incoherent-part}) was used. Again, the main contribution comes from the soft momentum region, $k\sim eT$, but now it is a smooth function that accumulates nonperturbatively large spectral weight, i.e., the incoherent part in this region contributes to the sum rules the amounts of order unity. Since $\rho_{\pm}^{({\rm LD})}\sim 1/(eT)$ in the soft momentum region, the naive estimate gives for the chirality flipping rate the following result:
\begin{equation}
\label{gamma-flip-LD}
\boxed{\Gamma_{\rm flip}^{7c,({\rm LD})}=\tilde{C}\frac{m_{e}^{2}}{T^{3}}\times (eT)^{4}\times \frac{1}{(eT)^{2}}=C\times\alpha\frac{m_{e}^{2}}{T}.}
\end{equation}
In Appendix~\ref{subapp-incoh} we explicitly calculate the integral (\ref{gamma-LD-LD}) and show that the estimate (\ref{gamma-flip-LD}) indeed holds with the constant $C$ equal to
\begin{eqnarray}
\label{C-const}
C&=&\frac{3}{8}\int_{0}^{1}\frac{dy}{1-y^{2}}\int_{0}^{\infty} \frac{\xi^{2}\,d\xi}{\left[\left(\xi+\frac{1}{4}\ln\frac{1+y}{1-y}+\frac{1}{2(1-y)}\right)^{2}+\frac{\pi^{2}}{16}\right]\left[\left(\xi-\frac{1}{4}\ln\frac{1+y}{1-y}+\frac{1}{2(1+y)}\right)^{2}+\frac{\pi^{2}}{16}\right]}\nonumber\\
&\approx& 0.24.
\end{eqnarray}

The rate given by Eqs.~(\ref{gamma-flip-LD}) and (\ref{C-const}) exactly coincides with the result coming from the Compton diagram calculated in the kinetic approach \cite{PaperI}. 
Let us discuss what is the connection between these two approaches.
At finite temperature, the imaginary part of the particle's self-energy emerges in the situations when the particle can take part in \textit{any process} involving the real particles, because there is a thermal bath of such particles (in vacuum at $T=0$, only in decay processes). In particular, even for soft electron with momentum $k\sim eT$, the imaginary part of the its self-energy is nonzero for spacelike momenta, because this virtual electron can take part in the processes  $e^{*}+\gamma\to e$, $e^{*}+\bar{e}\to \gamma$, where $e\,(\bar{e})$ and $\gamma$ are real (on-shell) particles.
Thus, the incoherent part of the self-energy appears when the on-shell particles run inside the loop.

\begin{figure}[h!]
	\centering
	$(a)$\includegraphics[width=5cm]{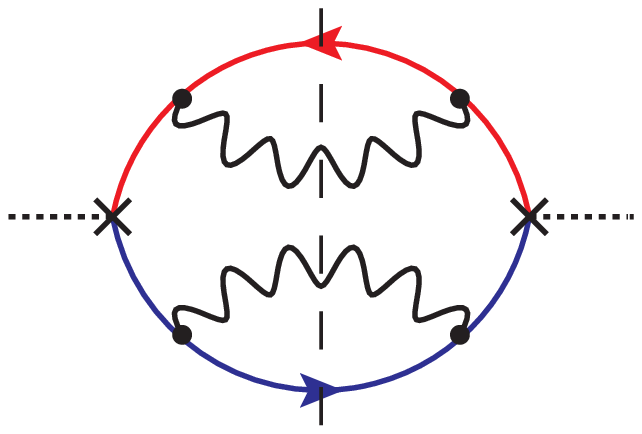} \hspace*{2cm}
	$(b)$\includegraphics[width=3cm]{Compton-t-2}\hspace*{1cm}
	\includegraphics[width=3cm]{Annihil-t-2}
	\caption{(a) The lowest order diagram in which the incoherent parts for both fermion lines arise. In this diagram, the loop momentum is off shell while all particles inside the self-energy blobs are on shell. Cutting this diagram, one gets the matrix elements of the Compton scattering and annihilation processes shown in panel (b). The contribution of such processes was calculated in the kinetic framework \cite{PaperI}. \label{fig-diagram-cut}}
\end{figure}

The lowest order diagram in which the two incoherent parts appear simultaneously is the three-loop diagram shown in Fig.~\ref{fig-diagram-cut}(a). In both self-energy loops the particles are on shell. Cutting the corresponding lines, we reveal the matrix elements of the Compton or annihilation processes shown in Fig.~\ref{fig-diagram-cut}(b). This qualitatively explains the relation between the two approaches.

Thus, we recovered the result of Ref.~\cite{PaperI} in the linear response formalism. However, we will see below that there is also another contribution to the leading order in $\alpha$ which is calculated in the following subsections.


\paragraph{Overlap of the quasiparticle pole with the incoherent part.}

Finally, we have to calculate the contribution coming from the overlap of the quasiparticle pole in $\rho_{+}$ with the spectral density of the fermion of the negative helicity $\rho_{-}$ and vice versa. In the HTL effective theory such an overlap vanishes because the width of the pole is infinitely small and it is always located in the region $k^{0}>k$, where $\rho_{-}$ is identically zero. However, the HTL spectral density is modified by collisions in plasma which cause the finite width $\gamma_{e}$ of the pole and also wash out the region where $\rho_{-}$ vanishes. 

For soft momenta $k\sim eT$, the quasiparticle pole is displaced from the ``old shell'' $k^{0}=k$ by the value of order $eT$. However, for hard momenta, this displacement becomes smaller and asymptotically vanishes, $\epsilon_{+}(k)-k\approx m_{\rm th}^{2}/(2k)\to 0$. Since the decay width parametrically behaves like $\gamma_{e}\propto e^{2}\ln e^{-1} T$, the above mentioned overlap will be more significant in the region of hard momenta. In fact, the contribution from the soft momentum region can be estimated in a similar way as we did for the plasmino branch and it is of higher order in $\alpha$. Thus, in what follows we consider only the hard electron momenta. The corresponding contribution to the chirality flipping rate reads as
\begin{equation}
\label{flip-pole}
\Gamma_{\rm flip}^{7c,({\rm pole})}=\frac{6m_{e}^{2}}{\pi^{2} T^{3}}\int_{0}^{\infty}\frac{k^{2}dk}{{\rm cosh}^{2}\frac{k}{2T}}\int_{-\infty}^{\infty} dx\, \delta_{\gamma_{e}}(x) \rho_{-}(\epsilon_{+}(k)+x,k),
\end{equation}
where we used the fact that the hyperbolic cosine is the smooth function and neglected $x$ as well as the thermal corrections to the energy dispersion in its argument. The only thing we need is to calculate the spectral density $\rho_{-}$ far from its shell.
\begin{eqnarray}
\label{rho-far}
\rho_{-}(\epsilon_{+}(k)+x,k)&=&-2\Im \frac{1}{\Delta_{-}(\epsilon_{+}(k)+x+i0,k)}\nonumber\\
&\approx&-\frac{2}{[\Delta^{(0)}_{-}(\epsilon_{+}(k)+x,k)]^{2}}\Im \Sigma_{-}(\epsilon_{+}(k)+x+i0,k)\nonumber\\
&\approx&-\frac{1}{2k^{2}}\Im \Sigma_{-}(\epsilon_{+}(k)+x+i0,k),
\end{eqnarray}
where $1/\Delta^{(0)}_{-}(k^{0},k)=1/(k^{0}+k)$ is the negative helicity component of the free electron propagator and $\Sigma_{-}$ is the corresponding component of the retarded electron self-energy. The details of the calculation are given in Appendix~\ref{subapp-pole} and the general expression for the chirality flipping rate is given by Eq.~(\ref{gamma-flip-pole-general}). It is important to note that there are two regions in the phase space which give the main contributions to the final result.

The first one comes from the region of small momenta of the photon running in the loop in the self-energy diagram. It is given by
\begin{equation}
\Gamma_{\rm flip}^{7c,({\rm pole},\,{\rm soft}\, \gamma)}=\frac{3m_{e}^{2}\alpha}{2\pi^{3}T^{2}}\int_{0}^{\infty}\frac{dk}{{\rm cosh}^{2}\frac{k}{2T}}\int_{0}^{\Lambda}Q^{2}dQ\int_{0}^{\pi} d\cos\theta\int dQ^{0}\frac{\vphantom{\rho}^{*\!\!}\rho_{\mu\nu}(Q^{0},\mathbf{Q})}{Q^{0}}P^{\mu\nu}_{t}(\mathbf{k})\,\delta_{2\gamma_{e}}(Q^{0}-Q\cos\theta),\label{soft-gamma-12}
\end{equation} 
where $\Lambda\sim eT$ is the upper momentum cutoff which separates the soft photon momenta, and $\vphantom{\rho}^{*\!\!}\rho_{\mu\nu}(Q^{0},\mathbf{Q})$ is the spectral density of the photon in HTL approximation. Using the approximate formula (\ref{rho-mu-nu-soft}) valid for small momenta of the photon, we finally obtain the parametric dependence of this part
\begin{equation}
\Gamma_{\rm flip}^{7c,({\rm pole},\,{\rm soft}\, \gamma)}\approx \frac{3}{\pi^{2}}\frac{m_{e}^{2}}{T}\alpha\log\alpha^{-1}.
\end{equation}
It is worth noting that this logarithmic enhancement is the regularized version of the old (fictitious) logarithmic divergence on the zero photon momentum. We will see further that it will be canceled by the corresponding contribution from the diagram in Fig.~\ref{fig-1-order}(c) (the ``vertex correction'' diagram).

\begin{figure}[h!]
	\centering
	\includegraphics[width=5.5cm]{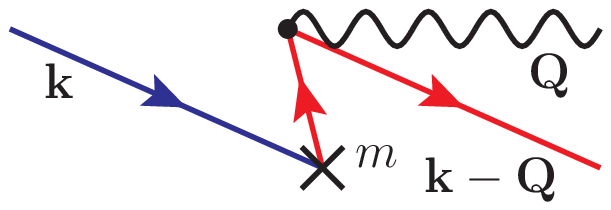}\hfill 
	\includegraphics[width=5cm]{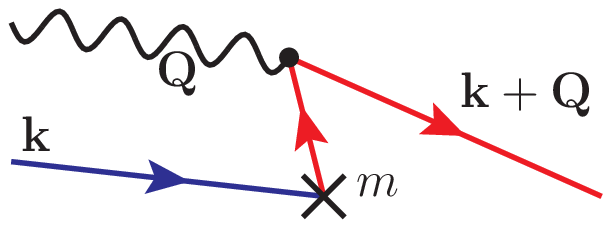}\hfill 
	\includegraphics[width=5cm]{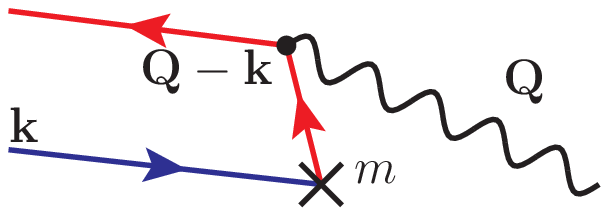}
	\caption{Nearly collinear $1\leftrightarrow 2$ processes in plasma with the chirality flip of the incoming electron (different colors show different chirality of quasiparticles). In these processes, quasiparticles have modified dispersion relations and finite width due to the thermal corrections. The latter shift the energies from free-particle shells and simultaneously allow for a slight violation of the energy conservation. The interplay of these two factors gives rise to the finite contribution of $1\leftrightarrow 2$ processes to the chirality flipping rate.   \label{fig-collinear-processes}}
\end{figure}

There is also another important contribution which comes from the region when the electron quasiparticle as well as photon and electron running in the loop in the self-energy diagram all are hard and nearly collinear. The corresponding expression for the chirality flipping rate is given by
\begin{multline}
\label{gamma-flip-collinear-general}
\Gamma_{\rm flip}^{7c,({\rm pole},\,{\rm collinear})}=\frac{3m_{e}^{2}\alpha}{T^{3}}\int_{0}^{\infty}\frac{dk}{{\rm cosh}^{2}\frac{k}{2T}}\int\frac{d^{3}\mathbf{Q}}{(2\pi)^{3}}\sum_{\lambda',\lambda''=\pm}\frac{\lambda''}{Q}\left({\rm coth}\frac{\lambda'' Q}{2T}+{\rm tanh}\frac{k-\lambda''Q}{2T}\right)\\
\times \delta_{2\gamma_{e}}\big(\epsilon_{+}(k)-\lambda'\epsilon_{+}(\mathbf{k}-\mathbf{Q})-\lambda''\omega_{t}(Q)\big) \left(1+\lambda'\frac{\mathbf{k}\cdot(\mathbf{k}-\mathbf{Q})}{k|\mathbf{k}-\mathbf{Q}|}\right).
\end{multline}
The combination in the argument of the Lorentz-function shows that for different choices of the signs $\lambda',\,\lambda''$ we would get different $1\leftrightarrow 2$ processes contributing to the chirality flip. Three of them are shown in Fig.~\ref{fig-collinear-processes}, the forth one, $\lambda'=\lambda''=-1$ is not realized.

The Lorentz function takes into account the approximate energy conservation law in the collision event. The amount for which this conservation law can be violated is determined by the decay width of the electrons. But since the decay width $\gamma_{e}\sim T\alpha\log\alpha^{-1}$ is much less than the energies of the quasiparticles, the process has to be very close to collinear so that the difference of the energies was small.
In Appendix~\ref{subapp-pole} we consider in detail the processes of nearly collinear bremsstrahlung $e_{L,R}(\mathbf{k})\to e_{R,L}(\mathbf{k}-\mathbf{Q})+\gamma(\mathbf{Q})$,  absorption of the photon $e_{L,R}(\mathbf{k})+\gamma(\mathbf{Q})\to e_{R,L}(\mathbf{k}+\mathbf{Q})$, and annihilation $e_{L,R}(\mathbf{k})+ \bar{e}_{L,R}(\mathbf{Q}-\mathbf{k})\to \gamma(\mathbf{Q})$. The final result reads as
\begin{multline}
\label{gamma-flip-collinear-full}
\Gamma_{\rm flip}^{7c,({\rm pole},\,{\rm collinear})}=\frac{3m_{e}^{2}\alpha}{\pi^{2}T}\int_{0}^{\infty}\frac{x\,dx}{{\rm sinh}^{2}x}\int_{-1/2}^{\infty}dy\frac{1+2y(y+1)}{y(y+1)}\\
\times\left[{\rm tanh\,}x(y+1)-{\rm tanh\,}xy\right]\frac{2}{\pi}{\rm arctan}\frac{8\gamma_{e}Tx}{m_{\gamma}^{2}+m_{\rm th}^{2}/(y(y+1))}.
\end{multline}
Here $m_{\gamma}=eT/\sqrt{6}$ is the photon thermal mass.

\subsubsection{Diagram in Fig.~\ref{fig-1-order}(c)}

So far we calculated only the contribution of the diagram in Fig.~\ref{fig-SE-ins-2}(c) which is a result of resummation of an infinite number of diagrams starting from Figs.~\ref{fig-1-order}(a) and \ref{fig-1-order}(b). Now, we are going to compute the diagram in Fig.~\ref{fig-1-order}(c) since it also gives the contribution of the same order. The corresponding expression is given by Eq.~(\ref{diag3}) where we must replace the free fermion and photon propagators with the full renormalized ones containing the thermal corrections.
Using the spectral representations for the electron and photon propagators, we obtain Eq.~(\ref{diag3-corrected}) and the expression for the imaginary part of the retarded Green's function is given by Eq.~(\ref{diag3-cor-Im}). 

Like in the previous case of the self-energy corrected diagram, here we also have two interesting regions in the phase space. For the details of calculation we refer the reader to Appendix~\ref{subapp-3d} and here only report the results.

The first nontrivial result comes from the region of small photon momenta. The corresponding contribution to the chirality flipping rate equals to
\begin{eqnarray}
\label{gamma-flip-3-soft-photon}
\Gamma_{\rm flip}^{4c,({\rm soft\,}\gamma)}&=&-\frac{3m_{e}^{2}\alpha}{2\pi^{3}T^{2}}\int_{0}^{\infty}\frac{dk}{{\rm cosh}^{2}\frac{k}{2T}}\nonumber\\
&\times&\int_{0}^{\Lambda}Q^{2}dQ\int_{0}^{\pi} d\cos\theta\int dQ^{0}\frac{\vphantom{\rho}^{*\!\!}\rho_{\mu\nu}(Q^{0},\mathbf{Q})}{Q^{0}}P^{\mu\nu}_{t}(\mathbf{k})\,\delta_{2\gamma_{e}}(Q^{0}-Q\cos\theta).
\end{eqnarray}
This expression has exactly the same form and an opposite sign compared to Eq.~(\ref{soft-gamma-12}) emerging from the first two diagrams in the case of the soft photon momenta. Therefore, in the final answer they exactly cancel each other.

The second important result emerges when the momenta of all three particles are hard. The corresponding expression for the retarded Green's function reads as
\begin{eqnarray}
\Im G^{\rm ret}_{3}&=&-\frac{\pi e^{2}}{32}\int\frac{d^{3}\mathbf{k}}{(2\pi)^{3}}\int\frac{d^{3}\mathbf{Q}}{(2\pi)^{3}}\sum_{\lambda,\lambda',\lambda''=\pm}\frac{\lambda\lambda'\lambda''}{kqQ}\left(1-\lambda\lambda'\frac{\mathbf{k}\cdot(\mathbf{k}-\mathbf{Q})}{k|\mathbf{k}-\mathbf{Q}|}\right)\nonumber\\
&\times& \mathcal{B}(\lambda\epsilon_{k}+\lambda'\epsilon_{q},\lambda''\omega_{Q})\mathcal{F}(\lambda\epsilon_{k},\lambda'\epsilon_{q})\delta_{2\gamma_{e}}\big(\lambda\epsilon_{k}+\lambda'\epsilon_{q}-\lambda'' \omega_{Q}\big),\label{diag3-cor-Im-coll}
\end{eqnarray}
where the functions $\mathcal{F}$ and $\mathcal{B}$ are defined in Eqs.~(\ref{F-funct}) and (\ref{B-funct}), respectively. The expression in the argument of the Lorentz function can be approximately zero if the momenta of colliding particles are almost collinear. Each of three possible choices of the relative signs corresponds to one of the three collinear processes considered in the previous subsection. In fact, the case $\lambda'=-\lambda,\ \lambda''=\lambda$ corresponds to the nearly collinear bremsstrahlung $e_{L,R}(\mathbf{k})\to e_{R,L}(\mathbf{k}-\mathbf{Q})+\gamma(\mathbf{Q})$, the case $\lambda'=\lambda''=-\lambda$ corresponds to the process of nearly collinear absorption of the photon $e_{L,R}(\mathbf{k})+\gamma(\mathbf{Q})\to e_{R,L}(\mathbf{k}+\mathbf{Q})$, and the case $\lambda'=\lambda''=\lambda$ corresponds to the process of nearly collinear annihilation $e_{L,R}(\mathbf{k})+ \bar{e}_{L,R}(\mathbf{Q}-\mathbf{k})\to \gamma(\mathbf{Q})$. These processes with the chirality flip are schematically depicted in Fig.~\ref{fig-collinear-processes}.

The contributions from all three processes can be combined together in the following expression:
\begin{equation}
\label{gamma-flip-3d-collinear-all}
\Gamma_{\rm flip}^{4c,({\rm collinear})}=-\frac{6m_{e}^{2}\alpha}{\pi^{2}T}\int_{0}^{\infty}\frac{x\,dx}{{\rm sinh}^{2}x}\int_{-1/2}^{\infty}dy\left[{\rm tanh\,}x(y+1)-{\rm tanh\,}xy\right]\frac{2}{\pi}{\rm arctan}\frac{8\gamma_{e}Tx}{m_{\gamma}^{2}+m_{\rm th}^{2}/(y(y+1))},
\end{equation}

\subsubsection{Chirality flipping rate from nearly collinear $1\leftrightarrow 2$ processes}
\label{sec-results}

Picking up all the results of the previous subsections, we obtain two main contributions to the chirality flipping rate: from the incoherent part of the electron spectral function which encodes the $2\leftrightarrow 2$ processes allowed even with zero decay width [see Eq.~(\ref{gamma-flip-LD})] and from the nearly collinear $1\leftrightarrow 2$ processes in plasma which are possible due to the finite lifetime of the quasiparticles. Here we summarize the second contribution.

Taking together Eqs.~(\ref{gamma-flip-collinear-full}) and (\ref{gamma-flip-3d-collinear-all}), we get 
\begin{equation}
\label{chirality-flip-coll-final}
\Gamma_{\rm flip}^{({\rm collinear})}=\frac{3m_{e}^{2}\alpha}{\pi^{2}T}\int_{0}^{\infty}\frac{x\,dx}{{\rm sinh}^{2}x}\int_{-1/2}^{\infty}dy\frac{{\rm tanh\,}x(y+1)-{\rm tanh\,}xy}{y(y+1)}\frac{2}{\pi}{\rm arctan}\frac{8\gamma_{e}Tx}{m_{\gamma}^{2}+m_{\rm th}^{2}/(y(y+1))}.
\end{equation}
Let us extract the leading parametric behavior of this expression. The arctangent can be replaced by  $(\pi/2)\, {\rm sign}(y)$ if $|xy|\gtrsim (8\gamma_{e}T/m_{\rm th}^{2})^{-1}\sim 1/\ln e^{-1}$. In this case, the $y$ integral is dominated by the lowest possible value of $|y|$, because it is logarithmically divergent on the lower bound. Therefore, we have
\begin{equation}
\label{chirality-flip-coll-approx}
\Gamma_{\rm flip}^{({\rm collinear})}\approx\frac{3m_{e}^{2}\alpha}{\pi^{2}T}\int_{0}^{\infty}\frac{x\,{\rm tanh\,}x\,dx}{{\rm sinh}^{2}x} 2\ln\frac{8\gamma_{e}T}{m_{\rm th}^{2} x}\approx \frac{3}{4}\frac{m_{e}^{2}}{T}\alpha\ln\frac{8\gamma_{e}T}{m_{\rm th}^{2}}+\mathcal{O}(\alpha).
\end{equation}
Comparing this result with Eq.~(\ref{gamma-flip-log-divergent}), we conclude that this is the same logarithmic contribution where the divergence is now regularized due to the thermal corrections. Indeed, Eq.~(\ref{gamma-flip-log-divergent}) contains  $\ln(k_{\rm max}/k_{\rm min})$, where $k_{\rm max}$ and $k_{\rm min}$ are the characteristic maximal and minimal possible values of the electron's momentum for which the collinear processes can occur. Naturally, the maximal value is cut by the temperature because the number of electrons with higher momentum is exponentially suppressed. The minimal value can be determined by the condition that the energy of the electron deviates from the old shell by the value not greater than the decay width, $\Delta\epsilon=\epsilon_{+}(k)-k\lesssim \gamma_{e}$. This immediately gives $k_{\rm min}\sim m_{\rm th}^{2}/\gamma_{e} \sim T/\ln e^{-1}$. Now, we can see that this lower bound cannot be deduced by some heuristic arguments and only by consistent taking into account of thermal effects. It is important to note that this contribution hardly can be derived in the kinetic approach because of the uncertainty of the type $0\times \infty$ which was discussed in Sec.~\ref{sec-statement}.

Assuming that the electron damping rate is given by Eq.~(\ref{decay-width}), we numerically integrate Eq.~(\ref{chirality-flip-coll-final}) and determine few first terms in the asymptotic decomposition over the inverse logarithm of the coupling constant
\begin{equation}
\label{gamma-flip-coll-numerical}
\Gamma_{\rm flip}^{({\rm collinear})}=\frac{m_{e}^{2}}{T}\alpha \left[\frac{3}{4}\ln\ln\alpha^{-1}-0.49+\frac{1.12}{\ln\alpha^{-1}}+\mathcal{O}(1/(\ln\alpha^{-1})^{2})\right].
\end{equation}
Combining Eqs.~(\ref{gamma-flip-LD}) and (\ref{gamma-flip-coll-numerical}), we get the chirality flipping rate in the leading order in the coupling constant
\begin{equation}
\label{gamma-flip-final}
\boxed{\Gamma_{\rm flip}=\frac{m_{e}^{2}}{T}\alpha \left[\frac{3}{4}\ln\ln\alpha^{-1}-0.25+\frac{1.12}{\ln\alpha^{-1}}+\mathcal{O}(1/(\ln\alpha^{-1})^{2})\right].}
\end{equation}

\section{Conclusion}
\label{sec-concl}

In this work we calculated the rate of chirality flipping processes in hot QED plasma with the chiral imbalance $\mu_{5}\neq 0$.
The final expression is given by Eq.~\eqref{gamma-flip-final}.
Numerically, this gives the following equation for the dissipation of the axial charge
\begin{equation}
\label{rate-eq-final}
\boxed{\dot{q}_5\approx -1.17\,\alpha\frac{m_{e}^{2}}{T}\times q_{5}.}
\end{equation}
This rate is  $\approx 1440$ times larger than the naive estimate~\eqref{eq:flip_naive}, cf. Refs.~\cite{Boyarsky:2011uy,Grabowska:2014efa,Manuel:2015zpa,Pavlovic:2016mxq}.

The reasons for this striking difference are as follows.
First, in plasma the nearly collinear $1\leftrightarrow 2$ processes with a slight violation of the energy conservation can take place because of the finite lifetime of the quasiparticles, which counteracts the suppression of the phase volume due to the modified dispersion relations of quasiparticles. Unlike the previously 
considered Compton processes, the collinear process is even 
nominally first order in $\alpha.$
Second, the tree-level matrix elements of some $2\leftrightarrow 2$ scattering processes have severe infrared divergences which 
are regularized in plasma by thermal corrections on the 
soft scale $q\sim eT$. In both cases, result in contributions 
to the chirality flipping rate which are of the order $\mathcal{O}(\alpha).$

The interest to the exact value of the chirality flipping rate has increased with the emergence of the \textit{chiral magnetohydrodynamics (chiral MHD)}~\cite{Joyce:1997uy,Boyarsky:2011uy,Rogachevskii:2017uyc}, see also Refs.~\cite{Giovannini:2013oga,DelZanna:2018dyb}.
Hydrodynamical description of magnetized systems of relativistic fermions in weakly coupled magnetized plasmas \cite{Joyce:1997uy, Alekseev:1998ds,Giovannini:2013oga,Boyarsky:2011uy,Rogachevskii:2017uyc}; dynamics of quasiparticles in new materials such as graphene (see, e.g., Ref.~\cite{Miransky:2015ava}); and behavior of quark-gluon plasma formed in heavy-ion collisions (see, e.g., Ref.~\cite{Kharzeev:2015znc}), cannot be formulated solely in terms of the usual magnetohydrodynamic variables (flow velocity, magnetic field, density and pressure) appearing in the Navier-Stokes and the Maxwell equations.
The hydrodynamics of ultrarelativistic particles necessarily contains an additional degree of freedom corresponding to the space- and time-dependent axial chemical potential~\cite{Boyarsky:2011uy,Boyarsky:2015faa,Rogachevskii:2017uyc} that can be expressed via an axionlike dynamical degree of freedom~\cite{Frohlich:2000en,Frohlich:2002fg,Boyarsky:2015faa,Rogachevskii:2017uyc}, see also Ref.~\cite{Froehlich:2018oce}.
The dynamics of this degree of freedom is coupled to the magnetic helicity [via Eq.~\eqref{relation-Q5-N5}] and to the Maxwell equations via the chiral magnetic effect (see review in Ref.~\cite{Kharzeev:2013ffa} and references therein).
Chiral MHD is the subject of active research
\cite{Dvornikov:2011ey,Boyarsky:2012ex,Dvornikov:2012rk,Semikoz:2012ka,Tashiro:2012mf,Dvornikov:2013bca,Akamatsu:2013pjd,Grabowska:2014efa,Wagstaff:2014fla,Tomalak:2014uma,Xia:2014wla,Manuel:2015zpa,Yamamoto:2015ria,Hirono:2015rla,Sidorenko:2016vom,Long:2016uez,Kamada:2016cnb,Pavlovic:2016mxq,Gorbar:2016klv,Pavlovic:2016gac,Gorbar:2016qfh,Sen:2016jzl,Dvornikov:2016jth,Kamada:2016eeb,Fujita:2016igl,Hirono:2017wqx,Rogachevskii:2017uyc,Brandenburg:2017rcb,Schober:2017cdw,Hattori:2017usa,Gorbar:2017toh,Schober:2018wlo,Masada:2018swb,Shovkovy:2018tks,DelZanna:2018dyb,Schober:2018ojn,Dvornikov:2018tsi,Mace:2019cqo,Schober:2020ogz}.
However, its effects are operational only while the chirality flipping processes are slower than other relevant rates.
In particular, chiral cascade~\cite{Joyce:1997uy, Boyarsky:2011uy,Tashiro:2012mf,Hirono:2015rla,Dvornikov:2016jth,Gorbar:2016klv,Brandenburg:2017rcb,Schober:2018wlo} that may be responsible for the survival of the primordial magnetic fields, stops being operational once chirality flipping rate is faster than the rate $\Gamma_B$ of chirality transfer between fermionic ($N_5$) and magnetic ($\mathcal{H}$) components of the axial charge~\eqref{relation-Q5-N5}.
Recent microscopic studies of the axial charge dynamics~\cite{Figueroa:2017qmv,Figueroa:2017hun,Figueroa:2019jsi} suggest that the rate $\Gamma_B$ may be much faster than one can estimate from the classical (MHD-based) description (see, e.g., Refs.~\cite{Joyce:1997uy,Giovannini:1997gp,Giovannini:1997eg,Boyarsky:2011uy}).
Therefore, the range of applicability of chiral MHD in the hot plasmas, and, in particular, the evolution of primordial magnetic fields~\cite{PaperI} remains an open research question.

In particular, when MHD admits a transfer of magnetic energy from short- to long-wavelength modes of helical magnetic fields, increasing their chance to survive dissipation in the early Universe and survive until today~\cite{Joyce:1997uy, Boyarsky:2011uy,Tashiro:2012mf,Hirono:2015rla,Dvornikov:2016jth,Gorbar:2016klv,Brandenburg:2017rcb,Schober:2018wlo}.

\begin{acknowledgments}
  We are grateful to Artem Ivashko, Oleksandr Gamayun, Kyrylo Bondarenko, Alexander Monin, and Mikhail Shaposhnikov for valuable discussions.
  This project has received funding from the European Research Council (ERC) under the European Union's Horizon 2020 research and innovation programme (GA 694896) and from the Carlsberg Foundation. The work of O.~S. was supported by the Swiss National Science Foundation Grant No. 200020B\_182864.
\end{acknowledgments}

\appendix

\renewcommand{\theequation}{\thesection.\arabic{equation}}
\numberwithin{equation}{section}

\section{Application of the linear response formalism}
\label{app-4-fermion}

The operator $d\hat{Q}_{5}/dt$ has vanishing expectation value in massless QED and satisfies the Heisenberg equation (\ref{Heisenberg-eq}) in the case of nonzero mass. Since in this Appendix only the operator $\hat{Q}_{5}$ appears, we will omit the hat in its notation. Therefore, its expectation value in the perturbed state can be calculated using the linear response formalism \cite{Mahan-book}
\begin{equation}
\label{chirality_flip_1}
\LLangle\frac{dQ_{5}}{dt}\RRangle=-i\int_{-\infty}^{t}dt'\LLangle\Big[\frac{dQ_{5}(t)}{dt},H_{m}(t')\Big]\RRangle_{0}
=2m_{e}^{2}\int\!d^{3}\mathbf{x}\, d^{3}\mathbf{y}\int_{-\infty}^{t}dt'\Llangle\left[\bar{\psi}(t,\mathbf{x})\gamma^{5}\psi(t,\mathbf{x}),\bar{\psi}(t',\mathbf{y})\psi(t',\mathbf{y})\right]\Rrangle_{0},
\end{equation}
where all field operators are considered in the Heisenberg representation with respect to massless QED Hamiltonian $H_{0}$ and the averaging is over the states of massless theory. This is marked by the index $0$, which will be omitted in what follows.

We consider the process of chirality flip in the hot chirally asymmetric electron-positron plasma at high temperature $m_{e}\ll T\lesssim T_{\rm EW}$, where $T_{\rm EW}$ is the temperature when the electroweak crossover takes place. Therefore, the averaging should be performed at finite temperature and chemical potentials for left- and right-handed particles in the grand canonical ensemble with the Hamiltonian
\begin{equation}
\label{Hamiltonian-great-canonical}
K=H_{0}-\mu N-\mu_{5}Q_{5}=H_{0}-\mu_{R}Q_{R}-\mu_{L}Q_{L},
\end{equation}
where the last two terms correspond to the conservation of left- and right-handed particle numbers separately. Here, $\mu=(\mu_{R}+\mu_{L})/2$, $\mu_{5}=(\mu_{R}-\mu_{L})/2$, and $\mu_{R,L}$ are the chemical potentials for the right and left chiral fermions, respectively.
There is some subtlety in the calculation of this correlator which complicates the averaging procedure. In fact, the time evolution of the system is governed by the quantum Hamiltonian of massless QED $H_{0}$, and the thermodynamic averaging in the grand canonical ensemble has to be performed with extended Hamiltonian (\ref{Hamiltonian-great-canonical}). In order to overcome this inconsistency, we represent the quantum Hamiltonian in the form $H_{0}=K+\mu N+\mu_{5}Q_{5}$ and, taking into account that it commutes with the number operators, we separate the corresponding exponents:
\begin{equation}
\psi(t,\mathbf{x})=e^{iH_{0}t}\psi_{S}(\mathbf{x})e^{-iH_{0}t}=e^{iKt}e^{it(\mu N+\mu_{5}Q_{5})}\psi_{S}(\mathbf{x})e^{-it(\mu N+\mu_{5}Q_{5})}e^{-iKt},
\end{equation}
Then, we substitute this into Eq.~(\ref{chirality_flip_1}). Further simplifications could be done using the well-known formula:
\begin{equation}
e^{A}Ce^{-A}=C+[A,C]+\frac{1}{2!}[A,[A,C]]+\frac{1}{3!}[A,[A,[A,C]]]+\dotsb,
\end{equation}
where $A=it(\mu N+\mu_{5}Q_{5})$ and there are two choices for the $C$ operator, $C_{1}(\mathbf{x})=\bar{\psi}(\mathbf{x})\gamma^{5}\psi(\mathbf{x})$ and $C_{2}(\mathbf{x})=\bar{\psi}(\mathbf{x})\psi(\mathbf{x})$. Using the canonical anticommutation relations it is easy to obtain
\begin{equation}
[C_{1,2},N]=0,\quad [C_{1,2},Q_{5}]=2C_{2,1}.
\end{equation}
Therefore, all the commutators are calculated in a chain
\begin{equation}
[A,C_{1,2}]=-2i\mu_{5}tC_{2,1}, \quad [A,[A,C_{1,2}]]=(-2i\mu_{5}t)^{2}C_{1,2},\  \dots.
\end{equation}
Performing the summation of all series we obtain
\begin{equation}
e^{A}C_{1,2}e^{-A}=\cos(2\mu_{5}t)C_{1,2}-i\sin(2\mu_{5}t)C_{2,1}.
\end{equation}
In each case we have
\begin{multline}
e^{A}\bar{\psi}(\mathbf{x})\gamma^{5}\psi(\mathbf{x})e^{-A}=\cos(2\mu_{5}t)\bar{\psi}(\mathbf{x})\gamma^{5}\psi(\mathbf{x})-i\sin(2\mu_{5}t)\bar{\psi}(\mathbf{x})\psi(\mathbf{x})\\
=e^{-2i\mu_{5}t}\bar{\psi}(\mathbf{x})P_{R}\psi(\mathbf{x})-e^{2i\mu_{5}t}\bar{\psi}(\mathbf{x})P_{L}\psi(\mathbf{x}),
\end{multline}
\begin{multline}
e^{A}\bar{\psi}(\mathbf{y})\psi(\mathbf{y})e^{-A}=\cos(2\mu_{5}t')\bar{\psi}(\mathbf{y})\psi(\mathbf{y})-i\sin(2\mu_{5}t')\bar{\psi}(\mathbf{y})\gamma^{5}\psi(\mathbf{y})\\
=e^{-2i\mu_{5}t'}\bar{\psi}(\mathbf{y})P_{R}\psi(\mathbf{y})+e^{2i\mu_{5}t'}\bar{\psi}(\mathbf{y})P_{L}\psi(\mathbf{y}).
\end{multline}

Now we can move into the Heisenberg representation with respect to the full Hamiltonian $K$:
\begin{equation}
\Psi(t,\mathbf{x})=e^{iKt}\psi_{S}(\mathbf{x})e^{-iKt}.
\end{equation}

Introducing the notations
\begin{eqnarray}
W(t,\mathbf{x})&=&\bar{\Psi}(t,\mathbf{x})P_{R}\Psi(t,\mathbf{x})=\bar{\Psi}_{L}(t,\mathbf{x})\Psi_{R}(t,\mathbf{x}),\\ W^{\dagger}(t,\mathbf{x})&=&\bar{\Psi}(t,\mathbf{x})P_{L}\Psi(t,\mathbf{x})=\bar{\Psi}_{R}(t,\mathbf{x})\Psi_{L}(t,\mathbf{x}),
\end{eqnarray}
we rewrite our linear response result in the following form:
\begin{multline}
\LLangle\frac{dQ_{5}}{dt}\RRangle=2m_{e}^{2}\int\!d^{3}\mathbf{x}\,d^{3}\mathbf{y}\int_{-\infty}^{t}\!dt'\Llangle\big[e^{-2i\mu_{5}t}W(t,\mathbf{x})-e^{2i\mu_{5}t}W^{\dagger}(t,\mathbf{x}),\\
e^{-2i\mu_{5}t'}W(t',\mathbf{y})+e^{2i\mu_{5}t'}W^{\dagger}(t',\mathbf{y})\big]\Rrangle=I_{1}+I_{2},
\end{multline}
where
\begin{equation}
\label{current-1}
I_{1}=2m_{e}^{2}\int\!d^{3}\mathbf{x}\,d^{3}\mathbf{y}\int_{-\infty}^{\infty}\!dt'\theta(t-t')\Big\{e^{-2i\mu_{5}(t-t')}\Llangle\big[W(t,\mathbf{x}),W^{\dagger}(t',\mathbf{y})\big]\Rrangle
-e^{2i\mu_{5}(t-t')}\Llangle\big[W^{\dagger}(t,\mathbf{x}),W(t',\mathbf{y})\big]\Rrangle \Big\},
\end{equation}
\begin{equation}
I_{2}=2m_{e}^{2}\int\!d^{3}\mathbf{x}\,d^{3}\mathbf{y}\int_{-\infty}^{\infty}\!dt'\theta(t-t')\Big\{e^{-2i\mu_{5}(t+t')}\Llangle\big[W(t,\mathbf{x}),W(t',\mathbf{y})\big]\Rrangle
-e^{2i\mu_{5}(t+t')}\Llangle\big[W^{\dagger}(t,\mathbf{x}),W^{\dagger}(t',\mathbf{y})\big]\Rrangle \Big\}.
\end{equation}
The latter one contains operators which do not conserve the numbers of left- and right-handed particles. Therefore, the corresponding average equals to zero. The first term $I_{1}$ can be expressed through the following retarded Green's function:
\begin{eqnarray}
&&G_{\rm ret}(t-t',\mathbf{x}-\mathbf{y})=-i\theta(t-t')\Llangle\big[W(t,\mathbf{x}),W^{\dagger}(t',\mathbf{y})\big]\Rrangle,\\
&&G_{\rm ret}(\omega,\mathbf{x}-\mathbf{y})=\int_{-\infty}^{\infty}\!\!dt\ e^{i\omega t}G_{\rm ret}(t, \mathbf{x}-\mathbf{y}).
\end{eqnarray}
\begin{equation}
I_{1}=2im_{e}^{2}\int \!d^{3}\mathbf{x}\,d^{3}\mathbf{y}\left\{G_{\rm ret}(-2\mu_{5},\mathbf{x}-\mathbf{y})-G^{\dagger}_{\rm ret}(-2\mu_{5},\mathbf{x}-\mathbf{y})\right\}
=-4m^{2}_{e}V\int\!d^{3}\mathbf{x}\,\Im \left[G_{\rm ret}(-2\mu_{5},\mathbf{x})\right].
\end{equation}
Finally, in terms of the retarded Green's function introduced here, the rate of change of the chiral charge per unit volume is given by Eq.~(\ref{chirality-flipping-rate}).

\normalsize

\section{Free propagators and summation over Matsubara frequencies}
\label{app-Matsubara}

The propagator of the free fermion in the imaginary time formalism is given by
\begin{equation}
\mathcal{S}_{0}(\tau,\mathbf{x})=\int \frac{d^{3}\mathbf{k}}{(2\pi)^{3}}e^{i\mathbf{k}\cdot\mathbf{x}}T\sum_{m}e^{-i\omega_{m}\tau}\mathcal{S}_{0}(i\omega_{m},\mathbf{k}),
\end{equation}
where $\omega_{m}=(2m+1)\pi T$, $m\in \mathds{Z}$ is the fermionic Matsubara frequency and the propagator in frequency-momentum representation reads as
\begin{equation}
\label{propagator}
\mathcal{S}_{0}(i\omega_{n},\mathbf{k})=\left(
\begin{array}{cc}
\hat{0} & \mathcal{S}_{0L}(i\omega_{n},\mathbf{k}) \\
\mathcal{S}_{0R}(i\omega_{n},\mathbf{k}) & \hat{0}
\end{array}
\right)=\left(
\begin{array}{cc}
\hat{0} & \frac{i\omega_{n}+\mu_{L}-\mathbf{k}\cdot\boldsymbol{\sigma}}{(i\omega_{n}+\mu_{L})^{2}-k^{2}} \\
\frac{i\omega_{n}+\mu_{R}+\mathbf{k}\cdot\boldsymbol{\sigma}}{(i\omega_{n}+\mu_{R})^{2}-k^{2}} & \hat{0}
\end{array}
\right),
\end{equation}
where $\sigma_{i}$ are the Pauli matrices. The left and right components of the propagator are decoupled because we consider the massless case. 

We use the chiral representation for the Dirac matrices
\begin{equation}
\label{gamma-matrices}
\gamma^{\mu}=\left(
\begin{array}{cc}
0 & \sigma^{\mu}\\
\tilde{\sigma}^{\mu} & 0
\end{array}
\right), \quad \gamma^{5}=\left(
\begin{array}{cc}
-\mathds{1} & 0\\
0 & \mathds{1}
\end{array}
\right), \quad \sigma^{\mu}=\tilde{\sigma}_{\mu}=(\mathds{1},\boldsymbol{\sigma}).
\end{equation}

For further convenience, it is also useful to decompose the chiral components of the propagator into two helicity components
\begin{equation}
\mathcal{S}_{0R,L}(i\omega_{n},\mathbf{k})=\sum_{\lambda=\pm}\frac{1}{i\omega_{n}+\mu_{R,L}-\lambda k}\frac{1\pm \lambda \boldsymbol{\sigma}\cdot\hat{\mathbf{k}}}{2}.
\end{equation}

The free photon propagator in the Feynman gauge has the following form:
\begin{equation}
\mathcal{D}_{0,\mu\nu}(\tau,\mathbf{x})=\int \frac{d^{3}\mathbf{Q}}{(2\pi)^{3}}e^{i\mathbf{Q}\cdot\mathbf{x}}T\sum_{p}e^{-i\Omega_{p}\tau}\frac{-g_{\mu\nu}}{(i\Omega_{p})^{2}-\mathbf{Q}^{2}},
\end{equation}
where $\Omega_{p}=2\pi p T$, $p\in \mathds{Z}$ are the bosonic Matsubara frequencies.

The summation over the Matsubara frequencies can be performed by means of the following formulas:
\begin{eqnarray}
\hspace*{-10mm}&&T\!\!\sum_{m=-\infty}^{+\infty}\!\!\frac{1}{i\omega_{m}-a} \frac{1}{i\omega_{m}-b}=-\frac{1}{2}\frac{{\rm tanh}\frac{a}{2T}-{\rm tanh}\frac{b}{2T}}{a-b},\qquad T\!\!\sum_{m=-\infty}^{+\infty}\!\!\frac{1}{(i\omega_{m}-a)^{2}} =-\frac{1}{4T{\rm cosh}^{2}\frac{a}{2T}},\\
\hspace*{-10mm}&&T\!\!\sum_{p=-\infty}^{+\infty}\!\!\frac{1}{(i\Omega_{p})^{2}-Q^{2}} \frac{1}{i\Omega_{p}-a}=\frac{{\rm coth}\frac{a}{2T}-\frac{a}{Q}{\rm coth}\frac{Q}{2T}}{2(Q^{2}-a^{2})},\\
\hspace*{-10mm}&&T\!\!\sum_{p=-\infty}^{+\infty}\!\!\frac{1}{(i\Omega_{p})^{2}-Q^{2}} \frac{1}{(i\Omega_{p}-a)^{2}}=\frac{1}{4T(a^{2}-Q^{2}){\rm sinh}^{2}\frac{a}{2T}}+\frac{a\,{\rm coth}\frac{a}{2T}}{(a^{2}-Q^{2})^{2}}-\frac{(Q^{2}+a^{2}){\rm coth}\frac{Q}{2T}}{2Q(a^{2}-Q^{2})^{2}}.
\end{eqnarray}

\normalsize

\section{Diagrams in the first order in $\alpha$}
\label{app-first-order}

In this Appendix we calculate the first-order contribution to the chirality flipping rate from three diagrams shown in Fig.~\ref{fig-1-order}. Using the Feynman rules we write down the corresponding analytical expressions: 
\begin{eqnarray}
\label{diag1}
\mathcal{G}_{a}&=&-e^{2}\!\int\!\!\frac{d^{3}\mathbf{k}}{(2\pi)^{3}}\frac{d^{3}\mathbf{q}}{(2\pi)^{3}}T\sum_{m}T\sum_{p} \mathcal{D}_{\mu\nu}(i\Omega_{p},\mathbf{Q}) \\
&\times&{\rm tr}\left[\mathcal{S}_{0L}(i\omega_{m}-i\Omega_{n},\mathbf{k})\mathcal{S}_{0R}(i\omega_{m},\mathbf{k})\sigma^{\mu}\mathcal{S}_{0R}(i\omega_{m}-i\Omega_{p},\mathbf{q})\sigma^{\nu}\mathcal{S}_{0R}(i\omega_{m},\mathbf{k})\right],\nonumber\\
\label{diag2}
\mathcal{G}_{b}&=&-e^{2}\!\int\!\!\frac{d^{3}\mathbf{k}}{(2\pi)^{3}}\frac{d^{3}\mathbf{q}}{(2\pi)^{3}}T\sum_{m}T\sum_{p} \mathcal{D}_{\mu\nu}(i\Omega_{p},\mathbf{Q}) \\
&\times&{\rm tr}\left[\mathcal{S}_{0L}(i\omega_{m},\mathbf{k})\tilde{\sigma}^{\nu}\mathcal{S}_{0L}(i\omega_{m}-i\Omega_{p},\mathbf{q})\tilde{\sigma}^{\mu}\mathcal{S}_{0L}(i\omega_{m},\mathbf{k})\mathcal{S}_{0R}(i\omega_{m}+i\Omega_{n},\mathbf{k})\right],\nonumber\\
\mathcal{G}_{c}&=&-e^{2}\!\int\!\!\frac{d^{3}\mathbf{k}}{(2\pi)^{3}}\frac{d^{3}\mathbf{q}}{(2\pi)^{3}}T\sum_{m}T\sum_{p} \mathcal{D}_{\mu\nu}(i\Omega_{p},\mathbf{Q}) \label{diag3}\\
&\times&{\rm tr}\left[\mathcal{S}_{0L}(i\omega_{m}-i\Omega_{p}-i\Omega_{n},\mathbf{q})\tilde{\sigma}^{\nu}\mathcal{S}_{0L}(i\omega_{m}-i\Omega_{n},\mathbf{k})\mathcal{S}_{0R}(i\omega_{m},\mathbf{k})\sigma^{\mu}\mathcal{S}_{0R}(i\omega_{m}-i\Omega_{p},\mathbf{q})\right]\nonumber,
\end{eqnarray}
where $\mathbf{Q}=\mathbf{k}-\mathbf{q}$ is the photon momentum and we used the shorthand notations $\sigma^{\mu}=(\mathds{1},\boldsymbol{\sigma})$, $\tilde{\sigma}^{\mu}=(\mathds{1},-\boldsymbol{\sigma})$ for the four-vectors of the Pauli matrices. 

Calculating traces in Eqs.~(\ref{diag1})--(\ref{diag3}) and performing the summation over the Matsubara frequencies we obtain
\begin{eqnarray}
\hspace*{-10mm}&&\mathcal{G}_{a}=\frac{e^{2}}{8}\!\!\int\!\!\frac{d^{3}\mathbf{k}}{(2\pi)^{3}}\frac{d^{3}\mathbf{q}}{(2\pi)^{3}}\!\!\sum_{\lambda,\lambda'=\pm}\!\!\left\{\frac{\left({\rm coth}\!\frac{\lambda k-\lambda' q}{2T}-\frac{\lambda k-\lambda' q}{Q}{\rm coth}\frac{Q}{2T}\right)}{\lambda\lambda' k q (i\Omega_{n}+2\mu_{5}+2\lambda k)}\left[\frac{\left({\rm tanh}\!\frac{\lambda k+\mu_{R}}{2T}-{\rm tanh}\!\frac{\lambda' q+ \mu_{R}}{2T}\right)}{(i\Omega_{n}+2\mu_{5}+2\lambda k)}-\frac{1}{2T {\rm cosh}^{2}\!\!\left[\frac{\lambda k+\mu_{R}}{2T}\right]}\right]\right.\nonumber\\
\label{diag1-full}
\hspace*{-10mm}&&-\frac{\left({\rm tanh}\!\frac{\lambda k+\mu_{R}}{2T}-{\rm tanh}\!\frac{\lambda' q+ \mu_{R}}{2T}\right)}{k q (i\Omega_{n}+2\mu_{5}+2\lambda k)}\left[\frac{(\lambda k-\lambda' q){\rm coth}\!\frac{\lambda k-\lambda' q}{2T}-\frac{k^{2}+q^{2}-k q(\lambda\lambda'+\cos\theta)}{Q}{\rm coth}\frac{Q}{2T}}{(1-\lambda\lambda'\cos\theta)k q}-\frac{1}{2T \lambda\lambda' {\rm sinh}^{2}\!\!\left[\frac{\lambda k-\lambda' q}{2T}\right]}\right] \\
\hspace*{-10mm}&&\left.+\frac{(1-\lambda\lambda'\cos\theta)\left({\rm tanh}\!\frac{\lambda k-\mu_{L}}{2T}+{\rm tanh}\!\frac{\lambda' q+ \mu_{R}}{2T}\right)}{Q(i\Omega_{n}+2\mu_{5}+2\lambda k)^{2}}\left[\frac{{\rm coth}\!\frac{2\mu_{5}+\lambda k+\lambda' q}{2T}-{\rm coth}\frac{Q}{2T}}{i\Omega_{n}+2\mu_{5}+\lambda k+\lambda' q-Q}-\frac{{\rm coth}\!\frac{2\mu_{5}+\lambda k+\lambda' q}{2T}+{\rm coth}\frac{Q}{2T}}{i\Omega_{n}+2\mu_{5}+\lambda k+\lambda' q+Q}\right]\right\}.\nonumber
\end{eqnarray}

\begin{eqnarray}
\hspace*{-10mm}&&\mathcal{G}_{b}=\frac{e^{2}}{8}\!\!\int\!\!\frac{d^{3}\mathbf{k}}{(2\pi)^{3}}\frac{d^{3}\mathbf{q}}{(2\pi)^{3}}\!\!\sum_{\lambda,\lambda'=\pm}\!\!\left\{\frac{\left({\rm coth}\!\frac{\lambda k-\lambda' q}{2T}-\frac{\lambda k-\lambda' q}{Q}{\rm coth}\frac{Q}{2T}\right)}{\lambda\lambda' k q (i\Omega_{n}+2\mu_{5}+2\lambda k)}\left[\frac{\left({\rm tanh}\!\frac{\lambda k-\mu_{L}}{2T}-{\rm tanh}\!\frac{\lambda' q- \mu_{L}}{2T}\right)}{(i\Omega_{n}+2\mu_{5}+2\lambda k)}-\frac{1}{2T {\rm cosh}^{2}\!\!\left[\frac{\lambda k-\mu_{L}}{2T}\right]}\right]\right.\nonumber\\
\label{diag2-full}
\hspace*{-10mm}&&-\frac{\left({\rm tanh}\!\frac{\lambda k-\mu_{L}}{2T}-{\rm tanh}\!\frac{\lambda' q- \mu_{L}}{2T}\right)}{k q (i\Omega_{n}+2\mu_{5}+2\lambda k)}\left[\frac{(\lambda k-\lambda' q){\rm coth}\!\frac{\lambda k-\lambda' q}{2T}-\frac{k^{2}+q^{2}-k q(\lambda\lambda'+\cos\theta)}{Q}{\rm coth}\frac{Q}{2T}}{(1-\lambda\lambda'\cos\theta)k q}-\frac{1}{2T \lambda\lambda' {\rm sinh}^{2}\!\!\left[\frac{\lambda k-\lambda' q}{2T}\right]}\right] \\
\hspace*{-10mm}&&\left.+\frac{(1-\lambda\lambda'\cos\theta)\left({\rm tanh}\!\frac{\lambda k+\mu_{R}}{2T}+{\rm tanh}\!\frac{\lambda' q- \mu_{L}}{2T}\right)}{Q(i\Omega_{n}+2\mu_{5}+2\lambda k)^{2}}\left[\frac{{\rm coth}\!\frac{2\mu_{5}+\lambda k+\lambda' q}{2T}-{\rm coth}\frac{Q}{2T}}{i\Omega_{n}+2\mu_{5}+\lambda k+\lambda' q-Q}-\frac{{\rm coth}\!\frac{2\mu_{5}+\lambda k+\lambda' q}{2T}+{\rm coth}\frac{Q}{2T}}{i\Omega_{n}+2\mu_{5}+\lambda k+\lambda' q+Q}\right]\right\}.\nonumber
\end{eqnarray}

\begin{eqnarray}
\label{diag3-full}
\hspace*{-10mm}&&\mathcal{G}_{c}=\frac{e^{2}}{4}\!\!\int\!\!\frac{d^{3}\mathbf{k}}{(2\pi)^{3}}\frac{d^{3}\mathbf{q}}{(2\pi)^{3}}\!\!\sum_{\lambda,\lambda'=\pm}\!\!\frac{1}{(i\Omega_{n}+2\mu_{5}+2\lambda k)(i\Omega_{n}+2\mu_{5}+2\lambda' q)} \nonumber\\
\hspace*{-10mm}&&\times\left\{\frac{\left({\rm tanh}\!\frac{\lambda k-\mu_{L}}{2T}-{\rm tanh}\!\frac{\lambda' q- \mu_{L}}{2T}+{\rm tanh}\!\frac{\lambda k+\mu_{R}}{2T}-{\rm tanh}\!\frac{\lambda' q+ \mu_{R}}{2T}\right)}{\lambda\lambda' k q (1-\lambda\lambda'\cos\theta)}\left[{\rm coth}\!\frac{\lambda k-\lambda' q}{2T}-\frac{\lambda k-\lambda' q}{Q}{\rm coth}\frac{Q}{2T}\right]
\right.\\
\hspace*{-10mm}&&\left.+\frac{\left({\rm tanh}\!\frac{\lambda k-\mu_{L}}{2T}+{\rm tanh}\!\frac{\lambda' q+ \mu_{R}}{2T}+{\rm tanh}\!\frac{\lambda k+\mu_{R}}{2T}+{\rm tanh}\!\frac{\lambda' q- \mu_{L}}{2T}\right)}{Q}\left[\frac{{\rm coth}\!\frac{2\mu_{5}+\lambda k+\lambda' q}{2T}-{\rm coth}\frac{Q}{2T}}{i\Omega_{n}+2\mu_{5}+\lambda k+\lambda' q-Q}-\frac{{\rm coth}\!\frac{2\mu_{5}+\lambda k+\lambda' q}{2T}+{\rm coth}\frac{Q}{2T}}{i\Omega_{n}+2\mu_{5}+\lambda k+\lambda' q+Q}\right]\right\}.\nonumber
\end{eqnarray}

Further, we have to perform the analytic continuation $i\Omega_{n}+2\mu_{5}\to i\delta$, $\delta\to 0^{+}$ and to take the imaginary part after that. In the majority of terms, the external frequency $i\Omega_{n}$ enters the expressions in the combinations $1/(i\Omega_{n}+2\mu_{5}+2\lambda k)$, $1/(i\Omega_{n}+2\mu_{5}+2\lambda k)^{2}$, and $(i\Omega_{n}+2\mu_{5}+2\lambda' q)$. After analytic continuation and taking of the imaginary part they will contribute $-\pi \delta(2\lambda k)$, $\pi \delta'(2\lambda k)$ and $-\pi \delta(2\lambda' q)$, correspondingly. It should be noted that in a spherical coordinate system the measure has the form $d^{3}k=k^2\sin \theta d\theta d\varphi$. It is easy to see that in all expressions the fraction $1/(i\Omega_{n}+2\mu_{5}+2\lambda k)$ is multiplied by the regular function which behaves like $\sim k$, when $k\to 0$, the fraction $1/(i\Omega_{n}+2\mu_{5}+2\lambda k)^{2}$ is multiplied by function $\sim k^{2}$ when $k\to 0$, and the fraction $1/(i\Omega_{n}+2\mu_{5}+2\lambda' q)$ is multiplied by function $\sim q^{2}$ when $q\to 0$. Therefore, by virtue of the identity $k^{n} \delta (k) \equiv 0$, $n \geq 1$, all such terms can be discarded and not considered in what follows. Then in each of the expressions (\ref{diag1-full})--(\ref{diag3-full}) only the last terms remain and they have a similar structure of the integrand and can be combined. Finally, we get the following expression for the sum of all three diagrams in Matsubara representation:
\begin{eqnarray}
\label{first-order-sum}
\mathcal{G}_{a}+\mathcal{G}_{b}+\mathcal{G}_{c}&=&\frac{\alpha}{16\pi^{3}}\int_{0}^{+\infty}\!\!k^{2}dk\int_{0}^{+\infty}\!\!q^{2}dq\int_{0}^{\pi}\!\!\sin\theta d\theta\nonumber\\
&\times&\sum_{\lambda,\lambda',\lambda''=\pm}\frac{\lambda'' \mathcal{F}(\lambda k,\lambda' q)\mathcal{B}(\lambda k+\lambda' q,\lambda'' Q)}{Q(i\Omega_{n}+2\mu_{5}+\lambda k+\lambda' q-\lambda'' Q)}\nonumber\\
&\times&\left[\frac{1-\lambda\lambda' \cos\theta}{(i\Omega_{n}+2\mu_{5}+2\lambda k)^{2}}+\frac{2}{(i\Omega_{n}+2\mu_{5}+2\lambda k)(i\Omega_{n}+2\mu_{5}+2\lambda' q)}\right],
\end{eqnarray}
where we introduced the following notations for the thermal factors:
\begin{equation}
\label{F-funct}
\mathcal{F}(k,q)={\rm tanh}\frac{k-\mu_{L}}{2T}+{\rm tanh}\frac{q-\mu_{L}}{2T}+{\rm tanh}\frac{k+\mu_{R}}{2T}+{\rm tanh}\frac{q+\mu_{R}}{2T},
\end{equation}
\begin{equation}
\label{B-funct}
\mathcal{B}(k,Q)={\rm coth}\frac{2\mu_{5}+k}{2T}-{\rm coth}\frac{Q}{2T}.
\end{equation}

There are four different ways to choose signs in Eq.~(\ref{first-order-sum}), $\lambda'=\pm \lambda$ and $\lambda''=\pm \lambda$. Below we consider each of them separately.

\subsection{Case $\lambda'=\lambda''=\lambda$}
Feynman parametrization allows us to combine two denominators into one:
\begin{equation}
\label{feynman}
\frac{1}{AB}=\int_{0}^{1}\frac{dx}{[Ax+B(1-x)]^{2}}, \quad \frac{1}{A^{2}B}=2\int_{0}^{1}\frac{x\,dx}{[Ax+B(1-x)]^{3}}.
\end{equation}
The contribution of the first two diagrams could be calculated using the second formula in (\ref{feynman}):
\begin{eqnarray}
\label{1case-first-second-1}
&&\mathcal{G}^{(1)}_{a}+\mathcal{G}^{(1)}_{b}=\frac{\alpha}{8\pi^{3}}\int_{0}^{+\infty}\!\!k^{2}dk\int_{0}^{+\infty}\!\!q^{2}dq\int_{0}^{\pi}\!\!\sin\theta d\theta\sum_{\lambda=\pm}\frac{\lambda \mathcal{F}(\lambda k,\lambda q)\mathcal{B}(\lambda (k+q),\lambda Q)(1-\cos\theta)}{Q}\nonumber\\
&&\times\int_{0}^{1}\!\!\frac{x\,dx}{[i\Omega_{n}+2\mu_{5}+\lambda k(1+x)+\lambda q(1-x)-\lambda Q(1-x)]^{3}}.
\end{eqnarray}

In elliptic coordinates
\begin{equation}
\label{elliptic-1}
\xi=\frac{k+q}{Q},\quad \eta=\frac{k-q}{Q},\quad \cos\theta=\frac{\xi^{2}+\eta^{2}-2}{\xi^{2}-\eta^{2}},\quad k^{2}q^{2}\sin\theta\, dk\,dq\,d\theta=\frac{Q^{5}}{8}(\xi^{2}-\eta^{2})dQ d\xi d\eta
\end{equation}
the expression takes the form
\begin{eqnarray}
\label{1case-first-second-2}
&&\mathcal{G}^{(1)}_{a}+\mathcal{G}^{(1)}_{b}=\frac{\alpha}{32\pi^{3}}\int_{0}^{+\infty}\!\!Q^{4}dQ\int_{1}^{+\infty}\!\!d\xi\int_{-1}^{1}\!\!d\eta(1-\eta^{2})\\
&&\times\sum_{\lambda=\pm}\lambda \mathcal{F}(\lambda Q(\xi+\eta)/2,\lambda Q(\xi-\eta)/2)\mathcal{B}(\lambda \xi Q,\lambda Q)\int_{0}^{1}\!\!\frac{x\,dx}{\{i\Omega_{n}+2\mu_{5}+\lambda Q[\xi-1+x(\eta+1)]\}^{3}}.\nonumber
\end{eqnarray}

Then, in order to reduce the power in the denominator we integrate by parts twice with respect to $Q$. Factor $Q^{4}$ provides the zero value on the lower boundary and $\mathcal{B}(\lambda\xi Q,\lambda Q)$ is exponentially decreasing on the upper boundary. After integration we obtain the following expression:
\begin{eqnarray}
\label{1case-first-second-3}
\hspace*{-10mm}&&\mathcal{G}^{(1)}_{a}+\mathcal{G}^{(1)}_{b}=\frac{\alpha}{64\pi^{3}}\int_{0}^{+\infty}\!\!dQ\int_{1}^{+\infty}\!\!d\xi\int_{-1}^{1}\!\!d\eta\!\int_{0}^{1}\!\!x\,dx\sum_{\lambda=\pm}\frac{\lambda}{i\Omega_{n}+2\mu_{5}+\lambda Q[\xi-1+x(\eta+1)]}\nonumber\\
\hspace*{-10mm}&&\times \frac{(1-\eta^{2})}{[\xi-1+x(\eta+1)]^{2}}\frac{d^{2}}{dQ^{2}}\left[Q^{4}\mathcal{F}(\lambda Q(\xi+\eta)/2,\lambda Q(\xi-\eta)/2)\mathcal{B}(\lambda \xi Q,\lambda Q)\right].
\end{eqnarray}

Now, we can perform the analytic continuation and take the imaginary part of the expression. In order to avoid the divergence, we move to the shifted point on the real axis, namely, $i\Omega_{n}\to -2\mu_{5}+\epsilon+i\delta$, $\delta\to 0^{+}$. 
\begin{eqnarray}
\label{1case-first-second-4}
\hspace*{-10mm}&&\Im [G^{(1)}_{{\rm ret},a}+G^{(1)}_{{\rm ret},b}]=-\frac{\alpha}{64\pi^{2}}\int_{0}^{+\infty}\!\!dQ\int_{1}^{+\infty}\!\!\!\!d\xi\int_{-1}^{1}\!\!d\eta\int_{0}^{1}\!\!x\,dx\frac{(1-\eta^{2})}{[\xi-1+x(\eta+1)]^{2}}\sum_{\lambda=\pm}\lambda\nonumber\\
\hspace*{-10mm}&&\times \frac{d^{2}}{dQ^{2}}\left[Q^{4}\mathcal{F}(\lambda Q(\xi+\eta)/2,\lambda Q(\xi-\eta)/2)\mathcal{B}(\lambda \xi Q,\lambda Q)\right] \delta(\epsilon+\lambda Q[\xi-1+x(\eta+1)]).
\end{eqnarray}

Integration over $x$ could be done with the help of the delta function. We consider the case of $\epsilon>0$ (the opposite sign will be discussed later), then, the nonzero contribution gives only the term with $\lambda=-1$. The delta function gives the Jacobian $J=\frac{1}{Q(\eta+1)}$ and the integration variable acquires the value  $x=\frac{\epsilon/Q+1-\xi}{\eta+1}$. Taking into account that $0\leq x\leq 1$, we obtain the integration region for the remaining variables: $1\leq\xi\leq 1+\epsilon/Q$ and $\xi+\eta\geq \epsilon/Q$. 
\begin{eqnarray}
\label{1case-first-second-5}
\hspace*{-10mm}&&\Im [G^{(1)}_{{\rm ret},a}+G^{(1)}_{{\rm ret},b}]=\frac{\alpha}{64\pi^{2}\epsilon^{2}}\int_{0}^{\epsilon/2}\!\!Q\,dQ\int_{-1}^{1}\!\!\frac{1-\eta}{1+\eta}d\eta\int_{\epsilon/Q-\eta}^{1+\epsilon/Q}\!\!\!\!\left(\frac{\epsilon}{Q}+1-\xi\right)d\xi \frac{d^{2}}{dQ^{2}}\left[Q^{4}\mathcal{F}\mathcal{B}\right]\nonumber\\
\hspace*{-10mm}&&+\frac{\alpha}{64\pi^{2}\epsilon^{2}}\int_{\epsilon/2}^{+\infty}\!\!\!Q\,dQ\int_{1}^{1+\epsilon/Q}\!\!\!\!\left(\frac{\epsilon}{Q}+1-\xi\right)d\xi\int_{\epsilon/Q-\xi}^{1}\!\!\frac{1-\eta}{1+\eta}d\eta \frac{d^{2}}{dQ^{2}}\left[Q^{4}\mathcal{F}\mathcal{B}\right].
\end{eqnarray}

Let us estimate the behavior of each term as $\epsilon\to 0$. In the first integral $Q<\epsilon/2$, therefore the functions could be replaced by their asymptotics $\mathcal{B}\sim1/Q$, $\mathcal{F}\sim {\rm const}$, and  $\frac{d^{2}}{dQ^{2}}\left[Q^{4}\mathcal{F}\cdot\mathcal{B}\right]\sim Q$. Integral over $\xi$ and $\eta$ gives the number of order unity and integral over $Q$ is proportional to $\sim\epsilon^{3}$. Finally, the first term in (\ref{1case-first-second-5}) vanishes as $\sim \epsilon$.

In the second term the main contribution is made by the region $\xi\sim 1$, $\eta\sim -1$. Therefore, we put in functions $\mathcal{B}$ and $\mathcal{F}$ exactly $\xi=1$, $\eta=-1$. The remaining integration over $\xi$ and $\eta$ is straightforward:
\begin{equation}
\int_{1}^{1+\epsilon/Q}\!\!\!\!\left(\frac{\epsilon}{Q}+1-\xi\right)d\xi\int_{\epsilon/Q-\xi}^{1}\!\!\frac{1-\eta}{1+\eta}d\eta=\frac{\epsilon^{2}}{2Q^{2}}\left(-2\ln\frac{\epsilon}{Q}+2\ln 2-1+\frac{2\epsilon}{3Q}\right)\simeq-\frac{\epsilon^{2}}{Q^{2}}\ln\frac{\epsilon}{Q}.
\end{equation}
After that in the leading order in $\epsilon$ the integration by parts over $Q$ could be done: 
\begin{equation}
\int_{0}^{+\infty}\!\!\frac{dQ}{Q}\frac{d^{2}}{dQ^{2}}\left[Q^{4}\mathcal{F}(0,-Q)\cdot\mathcal{B}(-Q,-Q)\right]=2\int_{0}^{+\infty}\!\!dQ\,Q\mathcal{F}(0,-Q)\cdot\mathcal{B}(-Q,-Q).
\end{equation}
Summarizing, the leading contribution is logarithmically divergent for $\epsilon\to 0$:
\begin{equation}
\Im [G^{(1)}_{{\rm ret},a}+G^{(1)}_{{\rm ret},b}]\simeq-\frac{\alpha}{32\pi^{2}}\ln\frac{\epsilon}{T}\int_{0}^{+\infty}\!\!dQ\,Q\mathcal{F}(0,-Q)\mathcal{B}(-Q,-Q)+{\rm finite}.
\end{equation}

Now, let us consider the contribution of the third diagram. There are three different denominators in Eq.~(\ref{first-order-sum}). It could be rewritten in the following form ($z=i\Omega_{n}+2\mu_{5}$):
\begin{eqnarray}
\hspace*{-10mm}&&\frac{1}{(z+2\lambda k)(z+2\lambda q)(z+\lambda (k+q-Q))}=\frac{1}{2(k-q)(k-q+Q)}\frac{1}{z+2\lambda k}\nonumber\\
\hspace*{-10mm}&&+\frac{1}{2(k-q)(k-q-Q)}\frac{1}{z+2\lambda q}+\frac{1}{Q^{2}-(k-q)^{2}}\frac{1}{z+\lambda (k+q-Q)}
\end{eqnarray}

The first two terms after the analytic continuation and taking of the imaginary part would be proportional to $\delta(k)$ and $\delta(q)$, correspondingly. They are multiplied by the function which behaves like  $\sim 1/k$ (or $\sim 1/q$) and the integration measure gives $k^{2}q^{2}$. Therefore, these terms vanish. The third term reads as 
\begin{equation}
\label{1case-third-1}
\Im  G^{(1)}_{{\rm ret},c}=-\frac{\alpha}{8\pi^{2}}\int_{0}^{+\infty}\!\!k^{2}dk\int_{0}^{+\infty}\!\!q^{2}dq\int_{0}^{\pi}\!\!\sin\theta d\theta\sum_{\lambda=\pm}\frac{\lambda \mathcal{F}(\lambda k,\lambda q)\mathcal{B}(\lambda (k+q),\lambda Q)}{Q[Q^{2}-(k-q)^{2}]}\delta(\epsilon+\lambda(k+q-Q)).
\end{equation}

In the elliptic coordinates (\ref{elliptic-1}) it could be rewritten as
\begin{multline}
\label{1case-third-2}
\Im  G^{(1)}_{{\rm ret},c}=-\frac{\alpha}{64\pi^{2}}\int_{0}^{+\infty}\!\!\!\!\!Q^{2}dQ\int_{1}^{+\infty}\!\!\!\!\!d\xi\int_{-1}^{1}\!\!d\eta\frac{\xi^{2}-\eta^{2}}{1-\eta^{2}}\\
\times\sum_{\lambda=\pm}\lambda \mathcal{F}\left(\frac{\lambda Q(\xi+\eta)}{2},\frac{\lambda Q(\xi-\eta)}{2}\right)\mathcal{B}(\lambda \xi Q,\lambda Q)\delta(\epsilon+\lambda Q(\xi-1)).
\end{multline}

The result is finite at $\epsilon\to 0$ and equals to
\begin{equation}
\label{1case-third-3}
\Im  G^{(1)}_{{\rm ret},3}=\frac{\alpha}{64\pi^{2}}\int_{0}^{+\infty}\!\!\!\!\!Q\,dQ\int_{-1}^{1}\!\!d\eta \mathcal{F}\left(-\frac{Q(1+\eta)}{2},-\frac{Q(1-\eta)}{2}\right)\mathcal{B}(-Q,-Q).
\end{equation}

The finite contributions from the first two diagrams also could be calculated. Then, the corresponding expression reads as
\begin{eqnarray}
\hspace*{-10mm}&&\Im \mathcal{G}^{(1)}_{\rm ret}=-\frac{\alpha}{32\pi^{2}}\left[\ln\frac{\epsilon}{2T}+2\right]\int_{0}^{+\infty}\!\!dQ\,Q\mathcal{F}(0,-Q)\mathcal{B}(-Q,-Q)\nonumber\\
\hspace*{-10mm}&&+\frac{\alpha}{32\pi^{2}}\int_{0}^{+\infty}\!\!dQ\,Q\ln\frac{Q}{T}\mathcal{F}(0,-Q)\mathcal{B}(-Q,-Q)\\
\hspace*{-10mm}&&+\frac{\alpha}{64\pi^{2}}\!\int_{\!-1}^{1}\!\!\frac{d\eta}{\eta+1}\int_{0}^{+\infty}\!\!dQ\,Q\left[2\mathcal{F}\left(-\frac{Q(1+\eta)}{2},-\frac{Q(1-\eta)}{2}\right)-(1-\eta)\mathcal{F}(0,-Q)\right]\mathcal{B}(-Q,-Q).\nonumber
\end{eqnarray}

If we take $\epsilon<0$ the contribution would be made by the terms with $\lambda=+1$. In ideal case $\epsilon=0$ the delta function would work for both $\lambda$ but only at the boundary of integration. As a result, the additional factor of $1/2$ is required. Therefore, we will take the half of the sum of the expressions for both $\lambda$. In summary, the final expression is the following:
\begin{eqnarray}
\hspace*{-10mm}&&\Im \mathcal{G}^{(1)}_{\rm ret}=\frac{\alpha}{64\pi^{2}}\left[\ln\frac{\epsilon}{2T}+2\right]\int_{0}^{+\infty}\!\!dQ\,Q\sum_{\lambda=\pm}\lambda\mathcal{F}(0,\lambda Q)\mathcal{B}(\lambda Q,\lambda Q)\nonumber\\
\hspace*{-10mm}&&-\frac{\alpha}{64\pi^{2}}\int_{0}^{+\infty}\!\!dQ\,Q\ln\frac{Q}{T}\sum_{\lambda=\pm}\lambda\mathcal{F}(0,\lambda Q)\mathcal{B}(\lambda Q,\lambda Q)\\
\hspace*{-10mm}&&-\frac{\alpha}{128\pi^{2}}\!\int_{\!-1}^{1}\!\!\frac{d\eta}{\eta+1}\int_{0}^{+\infty}\!\!dQ\,Q\sum_{\lambda=\pm}\lambda\left[2\mathcal{F}\left(\frac{\lambda Q(1+\eta)}{2},\frac{\lambda Q(1-\eta)}{2}\right)-(1-\eta)\mathcal{F}(0,\lambda Q)\right]\mathcal{B}(\lambda Q,\lambda Q).\nonumber
\end{eqnarray}

\subsection{Case $\lambda'=\lambda''=-\lambda$}
Using the Feynman parametrization (\ref{feynman}) we rewrite the expression of the first two diagrams in the following form:
\begin{eqnarray}
\label{2case-first-second-1}
\hspace*{-10mm}&&\mathcal{G}^{(2)}_{a}+\mathcal{G}^{(2)}_{b}=\frac{\alpha}{8\pi^{3}}\int_{0}^{+\infty}\!\!k^{2}dk\int_{0}^{+\infty}\!\!q^{2}dq\int_{0}^{\pi}\!\!\sin\theta d\theta\sum_{\lambda=\pm}\frac{(-\lambda) \mathcal{F}(\lambda k,-\lambda q)\mathcal{B}(\lambda (k-q),-\lambda Q)}{Q}\nonumber\\
\hspace*{-10mm}&&\times\int_{0}^{1}\!\!\frac{(1+\cos\theta)x\,dx}{[i\Omega_{n}+2\mu_{5}+\lambda k(1+x)-\lambda q(1-x)+\lambda Q(1-x)]^{3}}.
\end{eqnarray}

In elliptic coordinates
\begin{equation}
\label{elliptic-2}
\xi=\frac{k+Q}{q},\quad \eta=\frac{k-Q}{q},\quad \cos\theta=\frac{\xi\eta+1}{\xi+\eta},\quad k^{2}q^{2}\sin\theta\, dk\,dq\,d\theta=\frac{q^{5}}{8}(\xi^{2}-\eta^{2})dq d\xi d\eta
\end{equation}
the expression takes the form 
\begin{eqnarray}
\label{2case-first-second-2}
\hspace*{-10mm}&&\mathcal{G}^{(2)}_{a}+\mathcal{G}^{(2)}_{b}=-\frac{\alpha}{32\pi^{3}}\int_{0}^{+\infty}\!\!q^{4}dq\int_{1}^{+\infty}\!\!d\xi\int_{-1}^{1}\!\!d\eta(\xi+1)(\eta+1)\sum_{\lambda=\pm}\lambda \mathcal{F}(\lambda q(\xi+\eta)/2,-\lambda q)\nonumber\\
\hspace*{-10mm}&&\times\mathcal{B}(\lambda q(\xi+\eta-2)/2,-\lambda q(\xi-\eta)/2)\int_{0}^{1}\!\!\frac{x\,dx}{\{i\Omega_{n}+2\mu_{5}+\lambda q[\xi-1+x(\eta+1)]\}^{3}}.
\end{eqnarray}

Then, we integrate by parts twice with respect to $q$ and obtain the following expression:
\begin{eqnarray}
\label{2case-first-second-3}
\hspace*{-10mm}&&\mathcal{G}^{(2)}_{a}+\mathcal{G}^{(2)}_{b}=-\frac{\alpha}{64\pi^{3}}\int_{0}^{+\infty}\!\!dq\int_{1}^{+\infty}\!\!d\xi\int_{-1}^{1}\!\!d\eta\!\int_{0}^{1}\!\!x\,dx\sum_{\lambda=\pm}\frac{\lambda}{i\Omega_{n}+2\mu_{5}+\lambda q[\xi-1+x(\eta+1)]}\nonumber\\
\hspace*{-10mm}&&\times \frac{(\xi+1)(\eta+1)}{[\xi-1+x(\eta+1)]^{2}} \frac{d^{2}}{dq^{2}}\left[q^{4}\mathcal{F}(\lambda q(\xi+\eta)/2,-\lambda q)\mathcal{B}(\lambda q(\xi+\eta-2)/2,-\lambda q(\xi-\eta)/2)\right].
\end{eqnarray}

Expression for the third diagram could be rewritten in the similar form. In order to do this we use the identity ($z=i\Omega_{n}+2\mu_{5}$)
\begin{eqnarray}
\hspace*{-10mm}&&\frac{1}{(z+2\lambda k)(z-2\lambda q)(z+\lambda (k-q+Q))}=\frac{1}{2(k+q+Q)(k+q)}\frac{1}{z-2\lambda q}\nonumber\\
\hspace*{-10mm}&&-\frac{1}{2(k+q+Q)(k+q)}\frac{1}{z+2\lambda k}-\frac{1}{\lambda(k+q+Q)}\frac{1}{(z+2\lambda k)(z+\lambda (k-q+Q))}.
\end{eqnarray}

First two terms would be proportional to $\delta(q)$ and $\delta(k)$, correspondingly, and finally vanish. The third term could be rewritten with the help of (\ref{feynman}):
\begin{eqnarray}
\label{2case-third-1}
&&\mathcal{G}^{(2)}_{c}=\frac{\alpha}{8\pi^{3}}\int_{0}^{+\infty}\!\!k^{2}dk\int_{0}^{+\infty}\!\!q^{2}dq\int_{0}^{\pi}\!\!\sin\theta d\theta\sum_{\lambda=\pm}\frac{\mathcal{F}(\lambda k,-\lambda q)\mathcal{B}(\lambda (k-q),-\lambda Q)}{Q(k+q+Q)}\nonumber\\
&&\times\int_{0}^{1}\!\!\frac{dx}{[i\Omega_{n}+2\mu_{5}+\lambda k(1+x)-\lambda q(1-x)+\lambda Q(1-x)]^{2}}.
\end{eqnarray}

In elliptic coordinates (\ref{elliptic-2}) we integrate by parts in order to reduce the power of the denominator. After that we obtain 
\begin{eqnarray}
\label{2case-third-2}
\hspace*{-10mm}&&\mathcal{G}^{(2)}_{c}=\frac{\alpha}{32\pi^{3}}\int_{0}^{+\infty}\!\!dq\int_{1}^{+\infty}\!\!d\xi\int_{-1}^{1}\!\!d\eta\!\int_{0}^{1}\!\!dx\sum_{\lambda=\pm}\frac{\lambda}{i\Omega_{n}+2\mu_{5}+\lambda q[\xi-1+x(\eta+1)]}\\
\hspace*{-10mm}&&\times \frac{(\xi+\eta)}{[\xi-1+x(\eta+1)](\xi+1)} \frac{d}{dq}\left[q^{3}\mathcal{F}(\lambda q(\xi+\eta)/2,-\lambda q)\mathcal{B}(\lambda q(\xi+\eta-2)/2,-\lambda q(\xi-\eta)/2)\right].\nonumber
\end{eqnarray}

Then, we do the analytic continuation and take the imaginary part of Eqs.~(\ref{2case-first-second-3}) and (\ref{2case-third-2}). After integrating over $x$ with the help of the delta function the region of integration over the remaining variables is constrained by the conditions $1\leq\xi\leq 1+\epsilon/q$ and $\xi+\eta\geq \epsilon/q$. It could be shown like in the previous subsection that the integral over the region $q<\epsilon/2$ vanishes as $\sim\epsilon$ for $\epsilon\to 0$. Therefore, we consider only the integral over the region $q>\epsilon/2$.
\begin{eqnarray}
\label{2case-first-second-4}
\hspace*{-10mm}&&\Im [G^{(2)}_{{\rm ret},a}+G^{(2)}_{{\rm ret},b}]=-\frac{\alpha}{64\pi^{2}\epsilon^{2}}\int_{\epsilon/2}^{+\infty}\!\!q\,dq\int_{1}^{1+\epsilon/q}\!\!d\xi\int_{\epsilon/q-\xi}^{1}\!\!d\eta \frac{(\xi+1)(\epsilon/q+1-\xi)}{\eta+1}\frac{d^{2}}{dq^{2}}\left[q^{4}\mathcal{F}\cdot\mathcal{B}\right],\\
\label{2case-third-3}
\hspace*{-10mm}&&\Im  G^{(2)}_{{\rm ret},c}=\frac{\alpha}{32\pi^{2}\epsilon}\int_{\epsilon/2}^{+\infty}\!\!dq\int_{1}^{1+\epsilon/q}\!\!d\xi\int_{\epsilon/q-\xi}^{1}\!\!d\eta \frac{(\xi+\eta)}{(\eta+1)(\xi+1)}\frac{d}{dq}\left[q^{3}\mathcal{F}\cdot\mathcal{B}\right].
\end{eqnarray}

These expressions contain the divergences of two types. The first corresponds to electron momentum $k=0$ and occurs at $\xi\sim 1$, $\eta\sim -1$. The second is connected with the photon momentum $Q=0$ at $\xi\sim\eta\sim 1$. We consider them separately. 

In the first case, we put $\xi=1$, $\eta=-1$ in the arguments of $\mathcal{B}$ and $\mathcal{F}$. Then, the integrals over $\xi$ and $\eta$ could be easily taken:
\begin{equation}
\int_{1}^{1+\epsilon/q}\!\!\!\!\left(\frac{\epsilon}{q}+1-\xi\right)(\xi+1)d\xi\int_{\epsilon/q-\xi}^{1}\!\!\frac{1}{1+\eta}d\eta=\frac{\epsilon^{2}}{2q^{2}}\left(-2\ln\frac{\epsilon}{q}+2\ln 2-1\right)+\mathcal{O}(\epsilon^{3})\simeq-\frac{\epsilon^{2}}{q^{2}}\ln\frac{\epsilon}{q}.
\end{equation}
\begin{equation}
\int_{1}^{1+\epsilon/Q}\!\!\!\!d\xi\int_{\epsilon/Q-\xi}^{1}\!\!\frac{\xi+\eta}{(\xi+1)(\eta+1)}d\eta=\frac{\epsilon}{q}+\mathcal{O}(\epsilon^{2}).
\end{equation}

It is obvious that the first two diagrams give the divergent contribution while the third diagram is finite.
\begin{equation}
\label{2case-sum-1}
\Im  G^{(2)}_{\rm ret}=\frac{\alpha}{64\pi^{2}}\ln\frac{\epsilon}{T}\int_{0}^{+\infty}\!\!\frac{dq}{q}\frac{d^{2}}{dq^{2}}\left[q^{4}\mathcal{F}(0,q)\mathcal{B}(q,q)\right]+{\rm finite}.
\end{equation}

Integrating by parts we obtain the final expression with infrared logarithmic divergence: 
\begin{equation}
\label{2case-sum-final}
\Im  G^{(2)}_{\rm ret}=\frac{\alpha}{32\pi^{2}}\ln\frac{\epsilon}{T}\int_{0}^{+\infty}\!\!dq\, q\mathcal{F}(0,q)\mathcal{B}(q,q)+{\rm finite}.
\end{equation}

In the second case, when $Q\to0$, we cannot put $\xi=\eta=1$ directly in the argument of $\mathcal{B}$, because it is singular: $\mathcal{B}\sim -\frac{4T}{q(\xi-\eta)}$. However, we can use its asymptotic in the region $\xi-\eta\ll T/q$. In all other nonsingular expressions we put $\xi=\eta=1$. 
\begin{equation}
\int_{1}^{1+\epsilon/q}\!\!\!\!d\xi\int_{1-\beta}^{1}\!\!d\eta\frac{\epsilon/q+1-\xi}{\xi-\eta}\simeq-\frac{\epsilon^{2}}{2q^{2}}\ln\frac{\epsilon}{q}+\mathcal{O}(\epsilon^{2}),\quad \int_{1}^{1+\epsilon/q}\!\!\!\!d\xi\int_{1-\beta}^{1}\!\!d\eta\frac{1}{\xi-\eta}\simeq-\frac{\epsilon}{q}\ln\frac{\epsilon}{q}+\mathcal{O}(\epsilon).
\end{equation}

And the corresponding expressions for the diagrams read
\begin{eqnarray}
\label{2case-first-second-5}
\hspace*{-10mm}&&\Im [G^{(2)}_{{\rm ret},a}+G^{(2)}_{{\rm ret},b}]=-\frac{\alpha T}{32\pi^{2}}\ln\frac{\epsilon}{T}\int_{0}^{+\infty}\!\!\frac{dq}{q}\frac{d^{2}}{dq^{2}}\left[q^{3}\mathcal{F}(-q,q)\right]+{\rm finite},\\
\label{2case-third-4}
\hspace*{-10mm}&&\Im  G^{(2)}_{{\rm ret},c}=\frac{\alpha T}{16\pi^{2}}\ln\frac{\epsilon}{T}\int_{0}^{+\infty}\!\!\frac{dq}{q}\frac{d}{dq}\left[q^{2}\mathcal{F}(-q,q)\right]+{\rm finite}.
\end{eqnarray}

Integrating by parts we see that the leading divergent parts cancel each other. Therefore, the infrared divergence connected with the zero photon momentum cancels out. After calculating the finite contributions the final expression reads as
\begin{eqnarray}
\hspace*{-10mm}&&\Im  G^{(2)}_{\rm ret}=\frac{\alpha}{64\pi^{2}}\left[\ln\frac{\epsilon}{2T}+1\right]\int_{0}^{+\infty}\!\!dq\,q\sum_{\lambda=\pm}\lambda\mathcal{F}(0,\lambda q)\mathcal{B}(\lambda q,\lambda q)\nonumber\\
\hspace*{-10mm}&&-\frac{\alpha}{64\pi^{2}}\int_{0}^{+\infty}\!\!dq\,q\ln\frac{q}{T}\sum_{\lambda=\pm}\lambda\mathcal{F}(0,\lambda q)\mathcal{B}(\lambda q,\lambda q)-\frac{\alpha}{128\pi^{2}}\!\int_{\!-1}^{1}\!\!\frac{d\eta}{\eta+1}\int_{0}^{+\infty}\!\!dq\,q\sum_{\lambda=\pm}\lambda \nonumber\\
\hspace*{-10mm}&&\times\left[(1-\eta)\mathcal{F}\left(-\frac{\lambda q(1+\eta)}{2},\lambda q\right)\mathcal{B}\left(\frac{\lambda q(1-\eta)}{2},\frac{\lambda q(1-\eta)}{2}\right)-2\mathcal{F}(0,\lambda q)\mathcal{B}(\lambda q,\lambda q)\right].
\end{eqnarray}

\subsection{Cases $\lambda'=-\lambda''=\pm\lambda$}
In this case it could be shown that the expression is finite even without the Feynman parametrization. We expand the expression (\ref{first-order-sum}) into the sum of simple fractions ($z=i\Omega_{n}+2\mu_{5}$):
\begin{eqnarray}
\hspace*{-10mm}&&\frac{1}{(z+2\lambda k)^{2}(z+\lambda k+\lambda'(q+Q))}=-\frac{1}{\lambda k-\lambda' (q+Q)}\frac{1}{(z+2\lambda k)^{2}}\nonumber\\
\hspace*{-10mm}&&-\frac{1}{(\lambda k-\lambda' (q+Q))^{2}}\left[\frac{1}{z+2 \lambda k}-\frac{1}{z+\lambda k+\lambda'(q+Q)}\right],
\end{eqnarray}
\begin{eqnarray}
\hspace*{-10mm}&&\frac{1}{(z+2\lambda k)(z+2\lambda' q)(z+\lambda k+\lambda' (q+Q))}=\frac{1}{2(\lambda k-\lambda'q)(\lambda k-\lambda'(q+Q))}\frac{1}{z+2\lambda k}\nonumber\\
\hspace*{-10mm}&&+\frac{1}{2(\lambda k-\lambda'q)(\lambda k-\lambda'(q-Q))}\frac{1}{z+2\lambda' q} +\frac{1}{Q^{2}-(\lambda k-\lambda'q)^{2}}\frac{1}{z+\lambda k+\lambda' (q+Q)}.
\end{eqnarray}

After analytic continuation and taking the imaginary part the only terms which would give the nonzero contributions are proportional to  $1/(i\Omega_{n}+2\mu_{5}+\lambda k+\lambda' (q+Q))$.
\begin{eqnarray}
\label{3case-1}
&&\mathcal{G}^{(3)}_{a}+\mathcal{G}^{(3)}_{b}+\mathcal{G}^{(3)}_{c}=\frac{\alpha}{16\pi^{3}}\int_{0}^{+\infty}\!\!k^{2}dk\int_{0}^{+\infty}\!\!q^{2}dq\int_{0}^{\pi}\!\!\sin\theta d\theta\sum_{\lambda,\lambda'=\pm}\frac{(-\lambda') \mathcal{F}(\lambda k,\lambda' q)\mathcal{B}(\lambda k+\lambda' q,-\lambda' Q)}{Q(i\Omega_{n}+2\mu_{5}+\lambda k+\lambda' (q+Q)}\nonumber\\
&&\times\left[\frac{1-\lambda\lambda' \cos\theta}{(\lambda k-\lambda' (q+Q))^{2}}+\frac{2}{Q^{2}-(\lambda k-\lambda'q)^{2}}\right].
\end{eqnarray}

In elliptic coordinates
\begin{equation}
\label{elliptic-3}
\xi=\frac{q+Q}{k},\quad \eta=\frac{q-Q}{k},\quad \cos\theta=\frac{\xi\eta+1}{\xi+\eta},\quad k^{2}q^{2}\sin\theta\, dk\,dq\,d\theta=\frac{k^{5}}{8}(\xi^{2}-\eta^{2})dk d\xi d\eta
\end{equation}
the expression has the following form:
\begin{eqnarray}
\label{3case-2}
&&\mathcal{G}^{(3)}=\frac{\alpha}{64\pi^{3}}\int_{0}^{+\infty}\!\!k^{2}dk\int_{1}^{+\infty}\!\!d\xi\int_{-1}^{1}\!\!d\eta\sum_{\lambda,\lambda'=\pm}\left[\frac{(1-\lambda\lambda'\eta)}{(\xi-\lambda\lambda')}+\frac{2\lambda\lambda'(\xi+\eta)}{(\xi-\lambda\lambda')(1-\lambda\lambda'\eta)}\right]\nonumber\\
&&\times\frac{(-\lambda') \mathcal{F}(\lambda k,\lambda' k(\xi+\eta)/2)\mathcal{B}(\lambda k+\lambda' k(\xi+\eta)/2,-\lambda' k(\xi-\eta)/2)}{(i\Omega_{n}+2\mu_{5}+\lambda k(1+\lambda\lambda' \xi))}.
\end{eqnarray}

The nonzero contribution is only in the case $\lambda'=-\lambda$ and it is finite:
\begin{equation}
\label{3case-final}
\Im  G^{(3)}_{\rm ret}=\frac{\alpha}{256\pi^{2}}\int_{0}^{+\infty}\!\!k\,dk\int_{-1}^{1}\!\!d\eta(1-\eta)\sum_{\lambda=\pm}\lambda\mathcal{F}(\lambda k,- \lambda k(1+\eta)/2)\mathcal{B}(\lambda k(1-\eta)/2, \lambda k(1-\eta)/2).
\end{equation}

Taking all results together, we get
\begin{eqnarray}
\label{result-first-order}
\hspace*{-10mm}&&\Im  G_{\rm ret}=-\frac{\alpha}{16\pi^{2}}\left[\ln\frac{\epsilon}{2T}+\frac{3}{2}\right]\left({\rm tanh}\frac{\mu_{R}}{2T}-{\rm tanh}\frac{\mu_{L}}{2T}\right)\left[\pi^{2}T^{2}+\frac{\mu_{R}^{2}+\mu_{L}^{2}}{2}\right]\\
\hspace*{-10mm}&&-\frac{\alpha}{32\pi^{2}}\int_{0}^{+\infty}\!\!dq\,q\ln\frac{q}{T}\sum_{\lambda=\pm}\lambda\mathcal{F}(0,\lambda q)\mathcal{B}(\lambda q,\lambda q)-\frac{\alpha}{64\pi^{2}}\!\int_{\!-1}^{1}\!\!\frac{d\eta}{\eta+1}\int_{0}^{+\infty}\!\!dq\,q\sum_{\lambda=\pm}\lambda \nonumber\\
\hspace*{-10mm}&&\times\left[\mathcal{F}\left(\frac{\lambda q(1+\eta)}{2},\frac{\lambda q(1-\eta)}{2}\right)+\mathcal{F}\left(-\frac{\lambda q (1+\eta)}{1-\eta},\frac{2\lambda q}{1-\eta}\right)-\frac{(3-\eta)}{2}\mathcal{F}(0,\lambda q)\right]\mathcal{B}(\lambda q,\lambda q).\nonumber
\end{eqnarray}
The corresponding value of the chirality flipping rate is given by Eq.~(\ref{gamma-flip-log-divergent}).

\section{Fermion self-energy and full propagator in HTL approximation}
\label{app-self-energy}

\begin{figure}[h!]
	\centering
	\includegraphics[width=6cm]{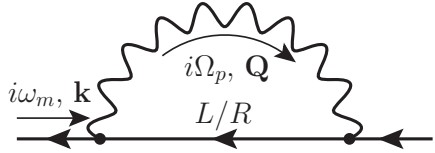}
	\caption{One-loop fermion self-energy. \label{fig-ap-self-energy}}
\end{figure}

The leading order fermion self-energy is given by the one-loop diagram shown in Fig.~\ref{fig-ap-self-energy}.

The analytical expression in the momentum space is the following:
\begin{equation}
\Sigma(i\omega_{m},\mathbf{k})=e^{2}T\sum_{p}\int\frac{d^{3}\mathbf{Q}}{(2\pi)^{3}}\frac{\gamma^{\mu}\mathcal{S}_{0}(i\omega_{m-p},\mathbf{k}-\mathbf{Q})\gamma_{\mu}}{(i\Omega_{p})^{2}-Q^{2}},
\end{equation}
where $\mathcal{S}_{0}$ is the free electron propagator in Matsubara representation given by Eq.~(\ref{propagator}). Taking into account the block structure of the propagator and Dirac gamma-matrices, we conclude that the self-energy also has the block structure and does not mix the left and right components. This is a consequence of the fact that the EM interaction respects the chiral symmetry. Therefore, one can consider the left and right components of the self-energy separately.

The leading contribution to the self-energy is captured by the HTL approximation when the external momentum is considered to be much less than the loop momentum and the former is neglected everywhere except the denominator. The final result has the following form \cite{LeBellac}:
\begin{equation}
\Sigma_{L,R}(i\omega_{m},\mathbf{k})=\frac{m_{\rm th}^{2}}{2}\int\frac{d\Omega_{\mathbf{v}}}{4\pi}\frac{1\pm \boldsymbol{\sigma}\cdot\mathbf{v}}{i\omega_{m}+\mu_{L,R}-\mathbf{k}\cdot\mathbf{v}},
\end{equation}
where the integration is performed over all possible directions of the unit vector $\mathbf{v}$ and $m_{\rm th}^{2}=e^{2}T^{2}/4$ is the electron asymptotic thermal mass (meaning that, for hard momenta, the energy dispersion of the electron quasiparticle takes the usual form $\epsilon=\sqrt{k^{2}+m^{2}_{\rm th}}$). One of the advantages of the HTL self-energy is its gauge invariance with respect to the Lorentz covariant gauges. 

Spatial isotropy leads to the simple matrix structure of the self-energy
\begin{equation}
\label{self-energy}
\Sigma_{L,R}(i\omega_{m},\mathbf{k})=\Sigma^{0}_{L,R}(i\omega_{m},\mathbf{k})+ \frac{\boldsymbol{\sigma}\cdot\mathbf{k}}{k}\Sigma^{1}_{L,R}(i\omega_{m},\mathbf{k}),
\end{equation}
where
\begin{equation}
\label{sigma-0}
\Sigma^{0}_{L,R}(i\omega_{m},\mathbf{k})=\frac{1}{2}{\rm tr}\Sigma_{L,R}(i\omega_{m},\mathbf{k})=\frac{m_{\rm th}^{2}}{2}\int\frac{d\Omega_{\mathbf{v}}}{4\pi}\frac{1}{i\omega_{m}+\mu_{L,R}-\mathbf{k}\cdot\mathbf{v}},
\end{equation}
\begin{equation}
\Sigma^{1}_{L,R}(i\omega_{m},\mathbf{k})=\pm\frac{1}{2k}{\rm tr}\left[(\boldsymbol{\sigma}\cdot\mathbf{k})\Sigma_{L,R}(i\omega_{m},\mathbf{k})\right]=-\frac{m_{\rm th}^{2}}{2k}+\frac{i\omega_{m}+\mu_{L,R}}{k}\Sigma^{0}_{L,R}(i\omega_{m},\mathbf{k}).
\end{equation}
Thus, the fermion self-energy in HTL approximation is determined by the single scalar function $\Sigma^{0}_{L,R}=\Sigma^{0}(i\omega_{m}+\mu_{L,R},\mathbf{k})$. Let us consider it as a function of the complex variable $k^{0}$ in a complex plane with the branch cut along the real axis from $k^{0}=-k$ to $k^{0}=k$. Then, the integration can be performed explicitly and we obtain:
\begin{equation}
\label{sigma-0-complex}
\Sigma^{0}(k^{0},\mathbf{k})=\frac{m_{\rm th}^{2}}{2}\int\frac{d\Omega_{\mathbf{v}}}{4\pi}\frac{1}{k^{0}-\mathbf{k}\cdot\mathbf{v}}=\frac{m_{\rm th}^{2}}{4k}\ln\frac{k^{0}+k}{k^{0}-k}.
\end{equation}

The full fermion propagator is determined by the Schwinger-Dyson equation
\begin{equation}
\label{SD-eq}
\mathcal{S}_{L,R}=\mathcal{S}_{0\,L,R}+\mathcal{S}_{0\,L,R}\Sigma_{L,R}\mathcal{S}_{L,R}
\end{equation}
whose solution reads as usual
\begin{equation}
\label{SD-solution}
\mathcal{S}_{L,R}=\left(\mathcal{S}_{0\, L,R}^{-1}-\Sigma_{L,R}\right)^{-1}.
\end{equation}
Substituting the HTL self-energy (\ref{self-energy}), we obtain the full propagator in the following form:
\begin{equation}
\label{full-prop-general}
\mathcal{S}_{L,R}(i\omega_{m},\mathbf{k})=\frac{i\omega_{m}+\mu_{L,R}-\Sigma^{0}_{L,R}\mp (\boldsymbol{\sigma}\cdot\mathbf{k})\left[1+\frac{m_{\rm th}^{2}}{2k^{2}}-\frac{i\omega_{m}+\mu_{L,R}}{k^{2}}\Sigma^{0}_{L,R}\right]}{\left(i\omega_{m}+\mu_{L,R}-\Sigma^{0}_{L,R}\right)^{2}-k^{2}\left(1+\frac{m_{\rm th}^{2}}{2k^{2}}-\frac{i\omega_{m}+\mu_{L,R}}{k^{2}}\Sigma^{0}_{L,R}\right)^{2}}.
\end{equation}

\begin{figure}[h!]
	\centering
	\includegraphics[width=8cm]{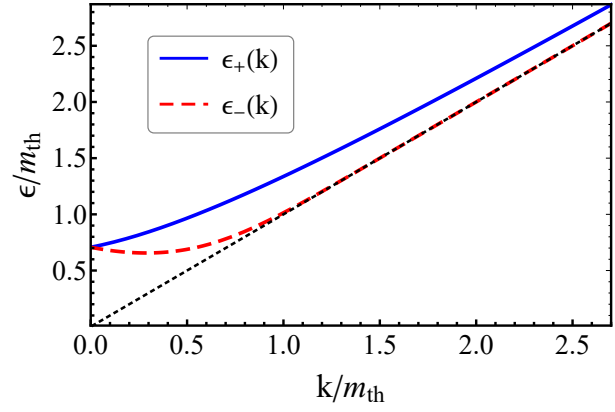}
	\caption{Energy dispersion for the electron quasiparticles: normal branch $\epsilon_{+}(k)$ (blue solid line) and plasmino branch $\epsilon_{-}(k)$ (red dashed line). The black dotted line shows the free massless electron dispersion in vacuum. \label{fig-app-dispersion}}
\end{figure}

It can be represented as a decomposition into the components with positive and negative helicity
\begin{equation}
\mathcal{S}_{L,R}(i\omega_{m},\mathbf{k})=\sum_{\lambda=\pm}\frac{1}{\Delta_{\lambda}(i\omega_{m}+\mu_{L,R},\mathbf{k})}\frac{1\mp \lambda\boldsymbol{\sigma}\cdot \hat{\mathbf{k}}}{2},
\end{equation}
where $\hat{\mathbf{k}}=\mathbf{k}/k$, 
\begin{equation}
\label{delta-denominator}
\Delta_{\lambda}(k^{0},\mathbf{k})=k^{0}-\Sigma^{0}(k^{0},\mathbf{k})-\lambda k \left(1+\frac{m_{\rm th}^{2}}{2k^{2}}-\frac{k^{0}}{k^{2}}\Sigma^{0}(k^{0},\mathbf{k})\right)
\end{equation}
is the denominator whose zeros determine the quasiparticle spectrum, and the function $\Sigma^{0}$ is given by Eq.~(\ref{sigma-0-complex}). The quasiparticle dispersion relations can be found from the requirement $\Delta_{\pm}(k^0,\mathbf{k})=0$. This gives
\begin{equation}
\epsilon_{\pm}(k)=\pm k\frac{\mathcal{A}_{\pm}(k)-1}{\mathcal{A}_{\pm}(k)+1},
\end{equation}
where $\mathcal{A}_{+}(k)=\mathcal{W}_{-1}(z)$, $\mathcal{A}_{-}(k)=\mathcal{W}_{0}(z)$, $z=-\exp\left(-4\frac{k^{2}}{m_{\rm th}^{2}}-1\right)$, and $\mathcal{W}_{0,-1}$ are the upper and lower branches of the Lambert W-function. Since $\mathcal{W}_{-1}(z)\leq-1$ for $z\in[-e^{-1};0)$ and $\mathcal{W}_{0}(z)\in [-1;0)$ for $z\in[-e^{-1};0)$, the values of $\epsilon_{\pm}$ are always positive. They are shown in Fig.~\ref{fig-app-dispersion}.

It is possible to find the asymptotic expressions for the dispersion relations for small and large momenta \cite{LeBellac}:
\begin{equation}
\epsilon_{\pm}(k\ll m_{\rm th})\approx \frac{m_{\rm th}}{\sqrt{2}}\pm \frac{k}{3},
\end{equation}
\begin{equation}
\label{asym-dispersion-large}
\epsilon_{+}(k\gg m_{\rm th})\approx k+\frac{m_{\rm th}^{2}}{2k}\approx \sqrt{k^{2}+m_{\rm th}^{2}}, \quad \epsilon_{-}(k\gg m_{\rm th})\approx k\left[1+2\exp\left(-4\frac{k^{2}}{m_{\rm th}^{2}}-1\right)\right].
\end{equation}

The spectral function $\rho_{\pm}=-2\Im \Delta_{\pm}^{-1}$ is given by Eq.~(\ref{spectral-function}), where the residues $Z_{\pm}(k)$ of the quasiparticle poles are given by
\begin{equation}
Z_{\pm}(k)=\frac{\epsilon_{\pm}^{2}-k^{2}}{m_{\rm th}^{2}}.
\end{equation}
For small momenta the spectral density is almost equally distributed between both poles,
\begin{equation}
Z_{\pm}(k\ll m_{\rm th})\approx \frac{1}{2}\pm\frac{k\sqrt{2}}{3 m_{\rm th}},
\end{equation}
however, for large momenta the ordinary branch survives and the plasmino branch is exponentially suppressed,
\begin{equation}
Z_{+}(k\gg m_{\rm th})\approx 1+\frac{m_{\rm th}^{2}}{4k^{2}}\left(1-\ln\frac{4k^{2}}{m_{\rm th}^{2}}\right),\quad Z_{-}(k\gg m_{\rm th})\approx \frac{4k^{2}}{m_{\rm th}^{2}}\exp\left(-\frac{4k^{2}}{m_{\rm th}^{2}}-1\right).
\end{equation}

Finally, the incoherent part of the spectral density has the form
\begin{equation}
\label{incoherent-part}
\rho_{\pm}^{LD}(k^{0},\mathbf{k})=\frac{\frac{\pi m_{\rm th}^{2}}{2k}\left(1\mp\frac{k^{0}}{k}\right)\theta(k^{2}-k^{2}_{0})}{\left[(k^{0}\mp k)\left(1\pm \frac{m_{\rm th}^{2}}{4k^{2}}\ln\left|\frac{k+k^{0}}{k-k^{0}}\right|\right)\mp\frac{m_{\rm th}^{2}}{2k}\right]^{2}+\left[\frac{\pi m_{\rm th}^{2}}{4k}\left(1\mp\frac{k^{0}}{k}\right)\right]^{2}}.
\end{equation}
\normalsize

\section{Details of calculation of the chirality flipping rate}
\label{app-details}

\subsection{Contribution from the plasmino branch}
\label{subapp-plasmino}

The plasmino branch $\epsilon_{-}(k)$ exists only in the soft momentum region $k\sim eT$. For larger momenta, the corresponding residue is exponentially suppressed. If the plasmino pole is multiplied by any other pole contribution, the following power counting takes place:
\begin{multline}
\label{soft-pole-pole}
\Gamma_{\rm flip}^{7c,(-,\pm)}\sim \frac{m_{e}^{2}}{T^{3}}\int_{\rm soft}\!\!\! d^{3}\mathbf{k} \int dk^{0}\delta_{\gamma_{e}}(k^{0}-\epsilon_{1})\delta_{\gamma_{e}}(k^{0}-\epsilon_{2})\sim \frac{m_{e}^{2}}{T^{3}}\int_{\rm soft}\!\!\! d^{3}\mathbf{k}\  \delta_{2\gamma_{e}}(\Delta\epsilon)\\
\sim \frac{m_{e}^{2}}{T^{3}} (eT)^{3} \frac{\gamma_{e}}{(\Delta\epsilon)^{2}}\sim \alpha^{3/2} \ln\alpha^{-1}\frac{m_{e}^{2}}{T},
\end{multline}
where $\epsilon_{1,2}$ are the energies of the poles at least one of which is of the plasmino branch, $\Delta\epsilon=|\epsilon_{1}-\epsilon_{2}|\sim eT$ for soft momenta. Similar contribution comes from the interference term between the plasmino pole and the Landau damping part
\begin{multline}
\label{soft-interference}
\Gamma_{\rm flip}^{7c,(-,{\rm LD})}\sim \frac{m_{e}^{2}}{T^{3}}\int_{\rm soft}\!\!\! d^{3}\mathbf{k} \int_{-k}^{k} dk^{0}\delta_{\gamma_{e}}(k^{0}-\epsilon_{-})\rho_{+}^{\rm LD}(k^{0},\mathbf{k})\sim \frac{m_{e}^{2}}{T^{3}}\int_{\rm soft}\!\!\! d^{3}\mathbf{k}\ \frac{\gamma_{e}/\pi}{(\epsilon_{-}-k)}\frac{1}{eT}\\
\sim \frac{m_{e}^{2}}{T^{3}} (eT)^{3} \frac{\gamma_{e}}{eT}\frac{1}{eT}\sim \alpha^{3/2} \ln\alpha^{-1}\frac{m_{e}^{2}}{T}.
\end{multline}
Therefore, in what follows we will omit the plasmino contribution in the spectral function $\rho_{\pm}$.

\subsection{Contribution of the incoherent part}
\label{subapp-incoh}

Further, let us calculate the contribution coming from the overlap between the incoherent parts of the spectral functions. Using expression (\ref{incoherent-part}), we have Eq.~(\ref{gamma-LD-LD}). Introducing new dimensionless integration variables $x$ and $y$ by the relations
\begin{equation}
k_{0}=m_{\rm th} xy,\qquad k=m_{\rm th} x,
\end{equation}
we rewrite Eq.~(\ref{gamma-LD-LD}) as
\begin{eqnarray}
\label{gamma-LD-LD-2}
\Gamma_{\rm flip}^{7c,({\rm LD})}&=&\frac{3m_{e}^{2}m_{\rm th}^{2}}{4\pi T^{3}}\int_{0}^{1}\frac{dy}{1-y^{2}}\int_{0}^{\infty} \frac{x^{5}\,dx}{{\rm cosh}^{2}\left(\frac{m_{\rm th}}{2T}xy\right)}\nonumber\\
&\times& \frac{1}{\left[\left(x^{2}+\frac{1}{4}\ln\frac{1+y}{1-y}+\frac{1}{2(1-y)}\right)^{2}+\frac{\pi^{2}}{16}\right]\left[\left(x^{2}-\frac{1}{4}\ln\frac{1+y}{1-y}+\frac{1}{2(1+y)}\right)^{2}+\frac{\pi^{2}}{16}\right]}.
\end{eqnarray}
It is obvious that the integrand is peaked in the region of $x\sim 1$ (i.e., the soft momentum region $k\sim eT$). In fact, for $x\ll 1$ we have the suppression of the integrand $\sim x^{5}$. On the contrary, for $x\gg 1$ we have $\sim \frac{\ln x}{x^{3}}$. Since that, we can safely replace the hyperbolic cosine with the unity. It starts to be important for $x\sim T/m_{\rm th}\propto e^{-1}\gg 1$. That is why neglecting it, we would bring the relative error of order
\begin{equation}
\int_{e^{-1}}^{\infty}\frac{\ln x}{x^{3}} \sim e^{2}\ln e^{-1},
\end{equation}
i.e., of higher order in $\alpha$. Then, the result for the chirality flipping rate is given by Eq.~(\ref{gamma-flip-LD}) with the constant $C$ given by Eq.~(\ref{C-const}).

\subsection{Contribution of the quasiparticle pole}
\label{subapp-pole}

The overlap between the quasiparticle pole and the incoherent part of the spectral function is determined by Eq.~(\ref{flip-pole}) where the spectral function far from the quasiparticle shell is needed. According to Eq.~(\ref{rho-far}), it is fully determined by the corresponding retarded fermion self-energy. As usual, we calculate the Matsubara self-energy and after that perform its analytic continuation to the real axis. The one-loop electron self-energy is shown in Fig.~\ref{fig-ap-self-energy}. Its component with the certain helicity $\lambda$ is given by
\begin{equation}
\Sigma_{\lambda}(i\omega_{m},\mathbf{k})=-e^{2}\int\frac{d^{3}\mathbf{Q}}{(2\pi)^{3}}T\sum_{p}\sum_{\lambda'=\pm} A^{\mu\nu}_{\lambda\lambda'}(\mathbf{k},\mathbf{k}-\mathbf{Q}) \vphantom{\mathcal{D}}^{*}\mathcal{D}_{\mu\nu}(i\Omega_{p},\mathbf{Q}) \Delta^{-1}_{\lambda'}(i\omega_{m-p},\mathbf{k}-\mathbf{Q}),
\end{equation}
where $\vphantom{\mathcal{D}}^{*}\mathcal{D}_{\mu\nu}$ is the full photon propagator and
\begin{equation}
A^{\mu\nu}_{\lambda\lambda'}(\mathbf{k},\mathbf{q})={\rm tr}\left[\frac{\mathds{1}+\lambda\boldsymbol{\sigma}\cdot\hat{\mathbf{k}}}{2}\sigma^{\mu}\frac{\mathds{1}+\lambda'\boldsymbol{\sigma}\cdot\hat{\mathbf{q}}}{2}\sigma^{\nu}\right].
\end{equation}
Here, $\hat{\mathbf{k}}=\mathbf{k}/k$. For further convenience, we list here the expressions for the components of $A^{\mu\nu}$
\begin{eqnarray}
A^{00}_{\lambda\lambda'}(\mathbf{k},\mathbf{q})&=&\frac{1}{2}\left(1+\lambda\lambda'\frac{\mathbf{k}\cdot\mathbf{q}}{kq}\right),\nonumber\\
A^{0i}_{\lambda\lambda'}(\mathbf{k},\mathbf{q})&=&A^{i0}_{\lambda\lambda'}(\mathbf{k},\mathbf{q})=\frac{1}{2}\left(\lambda\frac{k^{i}}{k}+\lambda'\frac{q^{i}}{q}\right),\nonumber\\
A^{ij}_{\lambda\lambda'}(\mathbf{k},\mathbf{q})&=&\frac{1}{2}\left[\delta^{ij}\left(1-\lambda\lambda'\frac{\mathbf{k}\cdot\mathbf{q}}{kq}\right)+\lambda\lambda'\frac{k^{i}q^{j}+k^{j}q^{i}}{kq}\right].\label{Amunu}
\end{eqnarray}
Using the spectral representation for the propagators
\begin{equation}
\label{spectral-prop}
\vphantom{\mathcal{D}}^{*}\mathcal{D}_{\mu\nu}(i\Omega_{p},\mathbf{Q})=\int\frac{dQ^{0}}{2\pi}\frac{\vphantom{\rho}^{*\!\!}\rho_{\mu\nu}(Q^{0},\mathbf{Q})}{i\Omega_{p}-Q^{0}}, \quad 
\Delta^{-1}_{\lambda'}(i\omega_{m-p},\mathbf{k}-\mathbf{Q})=\int\frac{dq^{0}}{2\pi}\frac{\rho_{\lambda'}(q^{0},\mathbf{k}-\mathbf{Q})}{i\omega_{m-p}-q^{0}},
\end{equation}
we can perform the summation over the Matsubara frequencies
\begin{equation}
\Sigma_{\lambda}(i\omega_{m},\mathbf{k})=\frac{e^{2}}{2}\int\frac{d^{4}Q}{(2\pi)^{4}}\int\frac{dq^{0}}{2\pi}\sum_{\lambda'=\pm}\frac{{\rm coth}\frac{Q^{0}}{2T}+{\rm tanh}\frac{q^{0}}{2T}}{i\omega_{m}-q^{0}-Q^{0}}\vphantom{\rho}^{*\!\!}\rho_{\mu\nu}(Q^{0},\mathbf{Q})\rho_{\lambda'}(q^{0},\mathbf{k}-\mathbf{Q})A^{\mu\nu}_{\lambda\lambda'}(\mathbf{k},\mathbf{k}-\mathbf{Q}).
\end{equation}
Then, performing the analytic continuation to the real axis, we get the spectral density in the form
\begin{multline}
\rho_{-}(\epsilon_{+}(k)+x,k)\approx \frac{e^{2}}{8k^{2}}\int\frac{d^{4}Q}{(2\pi)^{4}}\left({\rm coth}\frac{Q^{0}}{2T}+{\rm tanh}\frac{\epsilon_{+}(k)-Q^{0}}{2T}\right)\vphantom{\rho}^{*\!\!}\rho_{\mu\nu}(Q^{0},\mathbf{Q})\\
\times\sum_{\lambda'=\pm} \rho_{\lambda'}\big(\epsilon_{+}(k)+x-Q^{0},\mathbf{k}-\mathbf{Q}\big) A^{\mu\nu}_{-,\lambda'}(\mathbf{k},\mathbf{k}-\mathbf{Q}),
\end{multline}
which leads to the chirality flipping rate
\begin{multline}
\label{gamma-flip-pole-general}
\Gamma_{\rm flip}^{7c,({\rm pole})}=\frac{3m_{e}^{2}\alpha}{\pi T^{3}}\int_{0}^{\infty}\frac{dk}{{\rm cosh}^{2}\frac{k}{2T}}\int\frac{d^{4}Q}{(2\pi)^{4}}\left({\rm coth}\frac{Q^{0}}{2T}+{\rm tanh}\frac{\epsilon_{+}(k)-Q^{0}}{2T}\right)\vphantom{\rho}^{*\!\!}\rho_{\mu\nu}(Q^{0},\mathbf{Q})\\
\times\sum_{\lambda'=\pm}A^{\mu\nu}_{-,\lambda'}(\mathbf{k},\mathbf{k}-\mathbf{Q})\int_{-\infty}^{\infty}dx\,\delta_{\gamma_{e}}(x) \rho_{\lambda'}\big(\epsilon_{+}(k)+x-Q^{0},\mathbf{k}-\mathbf{Q}\big).
\end{multline}
This expression contains three different contributions which must be considered separately.

\paragraph{Soft electron in the loop.} For soft electrons, their spectral density $\rho_{\lambda'}$ is distributed in comparable amounts among the quasiparticle poles and incoherent part. On the other hand, the photon is obviously hard and we can substitute its spectral density in the form
\begin{equation}
\label{sp-dens-hard-photon}
\vphantom{\rho}^{*\!\!}\rho_{\mu\nu}(Q^{0},\mathbf{Q})=P_{\mu\nu}^{t}(\mathbf{Q})\sum_{\lambda''=\pm}\lambda'' \frac{\pi}{Q}\delta(Q^{0}-\lambda''\omega_{t}(Q)),
\end{equation}
where $\omega_{t}(Q)$ is the energy dispersion of hard transverse photon and
\begin{equation}
P_{\mu\nu}^{t}(\mathbf{Q})=\delta^{i}_{\mu}\delta^{j}_{\nu}\left(\delta_{ij}-\frac{Q_{i}Q_{j}}{Q^{2}}\right)
\end{equation}
is the transverse projector. We take only the part corresponding to the transverse photon, because it contains almost all the spectral weight at hard momenta \cite{LeBellac}. It is convenient to use the electron momentum $\mathbf{q}=\mathbf{k}-\mathbf{Q}$ as the integration variable. The nonzero contribution comes from $\lambda''=+1$, because in this case $\epsilon_{+}(k)-Q^{0}\approx q\cos\theta$ is small. Then, we obtain the following result:
\begin{equation}
\Gamma_{\rm flip}^{7c,({\rm pole},\,{\rm soft\,}e)}=\frac{3m_{e}^{2}\alpha}{4\pi^{3} T^{3}}\int_{\Lambda}^{\infty}\frac{dk}{k\,{\rm sinh}\frac{k}{T}}\int_{0}^{\Lambda}q^{2}dq\int_{0}^{\pi}d\cos\theta 
(1+\lambda'\cos\theta)\int_{-\infty}^{\infty}dx\,\delta_{\gamma_{e}}(x) \rho_{\lambda'}\big(q\cos\theta+x,\mathbf{q}\big),
\end{equation}
where $\Lambda \gtrsim eT$ is the separation scale between soft and hard momenta. We can roughly estimate this expression taking into account that in the soft momentum region the electron spectral function is not singular and is always of order $\sim 1/(eT)$. We get
\begin{equation}
\Gamma_{\rm flip}^{7c,({\rm pole},\,{\rm soft\,}e)}\propto \frac{m_{e}^{2}}{T^{3}}\alpha\times \frac{T}{\Lambda}\times \frac{(eT)^{3}}{(eT)}\lesssim \frac{m_{e}^{2}}{T^{3}}\alpha^{3/2},
\end{equation}
meaning that this contribution can be neglected.

\paragraph{Soft photon in the loop.}
If we restrict the $Q$ integration in Eq.~(\ref{gamma-flip-pole-general}) to the soft region, i.e., $Q^{0},\,Q\lesssim eT$, further simplifications can be done. First of all, the electron's momentum in the loop is hard and almost equal to the external one. Therefore, we can use the spectral density of the hard electron which leads to the following chirality flipping rate 
\begin{multline}
\label{gamma-flip-hard-e}
\Gamma_{\rm flip}^{7c,({\rm pole})}=\frac{6m_{e}^{2}\alpha}{T^{3}}\int_{0}^{\infty}\frac{dk}{{\rm cosh}^{2}\frac{k}{2T}}\int\frac{d^{4}Q}{(2\pi)^{4}}\left({\rm coth}\frac{Q^{0}}{2T}+{\rm tanh}\frac{\epsilon_{+}(k)-Q^{0}}{2T}\right)\\
\times\vphantom{\rho}^{*\!\!}\rho_{\mu\nu}(Q^{0},\mathbf{Q})\sum_{\lambda'=\pm} \delta_{2\gamma_{e}}\big(\epsilon_{+}(k)-Q^{0}-\lambda'\epsilon_{+}(\mathbf{k}-\mathbf{Q})\big) A^{\mu\nu}_{-,\lambda'}(\mathbf{k},\mathbf{k}-\mathbf{Q}).
\end{multline}
The Lorentz function (ensuring the approximate energy conservation) works only for $\lambda'=+1$ and the tensor $A^{\mu\nu}_{-+}(\mathbf{k},\mathbf{k}-\mathbf{Q})\approx P^{\mu\nu}_{t}(\mathbf{k})$. Finally, expanding the hyperbolic cotangent, we obtain Eq.~(\ref{soft-gamma-12}). In order to estimate that expression, we can use the following approximate formula
\begin{equation}
\label{rho-mu-nu-soft}
\frac{\rho_{\mu\nu}(Q^{0},\mathbf{Q})}{Q^{0}}\approx P^{\mu\nu}_{t}(\mathbf{Q})\frac{2\pi}{Q^{2}}\delta(Q^{0}), \qquad Q\ll eT.
\end{equation}
It becomes more and more exact as the momentum $Q$ decreases. Then, we get
\begin{eqnarray}
\Gamma_{\rm flip}^{7c,({\rm pole},\,{\rm soft}\, \gamma)}&=&\frac{6m_{e}^{2}\alpha}{\pi^{2}T}\int_{0}^{\Lambda}dQ\int_{0}^{\pi} d\cos\theta (1+\cos^{2}\theta) \delta_{2\gamma_{e}}(Q\cos\theta)\nonumber\\
&\approx& \frac{6m_{e}^{2}\alpha}{\pi^{2}T} \ln\frac{\Lambda}{2\gamma_{e}}\approx \frac{3}{\pi^{2}}\frac{m_{e}^{2}}{T} \alpha\ln\alpha^{-1}.
\end{eqnarray}

\paragraph{Nearly collinear processes.} If both, the electron and the photon running in the loop, are hard, we can use Eq.~(\ref{gamma-flip-hard-e}) and the photon spectral density (\ref{sp-dens-hard-photon}). This leads to Eq.~(\ref{gamma-flip-collinear-general}).
There are four ways of choosing the signs of $\lambda'$ and $\lambda''$ which correspond to different nearly collinear processes in plasma (one of them is forbidden though).

\subparagraph{The case $\lambda'=\lambda''=+1$} corresponds to the process of nearly collinear bremsstrahlung $e_{L,R}(\mathbf{k})\to e_{R,L}(\mathbf{k}-\mathbf{Q})+\gamma(\mathbf{Q})$, schematically shown by the first diagram in Fig.~\ref{fig-collinear-processes}. This process is possible if $Q<k$. We decompose
\begin{equation}
\mathbf{Q}=Q_{\parallel}\hat{\mathbf{k}}+\mathbf{Q}_{\perp}, \qquad \mathbf{k}-\mathbf{Q}=(k-Q_{\parallel})\hat{\mathbf{k}}-\mathbf{Q}_{\perp},
\end{equation}
and use the asymptotic expressions for the electron's and photon's dispersion relations
\begin{equation}
\epsilon_{+}(k)\approx k+\frac{m_{\rm th}^{2}}{2k},\qquad \omega_{t}(Q)\approx Q+\frac{m_{\gamma}^{2}}{2Q},
\end{equation}
where $m_{\rm th}=eT/2$ and $m_{\gamma}=eT/\sqrt{6}$ are the corresponding thermal masses. Then, we obtain
\begin{multline}
\label{gamma-flip-collinear-brem}
\Gamma_{\rm flip}^{7c,({\rm pole},\,{\rm brem})}=\frac{3m_{e}^{2}\alpha}{2\pi^{2}T^{3}}\int_{0}^{\infty}\frac{dk}{{\rm cosh}^{2}\frac{k}{2T}}\int_{0}^{k}\frac{dQ_{\parallel}}{Q_{\parallel}}\left({\rm coth}\frac{Q_{\parallel}}{2T}+{\rm tanh}\frac{k-Q_{\parallel}}{2T}\right)\\
\times\int_{0}^{\infty}Q_{\perp}\,dQ_{\perp} \delta_{2\gamma_{e}}\Big(\frac{Q_{\perp}^{2}}{2}\frac{k}{Q_{\parallel}(k-Q_{\parallel})}+\frac{m_{\gamma}^{2}}{2Q_{\parallel}}+\frac{m_{\rm th}^{2}Q_{\parallel}}{2k(k-Q_{\parallel})}\Big).
\end{multline}
In this expression, the integration over $Q_{\perp}$ can be easily done. Finally, introducing the new variables
\begin{equation}
Q_{\parallel}=2Tx,\qquad k=2Tx(y+1),
\end{equation}
and using the identity
\begin{equation}
\label{identity}
\frac{{\rm coth\,}a+{\rm tanh\,}b}{{\rm cosh}^{2}(a+b)}=\frac{{\rm tanh\,}(a+b)-{\rm tanh\,}b}{{\rm sinh}^{2}a},
\end{equation}
with $a=x$ and $b=xy$, we obtain the following result for the chirality flipping rate:
\begin{equation}
\label{gamma-flip-collinear-brem-2}
\Gamma_{\rm flip}^{7c,({\rm pole},\,{\rm brem})}=\frac{3m_{e}^{2}\alpha}{\pi^{2}T}\int_{0}^{\infty}\frac{x\,dx}{{\rm sinh}^{2}x}\int_{0}^{\infty}dy\frac{y}{y+1}\left[{\rm tanh\,}x(y+1)-{\rm tanh\,}xy\right]
\frac{2}{\pi}{\rm arctan}\frac{8\gamma_{e}Tx}{m_{\gamma}^{2}+m_{\rm th}^{2}/(y(y+1))}.
\end{equation}

\subparagraph{The case $\lambda'=+1,\ \lambda''=-1$} corresponds to the process of nearly collinear absorption of the photon $e_{L,R}(\mathbf{k})+\gamma(\mathbf{Q})\to e_{R,L}(\mathbf{k}+\mathbf{Q})$, schematically shown by the second diagram in Fig.~\ref{fig-collinear-processes}. This process is possible for any $Q$ and $k$. Decomposing
\begin{equation}
\mathbf{Q}=Q_{\parallel}\hat{\mathbf{k}}+\mathbf{Q}_{\perp}, \qquad \mathbf{k}+\mathbf{Q}=(Q_{\parallel}+k)\hat{\mathbf{k}}+\mathbf{Q}_{\perp},
\end{equation}
we obtain
\begin{multline}
\label{gamma-flip-collinear-abs}
\Gamma_{\rm flip}^{7c,({\rm pole},\,{\rm abs})}=\frac{3m_{e}^{2}\alpha}{2\pi^{2}T^{3}}\int_{0}^{\infty}\frac{dk}{{\rm cosh}^{2}\frac{k}{2T}}\int_{0}^{\infty}\frac{dQ_{\parallel}}{Q_{\parallel}}\left({\rm coth}\frac{Q_{\parallel}}{2T}-{\rm tanh}\frac{k+Q_{\parallel}}{2T}\right)\\
\times\int_{0}^{\infty}Q_{\perp}\,dQ_{\perp} \delta_{2\gamma_{e}}\Big(\frac{Q_{\perp}^{2}}{2}\frac{k}{Q_{\parallel}(Q_{\parallel}+k)}+\frac{m_{\gamma}^{2}}{2Q_{\parallel}}+\frac{m_{\rm th}^{2}Q_{\parallel}}{2k(Q_{\parallel}+k)}\Big).
\end{multline}
Again, we integrate over $Q_{\perp}$, introduce the new variables
\begin{equation}
Q_{\parallel}=2Tx,\qquad k=2Tx y,
\end{equation}
and use identity (\ref{identity}) with $a=x$ and $b=-x(y+1)$ to get
\begin{equation}
\label{gamma-flip-collinear-abs-2}
\Gamma_{\rm flip}^{7c,({\rm pole},\,{\rm abs})}=\frac{3m_{e}^{2}\alpha}{\pi^{2}T}\int_{0}^{\infty}\frac{x\,dx}{{\rm sinh}^{2}x}\int_{0}^{\infty}dy\frac{y+1}{y}\left[{\rm tanh\,}x(y+1)-{\rm tanh\,}xy\right]\frac{2}{\pi}{\rm arctan}\frac{8\gamma_{e}Tx}{m_{\gamma}^{2}+m_{\rm th}^{2}/(y(y+1))}.
\end{equation}

\subparagraph{The case $\lambda'=-1,\ \lambda''=+1$} corresponds to the process of nearly collinear annihilation $e_{L,R}(\mathbf{k})+ \bar{e}_{L,R}(\mathbf{Q}-\mathbf{k})\to \gamma(\mathbf{Q})$, schematically shown by the second diagram in Fig.~\ref{fig-collinear-processes}. This process is possible if $Q>k$. Decomposing
\begin{equation}
\mathbf{Q}=Q_{\parallel}\hat{\mathbf{k}}+\mathbf{Q}_{\perp}, \qquad \mathbf{k}-\mathbf{Q}=-(Q_{\parallel}-k)\hat{\mathbf{k}}-\mathbf{Q}_{\perp},
\end{equation}
we obtain
\begin{multline}
\label{gamma-flip-collinear-annihil}
\Gamma_{\rm flip}^{7c,({\rm pole},\,{\rm annih})}=\frac{3m_{e}^{2}\alpha}{2\pi^{2}T^{3}}\int_{0}^{\infty}\frac{dk}{{\rm cosh}^{2}\frac{k}{2T}}\int_{k}^{\infty}\frac{dQ_{\parallel}}{Q_{\parallel}}\left({\rm coth}\frac{Q_{\parallel}}{2T}-{\rm tanh}\frac{Q_{\parallel}-k}{2T}\right)\\
\times\int_{0}^{\infty}Q_{\perp}\,dQ_{\perp} \delta_{2\gamma_{e}}\Big(\frac{Q_{\perp}^{2}}{2}\frac{k}{Q_{\parallel}(Q_{\parallel}-k)}-\frac{m_{\gamma}^{2}}{2Q_{\parallel}}+\frac{m_{\rm th}^{2}Q_{\parallel}}{2k(Q_{\parallel}-k)}\Big).
\end{multline}
Again, we integrate over $Q_{\perp}$, introduce the new variables
\begin{equation}
Q_{\parallel}=2Tx,\qquad k=2Tx y,
\end{equation}
and use identity (\ref{identity}) with $a=x$ and $b=-x(1-y)$ to get
\begin{equation}
\label{gamma-flip-collinear-annihil-2}
\Gamma_{\rm flip}^{7c,({\rm pole},\,{\rm annih})}=\frac{3m_{e}^{2}\alpha}{\pi^{2}T}\int_{0}^{\infty}\frac{x\,dx}{{\rm sinh}^{2}x}\int_{0}^{1}dy\frac{1-y}{y}\left[{\rm tanh\,}xy+{\rm tanh\,}x(1-y)\right]\frac{2}{\pi}{\rm arctan}\frac{8\gamma_{e}Tx}{-m_{\gamma}^{2}+m_{\rm th}^{2}/(y(1-y))}.
\end{equation}

Finally, all three cases can be combined and the chirality flipping rate coming from the collinear processes is given by Eq.~(\ref{gamma-flip-collinear-full}).

\subsection{Contribution from the vertex correction diagram}
\label{subapp-3d}

Here, we calculate the diagram from Fig.~\ref{fig-1-order}(c) corresponding to the vertex renormalization. Its general expression is given by Eq.~(\ref{diag3}). Using the spectral representation (\ref{spectral-prop}) for the propagators, we substitute them into Eq.~(\ref{diag3}), perform the summation over the Matsubara frequencies and get
\begin{eqnarray}
\mathcal{G}_{4c}&=&-\frac{e^{2}}{4}\int\frac{d^{3}\mathbf{k}}{(2\pi)^{3}}\int\frac{d^{3}\mathbf{Q}}{(2\pi)^{3}}\int \frac{dQ^{0}}{2\pi}\int dx^{0}_{1}dx^{0}_{2}dx^{0}_{3}dx^{0}_{4}\delta_{\gamma_{e}}(x^{0}_{1})\delta_{\gamma_{e}}(x^{0}_{2})\delta_{\gamma_{e}}(x^{0}_{3})\delta_{\gamma_{e}}(x^{0}_{4})\sum_{\lambda,\lambda'=\pm} \nonumber\\
&\times& \vphantom{\rho}^{*\!\!}\rho_{\mu\nu}(Q^{0},\mathbf{Q}) \tilde{A}^{\mu\nu}_{\lambda\lambda'}(\mathbf{k},\mathbf{q})\frac{1}{i\Omega_{n}+2\mu_{5}+\lambda(2\epsilon_{k}+x^{0}_{2}+x^{0}_{3})}\frac{1}{i\Omega_{n}+2\mu_{5}+\lambda'(2\epsilon_{q}+x^{0}_{1}+x^{0}_{4})} \nonumber\\
&\times&\bigg[\frac{\big({\rm coth}\frac{\lambda\epsilon_{k}-\lambda'\epsilon_{q}}{2T}-{\rm coth}\frac{Q^{0}}{2T}\big)\big({\rm tanh}\frac{\lambda\epsilon_{k}-\mu_{L}}{2T}-{\rm tanh}\frac{\lambda'\epsilon_{q}-\mu_{L}}{2T}+{\rm tanh}\frac{\lambda\epsilon_{k}+\mu_{R}}{2T}-{\rm tanh}\frac{\lambda'\epsilon_{q}+\mu_{R}}{2T}\big)}{Q^{0}-\lambda(\epsilon_{k}+x^{0}_{2})+\lambda'(\epsilon_{q}+x^{0}_{1})} \nonumber\\
&+&\frac{\big({\rm coth}\frac{\lambda\epsilon_{k}+\lambda'\epsilon_{q}+2\mu_{5}}{2T}-{\rm coth}\frac{Q^{0}}{2T}\big)\big({\rm tanh}\frac{\lambda\epsilon_{k}-\mu_{L}}{2T}+{\rm tanh}\frac{\lambda'\epsilon_{q}-\mu_{L}}{2T}+{\rm tanh}\frac{\lambda\epsilon_{k}+\mu_{R}}{2T}+{\rm tanh}\frac{\lambda'\epsilon_{q}+\mu_{R}}{2T}\big)}{i\Omega_{n}+2\mu_{5}+\lambda(\epsilon_{k}+x^{0}_{2})+\lambda'(\epsilon_{q}+x^{0}_{4})-Q^{0}}\bigg],\label{diag3-corrected}
\end{eqnarray}
where $\mathbf{q}=\mathbf{k}-\mathbf{Q}$ and
\begin{equation}
\tilde{A}^{\mu\nu}_{\lambda\lambda'}(\mathbf{k},\mathbf{q})={\rm tr}\left[\frac{\mathds{1}-\lambda\boldsymbol{\sigma}\cdot\hat{\mathbf{k}}}{2}\sigma^{\mu}\frac{\mathds{1}-\lambda'\boldsymbol{\sigma}\cdot\hat{\mathbf{q}}}{2}\tilde{\sigma}^{\nu}\right].
\end{equation}
In Eq.~(\ref{diag3-corrected}), we took into account only the normal branch of the fermionic spectrum with the dispersion $\epsilon_{+}(k)$. Also we neglected $x^{0}_{i}$ in the arguments of hyperbolic tangents and cotangents because these are smooth functions and $x^{0}_{i}\sim e^{2}T\ll T$. Thus, we derived the analog of Eq.~(\ref{diag3-full}) which takes into account the modified dispersion of the electrons and their finite width. Further, in order to find the retarded function, we must perform the analytic continuation $i\Omega_{n}\to -2\mu_{5}+i0$. After that we will take the imaginary part of it. First of all, we should mention that the imaginary part cannot emerge from the combinations like $[\lambda(2\epsilon_{k}+x^{0}_{2}+x^{0}_{3})+i0]^{-1}$ or $[\lambda'(2\epsilon_{q}+x^{0}_{1}+x^{0}_{4})+i0]^{-1}$, because in contrast to the case of free fermions, the real parts of these expressions never equal to zero. Indeed, the energy is always $\epsilon_{k,q}\gtrsim eT$ and is of order $T$ for hard fermions while $x^{0}_{i}\sim e^{2}T\ll eT$. Thus, the first term in the square brackets will not contribute to the final answer. That is why we have the following result:
\begin{eqnarray}
\Im G^{\rm ret}_{4c}&=&\frac{\pi e^{2}}{16}\int\frac{d^{3}\mathbf{k}}{(2\pi)^{3}}\int\frac{d^{3}\mathbf{Q}}{(2\pi)^{3}}\int \frac{dQ^{0}}{2\pi}\int dx^{0}_{1}dx^{0}_{2}\delta_{\gamma_{e}}(x^{0}_{1})\delta_{\gamma_{e}}(x^{0}_{2})\sum_{\lambda,\lambda'=\pm}\frac{\lambda\lambda'}{\epsilon_{k}\epsilon_{q}}\tilde{A}^{\mu\nu}_{\lambda\lambda'}(\mathbf{k},\mathbf{q}) \nonumber\\
&\times& \vphantom{\rho}^{*\!\!}\rho_{\mu\nu}(Q^{0},\mathbf{Q}) \mathcal{B}(\lambda\epsilon_{k}+\lambda'\epsilon_{q},Q^{0})\mathcal{F}(\lambda\epsilon_{k},\lambda'\epsilon_{q})\delta\big(\lambda(\epsilon_{k}+x^{0}_{1})+\lambda'(\epsilon_{q}+x^{0}_{2})-Q^{0}\big),\label{diag3-cor-Im}
\end{eqnarray}
where the functions $\mathcal{F}$ and $\mathcal{B}$ are defined in Eqs.~(\ref{F-funct}) and (\ref{B-funct}), respectively. For further convenience, we calculate the components of $\tilde{A}^{\mu\nu}_{\lambda\lambda'}$:
\begin{eqnarray}
\tilde{A}^{00}_{\lambda\lambda'}(\mathbf{k},\mathbf{q})&=&\frac{1}{2}\left(1+\lambda\lambda'\frac{\mathbf{k}\cdot\mathbf{q}}{kq}\right),\nonumber\\
\tilde{A}^{0i}_{\lambda\lambda'}(\mathbf{k},\mathbf{q})&=&-\tilde{A}^{i0}_{\lambda\lambda'}(\mathbf{k},\mathbf{q})=\frac{1}{2}\left(\lambda\frac{k^{i}}{k}+\lambda'\frac{q^{i}}{q}\right),\nonumber\\
\tilde{A}^{ij}_{\lambda\lambda'}(\mathbf{k},\mathbf{q})&=&-\frac{1}{2}\left[\delta^{ij}\left(1-\lambda\lambda'\frac{\mathbf{k}\cdot\mathbf{q}}{kq}\right)+\lambda\lambda'\frac{k^{i}q^{j}+k^{j}q^{i}}{kq}\right].\label{Atilde-munu}
\end{eqnarray}

Like in the previous subsection, we consider the contributions from the soft photon spectral density and from the nearly collinear processes separately.

\paragraph{Soft photon.} In the case when the photon momentum is soft $\mathbf{Q},\,Q^{0}\lesssim eT$, $\epsilon_{q}\approx \epsilon_{k}-Q\cos\theta$ and the delta function works only for $\lambda'=-\lambda$. Expanding the hyperbolic cotangent, we get
\begin{eqnarray}
\Im G^{\rm ret,\,{\rm soft\,}\gamma}_{4c}&=&-\frac{\pi e^{2}T}{8}\int\frac{d^{3}\mathbf{k}}{(2\pi)^{3}}\int\frac{d^{3}\mathbf{Q}}{(2\pi)^{3}}\int \frac{dQ^{0}}{2\pi}\frac{\vphantom{\rho}^{*\!\!}\rho_{\mu\nu}(Q^{0},\mathbf{Q})}{Q^{0}}P^{\mu\nu}_{t}(\mathbf{k}) \nonumber\\
&\times&\sum_{\lambda=\pm}\frac{\mathcal{F}(\lambda\epsilon_{k},-\lambda\epsilon_{k})}{\epsilon_{k}^{2}}\delta_{2\gamma_{e}}(\lambda Q\cos\theta-Q^{0}\big).\label{diag3-cor-Im-soft-gamma}
\end{eqnarray}

In order to extract the chirality flipping rate, we expand the integrand for small chemical potential
\begin{equation}
\mathcal{F}(\lambda\epsilon_{k},-\lambda\epsilon_{k})\approx \frac{1}{{\rm cosh}^{2}\frac{\epsilon_{k}}{2T}}\frac{2\mu_{5}}{T}.
\end{equation}
Taking into account that the photon spectral density $\vphantom{\rho}^{*\!\!}\rho_{\mu\nu}(Q^{0},\mathbf{Q})$ is the odd function of frequency $Q^{0}$, we finally obtain Eq.~(\ref{gamma-flip-3-soft-photon}).

\paragraph{Nearly collinear processes.} For the hard photon momenta, we use the spectral density given by Eq.~(\ref{sp-dens-hard-photon}) and Eq.~(\ref{diag3-cor-Im}) takes the form of Eq.~(\ref{diag3-cor-Im-coll}). This corresponds to the nearly collinear $1\leftrightarrow 2$ processes in plasma, the same as we considered in the previous subsection.

There are three independent ways to choose the signs of $\lambda, \lambda',\, \lambda''$ in Eq.~(\ref{diag3-cor-Im-coll}). In each of these cases, we decompose the vector $\mathbf{Q}$ into two components: parallel to the vector $\mathbf{k}$ and perpendicular to it. The latter must be soft $Q_{\perp}\lesssim eT$ in order to satisfy the approximate energy conservation in the collision. Then, the integration over $Q_{\perp}$ can be performed and we obtain the following expressions:
\begin{equation}
\Im G^{\rm ret,\,brem}_{4c}=\frac{\alpha}{64\pi^{2}}\int_{0}^{\infty}dk\int_{0}^{k}dQ_{\parallel}\sum_{\lambda=\pm}\lambda \mathcal{B}(\lambda Q_{\parallel},\lambda Q_{\parallel})\mathcal{F}(\lambda k,-\lambda(k-Q_{\parallel}))\frac{2}{\pi}{\rm arctan}\frac{4\gamma_{e}}{\frac{m_{\gamma}^{2}}{Q_{\parallel}}+\frac{m_{\rm th}^{2}Q_{\parallel}}{k(k-Q_{\parallel})}}.\label{diag3-cor-Im-brem}
\end{equation}
\begin{equation}
\Im G^{\rm ret,\,abs}_{4c}=-\frac{\alpha}{64\pi^{2}}\int_{0}^{\infty}dk\int_{0}^{\infty}dQ_{\parallel}\sum_{\lambda=\pm}\lambda \mathcal{B}(-\lambda Q_{\parallel},-\lambda Q_{\parallel})\mathcal{F}(\lambda k,-\lambda(k+Q_{\parallel}))\frac{2}{\pi}{\rm arctan}\frac{4\gamma_{e}}{\frac{m_{\gamma}^{2}}{Q_{\parallel}}+\frac{m_{\rm th}^{2}Q_{\parallel}}{k(k+Q_{\parallel})}}.\label{diag3-cor-Im-abs}
\end{equation}
\begin{equation}
\Im G^{\rm ret,\,annih}_{4c}=-\frac{\alpha}{64\pi^{2}}\int_{0}^{\infty}dk\int_{k}^{\infty}dQ_{\parallel}\sum_{\lambda=\pm}\lambda \mathcal{B}(\lambda Q_{\parallel},\lambda Q_{\parallel})\mathcal{F}(\lambda k,\lambda(Q_{\parallel}-k))
\frac{2}{\pi}{\rm arctan}\frac{4\gamma_{e}}{-\frac{m_{\gamma}^{2}}{Q_{\parallel}}+\frac{m_{\rm th}^{2}Q_{\parallel}}{k(Q_{\parallel}-k)}}.\label{diag3-cor-Im-annihil}
\end{equation}

In order to extract from these expressions the chirality flipping rate, we expand the integrand for small $\mu_{5}$. The only quantities containing chemical potentials are the thermal distribution functions  $\mathcal{B}$ and $\mathcal{F}$. Therefore, we have
\begin{eqnarray}
\sum_{\lambda=\pm}\lambda \mathcal{B}(\lambda Q_{\parallel},\lambda Q_{\parallel})\mathcal{F}(\lambda a,\lambda b)&\approx& \sum_{\lambda=\pm}\lambda\left(-\frac{1}{{\rm sinh}^{2}\frac{Q_{\parallel}}{2T}}\frac{\mu_{5}}{T}\right)\left(2{\rm tanh\,}\lambda a+ 2{\rm tanh\,}\lambda b+\mathcal{O}(\mu_{5}/T)\right) \nonumber\\
&\approx& -\frac{4\mu_{5}}{T}\frac{{\rm tanh\,} a+ {\rm tanh\,} b}{{\rm sinh}^{2}\frac{Q_{\parallel}}{2T}}.
\end{eqnarray}
The chirality flipping rate can be found using Eq.~(\ref{eq-chirality-flip}).
Choosing the integration variables in the same way as we did in Sec.~\ref{subapp-pole}, we get the following results:
\begin{equation}
\label{gamma-flip-3d-coll-brem-abs}
\Gamma_{\rm flip}^{4c,({\rm brem})}=\Gamma_{\rm flip}^{4c,({\rm abs})}=-\frac{3m_{e}^{2}\alpha}{\pi^{2}T}\int_{0}^{\infty}\frac{x\,dx}{{\rm sinh}^{2}x}\int_{0}^{\infty}dy\left[{\rm tanh\,}x(y+1)-{\rm tanh\,}xy\right]\frac{2}{\pi}{\rm arctan}\frac{8\gamma_{e}Tx}{m_{\gamma}^{2}+m_{\rm th}^{2}/(y(y+1))},
\end{equation}
\begin{equation}
\label{gamma-flip-3d-coll-annih}
\Gamma_{\rm flip}^{4c,({\rm annih})}=\frac{3m_{e}^{2}\alpha}{\pi^{2}T}\int_{0}^{\infty}\frac{x\,dx}{{\rm sinh}^{2}x}\int_{0}^{1}dy\left[{\rm tanh\,}xy+{\rm tanh\,}x(1-y)\right]\frac{2}{\pi}{\rm arctan}\frac{8\gamma_{e}Tx}{-m_{\gamma}^{2}+m_{\rm th}^{2}/(y(1-y))}.
\end{equation}
Combining all three expressions together, we end up with Eq.~(\ref{gamma-flip-3d-collinear-all}).

\normalsize

\section{Impact of the Landau-Pomeranchuk-Migdal effect of the photon production rate and chirality flipping rate}
\label{app-LPM}

In this Appendix we compare the calculations of the photon production rate and the chirality flipping rate in hot Abelian plasma. The former was exhaustively studied in Refs.~\cite{Arnold:2001ba,Arnold:2002ja,Arnold:2002zm} while the latter is the subject of the present study.

\begin{figure}[h!]
	\centering
	$(a)$\includegraphics[width=7cm]{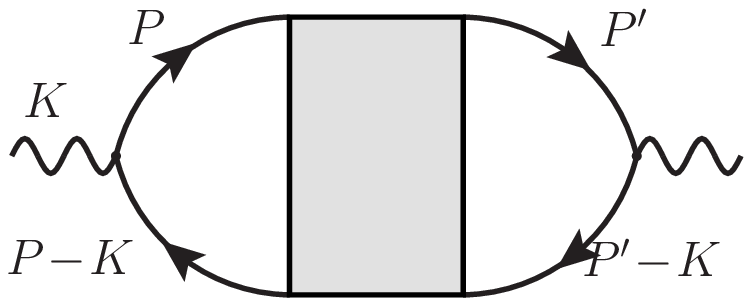} \hspace*{0.5cm}
	$(b)$\includegraphics[width=7cm]{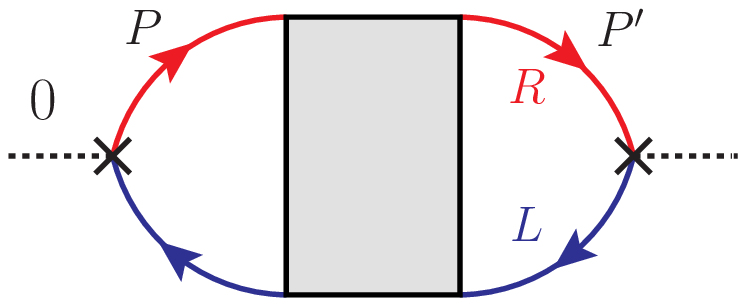}
	\caption{General form of the loop diagrams contributing to the photon production rate (a) and chirality flipping rate (b). The gray rectangle denotes the 4-point function which must be calculated perturbatively. There are three main differences between them: (i) the incoming momentum in diagram (a) is hard $K\sim T$ and on shell (corresponds to a real photon), while in diagram (b) it is equal to zero; 
	(ii) the vertices in diagram (a) are usual QED vertices containing the coupling constant $e$ while in diagram (b) these are just the mass insertion vertices with chiral projectors $P_{R,L}$ inside; (iii) fermions running in the loop in diagram (a) have the same chiralities on upper and lower branches, while in diagram (b) they always have opposite chiralities (shown by different colors). \label{fig-app-2-rates}}
\end{figure}

At first sight, the computation of these rates are very similar. The photon production rate is given by the following expression [see Eqs.~(2.1), (2.2), and (2.4) in Ref.~\cite{Arnold:2001ba}]:
\begin{equation}
\label{PPR}
d\Gamma_{\gamma}=\frac{d^{3}\mathbf{k}}{(2\pi)^{3}|\mathbf{k}|}(n_{B}(k^{0})+1)\sum\limits_{a=1,2}\epsilon^{\mu}_{(a)}(\mathbf{k}) \epsilon^{\nu}_{(a)}(\mathbf{k}) \Im D^{\rm ret}_{\mu\nu}(K),
\end{equation}
i.e., it is proportional to the imaginary part of the retarded correlator of two electromagnetic currents
\begin{equation}
D^{\rm ret}_{\mu\nu}(K)=i\int d^{4}X\,e^{-i K\cdot X} \theta(x^{0}) \llangle[j_{\mu}(X),\,j_{\nu}(0)] \rrangle.
\end{equation}
The latter object can be calculated perturbatively in the electromagnetic coupling $e$ and the general diagram contributing to it is just a fermionic loop with some 4-point function inserted into it. It is shown in Fig.~\ref{fig-app-2-rates}(a).

In the similar manner, the chirality flipping rate is given by Eq.~(\ref{eq-chirality-flip}) which also contain the imaginary part of the retarder correlation function (\ref{retarded-Green-function}). Diagrammatically, this object can be depicted as shown in Fig.~\ref{fig-app-2-rates}(b). Although diagrams (a) and (b) look similar, they have three main differences. First of all, the external legs of diagram (b) correspond to the flip of chirality due to the finite electron mass $m_{e}$ and therefore bring 0 momentum. As for diagram (a), its incoming momentum is just the photon momentum $K$ which is hard and on shell. Correspondingly, the vertices are also different. Whereas diagram (a) contains the usual QED vertex, diagram (b) has mass insertions with the chiral projectors $P_{R,L}$ inside. This implies the third difference: the fermions running in the loop (a) have the same chiralities, while in the loop (b) the chiralities in the upper and lower branches are opposite (this is shown by different colors).

\begin{figure}[h!]
	\centering
	$(a)$\includegraphics[width=5cm]{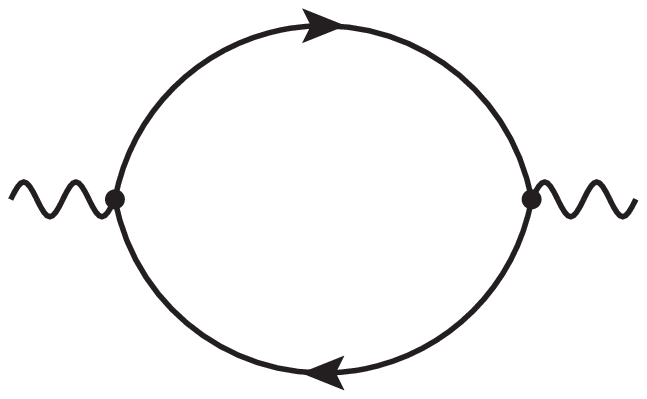} \hspace*{0.5cm}
	$(b)$\includegraphics[width=5cm]{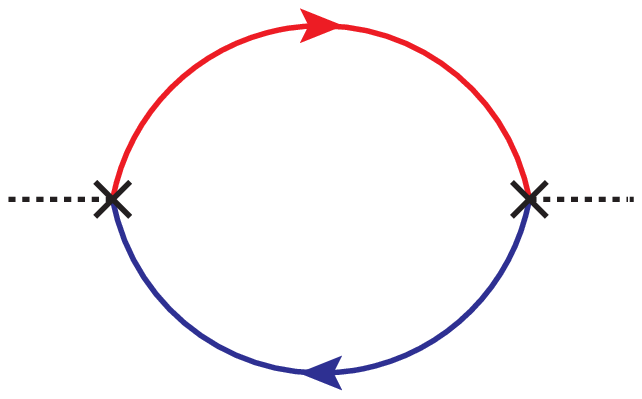}
	\caption{Lowest order diagrams in the perturbative series for the photon production rate (a) and chirality flipping rate (b). Both diagrams are real and thus do not contribute to the corresponding rates. \label{fig-app-lowest-orders}}
\end{figure}

Let us now analyze the perturbative calculation of the diagrams in Fig.~\ref{fig-app-2-rates}. The lowest order contribution is just an empty loop. This is shown in Fig.~\ref{fig-app-lowest-orders}. Imaginary parts of both diagrams vanish and the reasons are the following. The imaginary part of diagram (a) is connected by the optical theorem to the matrix element of the bremsstrahlung process $e\to e+\gamma$. In massless case it is allowed only for the strictly collinear momenta, however, such a process is forbidden by the angular momentum conservation. In the same way, the imaginary part of the diagram (b) corresponds to the spontaneous flip of the chirality by a free electron [shown in Fig.~\ref{fig-lowest}(a)] which is also forbidden by the angular momentum conservation.

\begin{figure}[h!]
	\centering
	$(a)$\includegraphics[width=5cm]{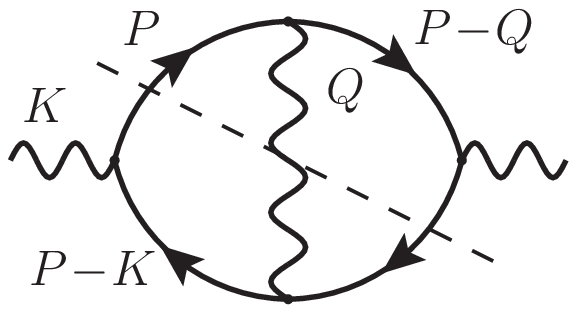} \hspace*{0.5cm}
	$(b)$\includegraphics[width=5cm]{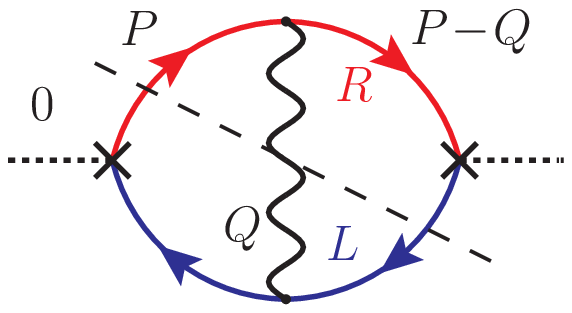}
	\caption{Lowest order nonvanishing diagrams in the perturbative series for the photon production rate (a) and chirality flipping rate (b). Both diagrams have finite imaginary parts for hard on-shell momentum $Q$ and their parametric behavior corresponds to their naive perturbative order. \label{fig-app-first-order}}
\end{figure}

First nontrivial contributions appear only in the next perturbative order. Here we must add one photon line to the diagrams in Fig.~\ref{fig-app-lowest-orders}. This photon line can be inserted into one of the fermion arcs or connect the two arcs with each other. ``Dressing'' of only one fermion leg can be absorbed into the self-energy correction which is indeed done in Ref.~\cite{Arnold:2001ba} as well as in the present paper. But here we would like to focus on the ``vertex correction'' diagrams shown in Figs.~\ref{fig-app-first-order} (a) and \ref{fig-app-first-order}(b). The imaginary part of the diagram (a) has a support from the region where the intermediate momentum $Q$ is hard and on-shell. This corresponds to $2\to 2$ Compton scattering process in plasma resulting in a photon with momentum $K$. As it is shown in Ref.~\cite{Arnold:2001ba}, the parametric dependence of this contribution corresponds to the naive perturbative order and is $\propto \alpha^{2}$. As for the chirality flipping rate diagram (b), it is also nontrivial for the on-shell hard momentum $Q$ and corresponds to $1\to 2$ collinear processes of bremsstrahlung (or cross-channels) with the chirality flip of a fermion, shown in Fig.~\ref{fig-lowest}(b). As we have shown in Appendix~\ref{app-first-order}, this diagram also has a fictitious IR divergence for $Q\to 0$ which is canceled by another two first-order diagrams with the ``dressing'' of one fermion leg. This cancellation is a consequence of the gauge invariance, since the photon with zero momentum corresponds to a pure gauge. Apart from this divergence, the diagram in Fig.~\ref{fig-app-first-order} is \textit{finite} and gives ${\rm O}(\alpha)$ contribution to the chirality flipping rate.

\begin{figure}[h!]
	\centering
	$(a)$\includegraphics[height=2.5cm]{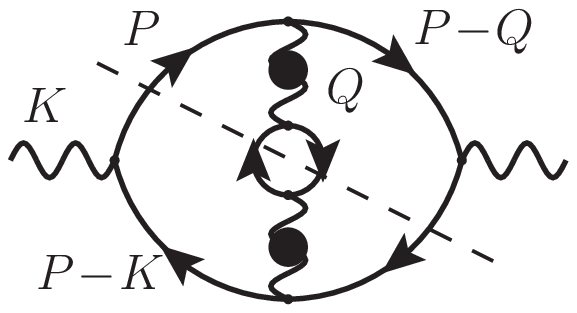} \hspace*{0.5cm}
	$(b)$\includegraphics[height=2.5cm]{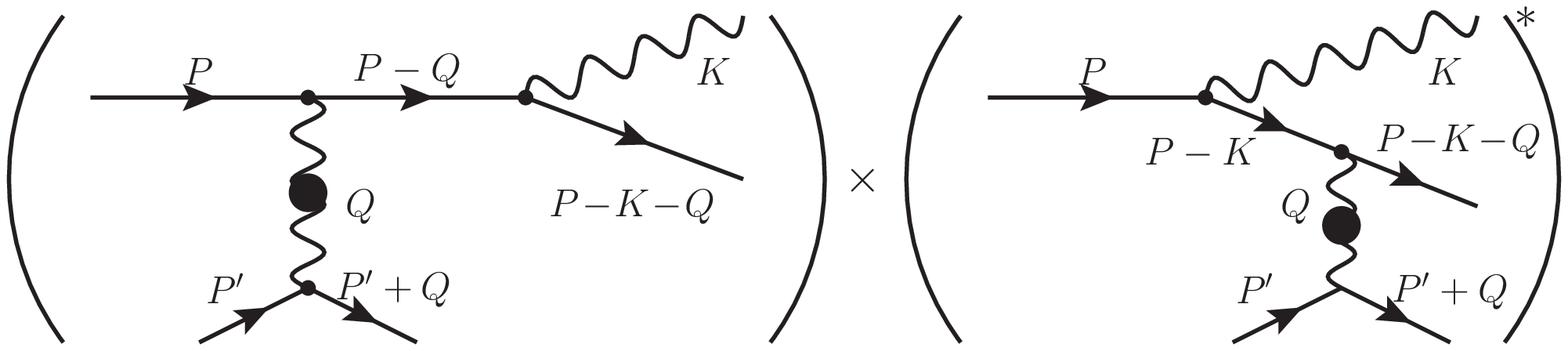}
	\caption{(a) Diagram for the photon production rate which is naively of order $\alpha^{3}$, however, for soft intermediate momenta $Q\sim eT$ satisfying $P\cdot Q={\rm O}(e^{2}T^{2})$ it is IR divergent and after regularization appears to be of order $\alpha^{2}$. (b) Physical process of collinear bremsstrahlung accompanied by the soft scattering in plasma which causes the imaginary part of the diagram (a). Importantly, this diagram contains the interference of two processes in which the scattering occurs before and after the photon emission. \label{fig-app-soft-photon}}
\end{figure}

It is important to note that in the calculation of the photon production rate some processes of the higher (naive) perturbative order due to accumulated IR divergences can also contribute to the leading ${\rm O}(\alpha^{2})$ order result. An example of such a diagram is shown in Fig.~\ref{fig-app-soft-photon}(a). As is discussed in Ref.~\cite{Arnold:2001ba}, physically this corresponds to the processes shown in Fig.~\ref{fig-app-soft-photon}(b), and the phase space region giving the dominant contribution to this diagram is $Q\sim eT$ (soft intermediate photon), $P\sim T$ (hard fermions), $K\cdot P={\rm O}(e^{2}T^{2})$, $Q\cdot P ={\rm O}(e^{2}T^{2})$ (nearly collinear bremsstrahlung). Figure~\ref{fig-app-soft-photon}(b) also shows that the contribution to the photon production rate from diagram (a) is caused by the \textit{interference} of the two processes where the soft scattering occurs before and after the photon emission. We would like to note that each of these processes separately is present in another diagram of the same order as (a), but with the soft photon line ``dressing'' only one of the fermion legs. Obviously, such processes would automatically be taken into account by using the full fermion propagators with the self-energy corrections. However, their interference is is not included in the self-energy. That is why keeping the diagram in Fig.~\ref{fig-app-soft-photon}(a) is extremely important. Moreover, as we will see below, diagrams with more soft photon insertions must be included since they contain the accumulating IR divergences and all contribute to the leading order result for the photon production rate. Such a resummation takes into account the interference effects in the multiple soft photon scatterings which is the essence of the Landau-Pomeranchuk-Migdal effect.

\begin{figure}[h!]
	\centering
	$(a)$\includegraphics[height=2.5cm]{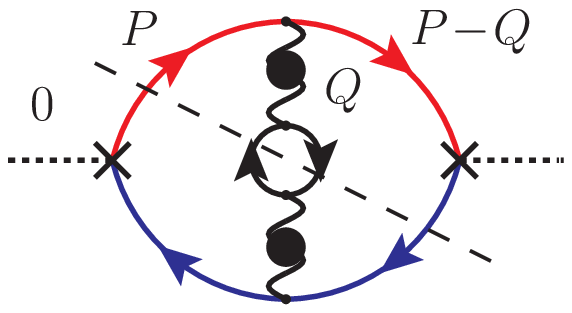} \hspace*{0.5cm}
	$(b)$\includegraphics[height=2.5cm]{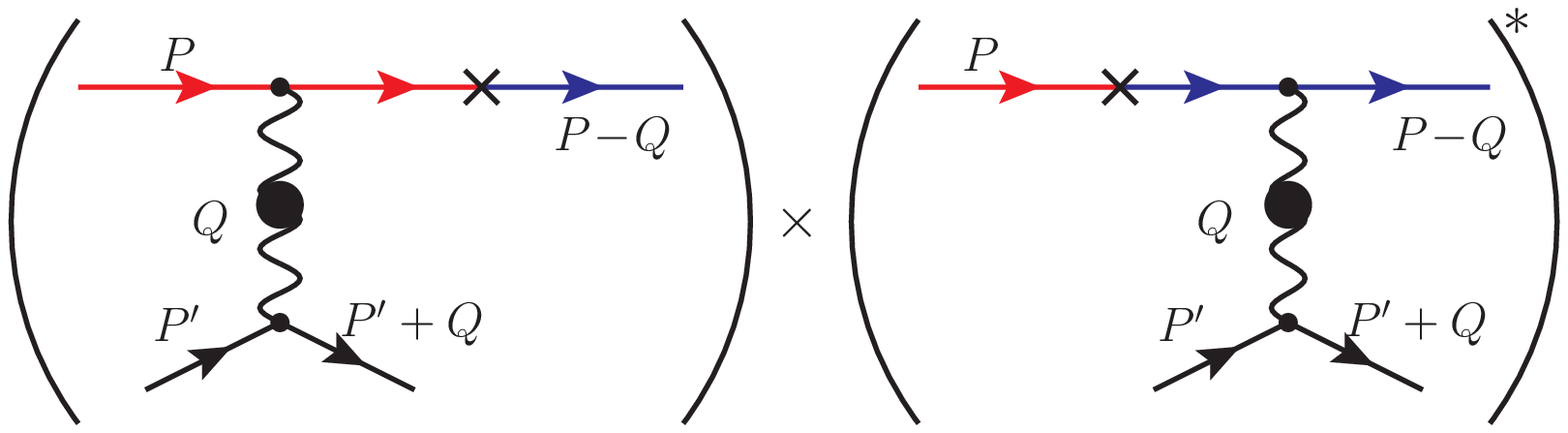}
	\caption{(a) Diagram for the chirality flipping rate with a soft photon line connecting the fermion arcs with opposite chiralities. (b) Physical process of chirality flip accompanied by the soft scattering in plasma which causes the imaginary part of the diagram (a). This diagram contains the interference of two processes in which the scattering occurs before and after the flip of chirality. \label{fig-app-soft-photon-cfr}}
\end{figure}

It is easy to understand that the fermion loop inside the photon line in Fig.~\ref{fig-app-soft-photon}(a) can be absorbed into the photon self-energy. In other words, in order to take into account this diagram (as well as an infinite number of higher order diagrams with multiple fermion loops in the same photon line) one should consider the diagram in Fig.~\ref{fig-app-first-order}(a) in which the intermediate photon line corresponds to the full propagator including the self-energy corrections. This is the right place to note that we do take into account the analogous diagram for the chirality flipping rate, i.e., the diagram in Fig.~\ref{fig-app-first-order}(b) with the dressed photon line, the corresponding calculation is shown in Appendix~\ref{subapp-3d}. Thus, we include the diagram depicted in Fig.~\ref{fig-app-soft-photon-cfr} (a), which contains the interference of the physical processes in which the soft scattering occurs before and after the flip of chirality, see Fig.~\ref{fig-app-soft-photon-cfr}(b). 
However, as we will see below, the higher order diagrams with multiple soft photon lines accumulate the divergences much slower than in the case of photon production and the ladder resummation of Ref.~\cite{Arnold:2001ba} is not needed for the computation of the chirality flipping rate.

\begin{figure}[h!]
	\centering
	$(a)$\includegraphics[width=8cm]{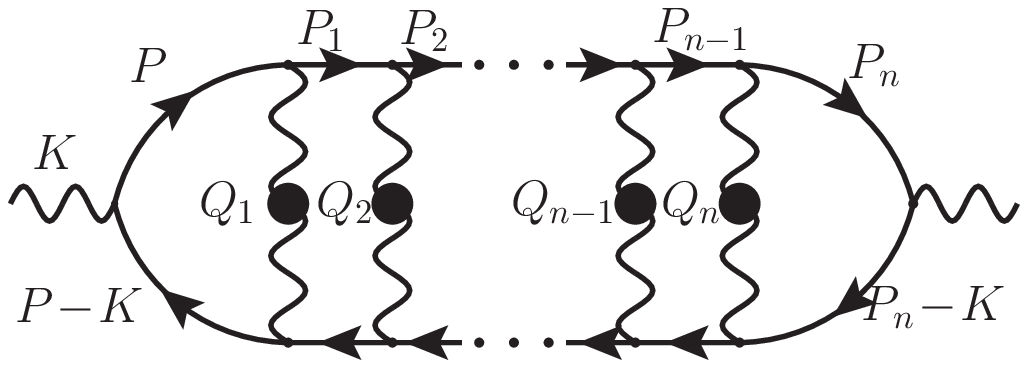} \hspace*{0.5cm}
	$(b)$\includegraphics[width=8cm]{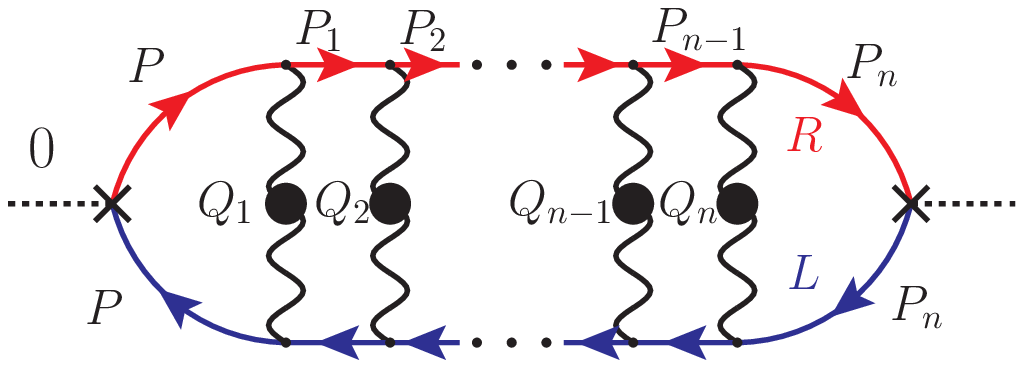}
	\caption{General diagrams with the ladder of soft photons contributing to the photon production rate (a) and chirality flipping rate (b). Each of the intermediate momenta is soft $Q_{i}\sim eT$ and satisfies $Q_{i}\cdot P={\rm O}(e^{2}T^{2})$. \label{fig-app-gen-order}}
\end{figure}

Let us consider the general case of multiple soft photon scatterings. In Ref.~\cite{Arnold:2001ba} it was shown that the dominating contribution is made by the ladder diagrams shown in Fig.~\ref{fig-app-gen-order}(a).

Let us try to understand why all these diagrams contribute to the photon production rate in the same parametric order. For this, we now take into account that all intermediate photon momenta are soft, $Q_{i}\sim eT$, the fermion momenta $P$ and $P_{n}-K$ are hard and on shell (these lines are cut when we take the imaginary part), other fermion lines $P_{1},\ \ldots,\ P_{n}$ are hard, nearly on shell (because of small $Q$) and nearly collinear to the external photon $P_{i}\cdot K={\rm O}(e^{2}T^{2})$. This implies also that $P_{i}\cdot Q_{j}={\rm O}(e^{2}T^{2})$. Now let us see how the parametric dependence is changed if we add one additional soft photon line into the ladder. We get
\begin{itemize}
    \item $e^{2}$ from two additional vertices;
    \item $\sim (eT)^{4}$ from the phase space of one additional $Q$;
    \item $1/Q^2\sim 1/(eT)^{2}$ from the additional soft photon propagator;
    \item $\gamma^{\mu}P_{i,\mu}/P_{i}^{2}\sim T/(P\cdot Q) \sim T/(eT)^{2}$ from \textit{each} of the two additional fermion propagators.
\end{itemize}
In total, all powers of $e$ cancel and we conclude that adding one soft photon line does not change the parametric dependence and the resummation of all ladder diagrams is required. This is the main result of Ref.~\cite{Arnold:2001ba}.

We would like to emphasize that we got such a result only due to the fact that \textit{both} additional fermion propagators which appear in the diagram are nearly on shell. Indeed, they bring the momentum $P_{i}=P-Q_{1}-Q_{2}-\ldots -Q_{i}$ and $P_{i}-K$. We know that $P$ and $P_{n}-K$ are on shell (this is where we cut our diagram). That is why $P_{i}^{2}=2P\cdot (Q_{1}+Q_{2}+\ldots +Q_{i})+ (Q_{1}+Q_{2}+\ldots +Q_{i})^{2}\sim (eT)^{2}$, the same for $(P_{i}-K)^{2}$.

Now, let us see what happens in the case of the chirality flipping rate, whose ladder diagram is shown in Fig.~\ref{fig-app-gen-order}. Here we have an important difference: the incoming momentum is 0 so that we always have pairs of fermion propagators with \textit{the same}
momenta but \textit{opposite} chiralities. As a result, if one propagator is nearly on shell, the other one as always is very far off shell, because the fermions of different chiralities have different dispersion relations, namely, in vacuum $p^{0}=\pm |\mathbf{p}|$ [see the discussion around Eq.~(\ref{propagator-free})].
Then, for the nearly on-shell fermion of one chirality we have
\begin{equation}
\mathcal{S}_{1}\sim\frac{1}{p^{0}-q^{0}-|\mathbf{p}-\mathbf{q}|}\simeq \frac{2p^{0}}{(P-Q)^{2}}=\frac{2p^{0}}{-2P\cdot Q+Q^{2}}\sim \frac{T}{(eT)^{2}}.
\end{equation}
Then, for the opposite chirality we would get
\begin{equation}
\mathcal{S}_{2}\sim\frac{1}{p^{0}-q^{0}+|\mathbf{p}-\mathbf{q}|}\sim\frac{1}{T}.
\end{equation}
Repeating the similar power counting as previously, we get that each new soft photon gives additional power of $e^{2}$. That is why for the leading order result the ladder resummation is not needed.

\begin{figure}[h!]
	\centering
	$(a)$\includegraphics[width=8cm]{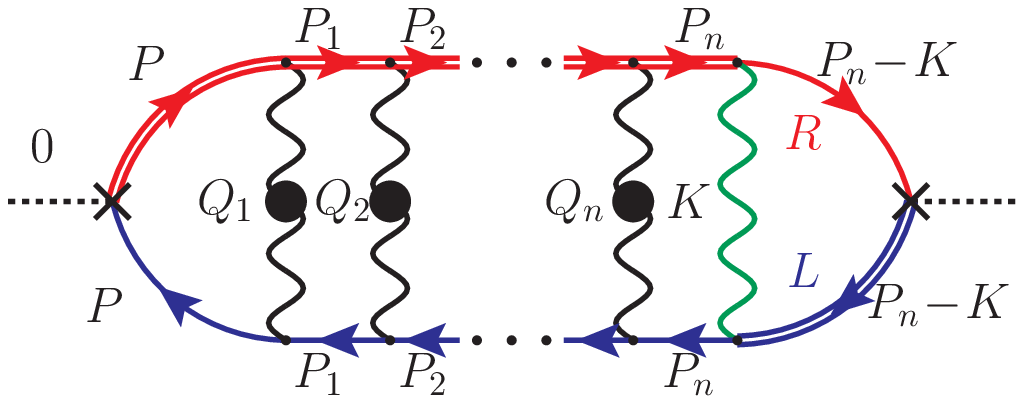} \hspace*{0.5cm}
	$(b)$\includegraphics[width=8cm]{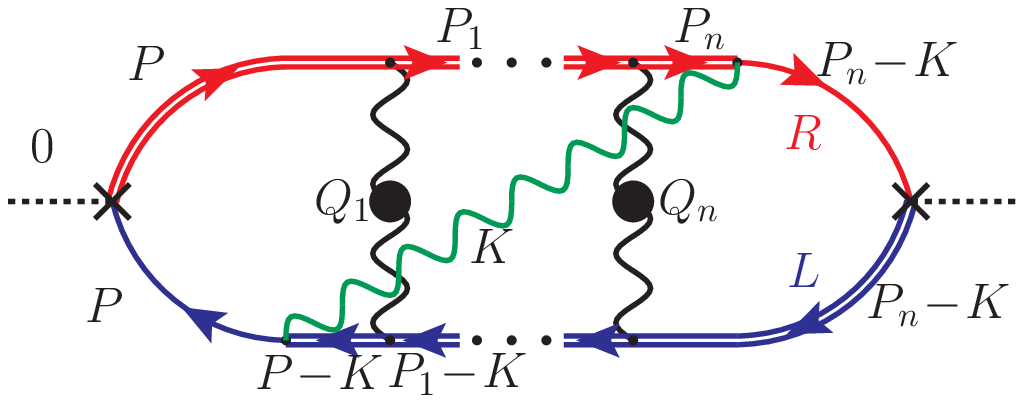}\\
	$(c)$\includegraphics[width=8cm]{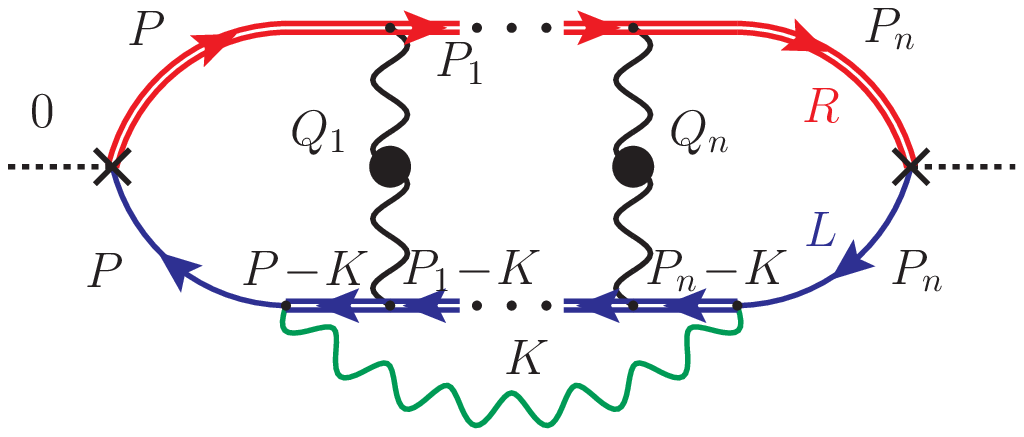}
	\caption{General diagrams with the ladder of soft photons and one hard photon $K$ (shown by the green wavy line) in three topologically different situations: (a) the hard photon line is a part of the ladder; (b) it crosses all soft photon lines in the ladder; (c) it embraces the soft photon lines. Special attention should be paid to the case when the hard momentum $K$ is nearly on shell and almost collinear to $P$, $K\cdot P={\rm O}(e^{2}T^{2})$, while each of the intermediate momenta is soft $Q_{i}\sim eT$ and satisfies $Q_{i}\cdot P={\rm O}(e^{2}T^{2})$. Nearly on-shell fermion propagators are shown by the double lines. \label{fig-app-gen-order-hard}}
\end{figure}

Also we should pay attention to the diagrams with one additional hard photon. Three topologically nontrivial classes of such diagrams are shown in Fig.~\ref{fig-app-gen-order-hard}. Obviously, the diagram (a) has exactly the same properties as previously discussed diagram in Fig.~\ref{fig-app-gen-order}(b). Indeed, it contains all fermionic propagators in pairs. Each pair consists of propagators with identical momenta and opposite chiralities. That is why only one of them can be nearly on shell and adding more soft lines makes the result higher order in $\alpha$. 

However, the diagrams shown in Figs.~\ref{fig-app-gen-order-hard} (b) and \ref{fig-app-gen-order-hard}(c) may be potentially more singular. In fact, if the hard photon momentum is on shell and nearly collinear to $P$, it is possible to get all fermion propagators on shell, except two. These on-shell propagators are shown by double lines in Fig.~\ref{fig-app-gen-order-hard}. Then, repeating the same power counting as previously, we conclude that the parametric behavior of such diagrams does not change of we add more soft lines. In other words, if the diagram with one soft photon line were important for us, then all the ladder would have to be resummed. Fortunately, this is not the case as we will see below.

\begin{figure}[h!]
	\centering
	$(a)$\includegraphics[width=7.5cm]{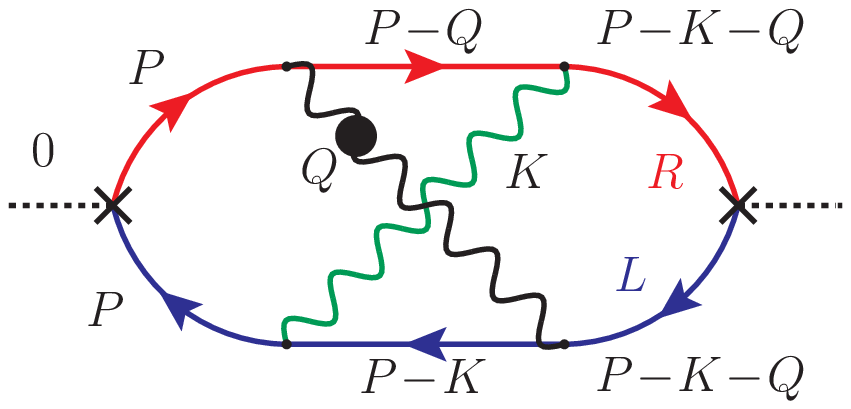} \hspace*{0.5cm}
	$(b)$\includegraphics[width=7.5cm]{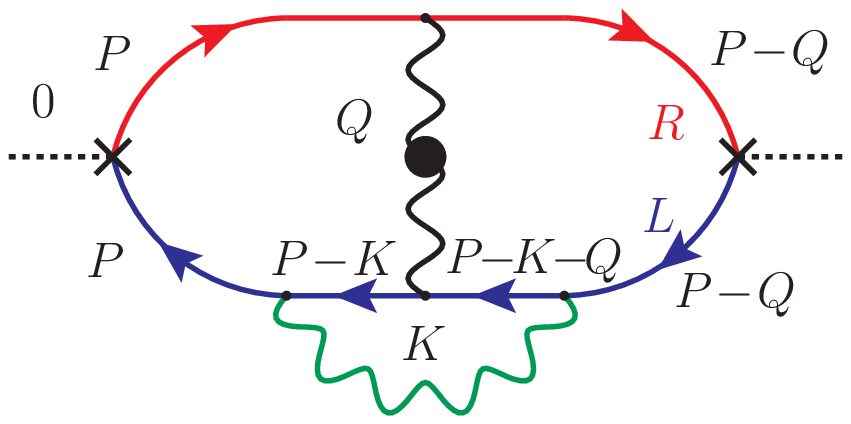} \\
	$(c)$\includegraphics[width=11.cm]{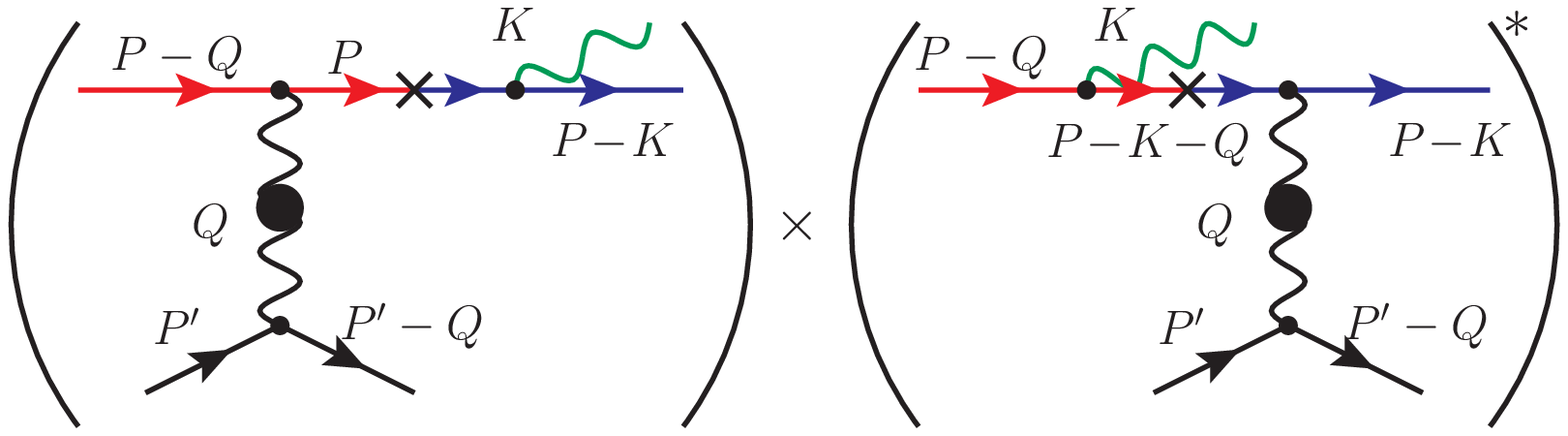}
	\caption{Panels (a) and (b) show first diagrams in the ladder series depicted in Figs.~\ref{fig-app-gen-order-hard} (b) and \ref{fig-app-gen-order-hard}(c), correspondingly. Their contributions to the chirality flipping rate have the parametric behavior $O(e^3)$ which is the subleading order compared to our result. (c) The interference of two processes of hard photon bremsstrahlung with the chirality flip and soft scattering in plasma which contributes to the diagram (a). \label{fig-app-crossed}}
\end{figure}

Let us first consider the diagram of type (b). The first term in the ladder series contains one soft photon line crossed by one hard line and is shown in Fig.~\ref{fig-app-crossed} (a). This diagram involves three summations over the Matsubara frequencies (of $P$, $K$, and $Q$) and 8 propagators. In each summation the Matsubara contour catches the quasiparticle pole of one of the propagators. In other words, upon summation we get three propagators on shell. Among different ways to choose these three propagators one leads to the most singular infrared behavior. Indeed, let us consider the photon $K$, the left fermion $P-K$, and the right fermion $P-Q$ to be on shell. Then, $k^{0}=|\mathbf{k}|$, $p^{0}-k^{0}=-|\mathbf{p}-\mathbf{k}|$, $p^{0}-q^{0}=|\mathbf{p}-\mathbf{q}|$. Note that we are left with the soft photon $Q$, the pair of left and right fermions with momentum $P$, and the pair of left and right fermions with momentum $(P-K-Q)$.  Then, the parametric dependence of the diagram in Fig.~\ref{fig-app-crossed}(a) can be estimated as follows:
\begin{equation}
\label{app-G-a}
    G^{(a)}\propto e^{4} \int d^{3}\mathbf{p}\ d^{3}\mathbf{k}\ d^{3}\mathbf{q}\ D(Q)\times \frac{1}{P^{2}} \times \frac{1}{(P-K-Q)^{2}},
\end{equation}
where $D(Q)$ is the full propagator of the soft photon with momentum $Q\sim eT$ which behaves like $D(Q)\sim 1/(e^{2}T^{2})$ \cite{LeBellac}. Especially interesting is the case when the momenta $\mathbf{p}$ and $\mathbf{k}$ are nearly collinear with the opening angle $\theta \sim e$ (the phase space for $\mathbf{k}$ is then $\sim e^{2}T^{3}$. It is easy to see that in this case the momentum $P$ satisfies $P^{2}\simeq -\frac{p^{2}k}{k-p}\theta^{2}\sim e^{2}T^{2}$ and $P\cdot Q={\rm O}(q^2, \theta^{2})\sim e^{2} T^{2}$. From this immediately follows $(P-K-Q)^{2}\sim e^{2}T^{2}$. To say, this is exactly the same region of the phase space which was considered in Ref.~\cite{Arnold:2001ba} and the corresponding physical processes are shown in Fig.~\ref{fig-app-crossed}(c). Finally, the parametric behavior of the diagram in Fig.~\ref{fig-app-crossed}(a) is given by\footnote{This estimate neglects the logarithmic dependence on the coupling constant.}
\begin{equation}
\label{app-G-a-estimate}
    G^{(a)}\propto e^{4}\times T^{3} \times e^{2}T^{3}\times (eT)^{3} \times \frac{1}{(eT)^{2}}\times \frac{1}{e^{2}T^{2}}\times\frac{1}{e^{2}T^{2}}\propto e^{3}\propto \alpha^{3/2}.
\end{equation}
Therefore, this diagram, as well as the whole ladder is of higher parametric order in $\alpha$ than our leading order result for the chirality flipping rate.

Let us now consider the ladder shown in Fig.~\ref{fig-app-gen-order-hard} (c). The first diagram from this series is given by Fig.~\ref{fig-app-crossed} (b). Its parametric behavior can be estimated in the similar way as written above. In fact, this diagram differs from the diagram (a) by the absence of the off-shell right fermion propagator $(P-K-Q)$ and by the presence of the off-shell left fermion propagator $(P-Q)$. Thus, we get the following estimate:
\begin{equation}
    G^{(b)}\propto e^{4} \int d^{3}\mathbf{p}\ d^{3}\mathbf{k}\ d^{3}\mathbf{q} \ D(Q)\times \frac{1}{P^{2}}\times  \frac{1}{p^{0}-k^{0}-q^{0}+|\mathbf{p}-\mathbf{k}-\mathbf{q}|}\times \frac{1}{p^{0}-q^{0}+|\mathbf{p}-\mathbf{q}|}.
\end{equation}
In the same phase space region as described before Eq.~(\ref{app-G-a}), the denominator $(p^{0}-k^{0}-q^{0}+|\mathbf{p}-\mathbf{k}-\mathbf{q}|)={\rm O}(e^{2}T)$ while $(p^{0}-q^{0}+|\mathbf{p}-\mathbf{q}|)\sim 2|\mathbf{p}|={\rm O}(T)$. Then, the same calculation as (\ref{app-G-a-estimate}) gives
\begin{equation}
 G^{(b)}\propto e^{4}\times T^{3} \times e^{2}T^{3}\times (eT)^{3} \times \frac{1}{(eT)^{2}}\times \frac{1}{e^{2}T^{2}}\times\frac{1}{e^{2}T}\times \frac{1}{T}\propto e^{3}\propto \alpha^{3/2}.
\end{equation}
Thus, the first term in the ladder series, as well as the full series of Fig.~\ref{fig-app-gen-order-hard}(c) gives a subleading contribution to the chirality flipping rate.

General conclusion would be the following. Contrary to the case of the hard photon production rate, for the chirality flipping diagrams adding more photon lines connecting the fermion arcs with opposite chirality results in corrections which are subleading in the small coupling constant $\alpha$. On the other hand, if the photon lines start and finish at the same (upper or lower) fermion arc with a fixed chirality, this results in accumulating divergences which contribute to the leading-order result. The resummation of such divergences can be completely absorbed into the electron self-energy, which has been systematically investigated to the leading order in $\alpha$ elsewhere.

\end{document}